**Investigating the growth cycles of titania and carbonaceous nano dusty plasmas**

by

Bhavesh Ramkorun

A dissertation submitted to the Graduate Faculty of
Auburn University
in partial fulfillment of the
requirements for the Degree of
Doctor of Philosophy

Auburn, Alabama
August 9, 2025

Keywords: Capacitively Coupled Plasma, Monodisperse Nanoparticles, Magnetic Fields, Fast
Fourier Transform, Electron Magnetization, Axially Magnetized Ring Magnets

Copyright 2025 by Bhavesh Ramkorun

Approved by

Edward Thomas Jr., Chair, COSAM Dean and Professor of Physics
Ryan B. Comes, co-Chair, Adjunct Professor of Physics
Saikat C. Thakur, Associate Research Professor of Physics
Stuart Loch, Alumni Professor of Physics
Minseo Park, Professor of Physics
Christian R. Goldsmith, Professor of Chemistry and Biochemistry

Abstract

Understanding and controlling dust formation in plasmas — especially from gaseous chemical precursors is a key technological goal, and the overarching theme of this dissertation. Solid nanoparticles (i.e. nano dust) which range in size from $\sim$ 1 to 500 nm can spontaneously grow from reactive gaseous precursors in non-thermal plasmas. This dissertation studies the particle size, and growth time with and without a background magnetic field in the plasma. Traditionally, studies have focused on the growth of either carbonaceous or silicate dust from either acetylene or silane, respectively. However, recently, there have been a shift towards studying new kinds of dust. For example, recent studies have grown polymers and metallic dust in nonthermal plasmas. This dissertation first introduces and studies the growth of titanium dioxide (a.k.a. titania) dust from the metal-organic vapor precursor of titanium tetraisopropoxide (TTIP). The as-grown materials are of amorphous structure but high temperature annealing crystallizes the samples into anatase and subsequently rutile. Then, the growth of titania dusty plasma is and compared to the growth of carbonaceous dust from acetylene, in argon plasma. They are grown during the presence and absence of weak magnetic fields of $\sim$ 500 Gauss. Both kinds of dust growth exhibit a growth cycle, which had already been shown for various nano dusty plasma. This occurs because once the dust accumulates a critical radius and mass, they move away from the central region of the plasma, allowing a new generation of growth begins. Ultimately, the new generation of growth also moves away leading to a continuous cycle of particle formation and transport as long as the plasma is on. However, with the presence of the magnetic field, the cycle time decreases, and the spatial distribution of the dust cloud appears differently. For example, titania dust are more concentrated in the middle of the plasma where the field strength is higher but carbonaceous dust appear to move away from the high magnetic field region. Finally, we focus on the growth cycle time of carbonaceous dust and how it decreases with a gradual increase in magnetic field, varying from $\sim$ 20 - 1000

Gauss. We particularly noticed a minimum at ∼ 330 Gauss, which coincides with electron magnetization in the plasma.

To understand the physical factors contributing to observed changes in the growth cycle, we study the growth rate of carbonaceous dust, and plasma potential of the background plasma, as a function of magnetic field. The former is done via scanning electron microscope measurement of the dust size distribution throughout the first growth cycle, and the latter is done using emissive probe measurements in the plasma. Our results suggest that both the growth rate and the plasma potential of the background plasma decreases during weak magnetic fields. Therefore, we conclude that the physics of the plasma changes during the presence of the weak magnetic fields; specifically, the magnitude of the electric field decreases causing smaller dust particles to be levitated in the plasma. However, we also have initial results from optical emission spectroscopy (OES) of the dusty plasma which suggest that the chemistry governing dust formation is changing. For example, we see that the amount of time needed to reach the first maximum peak in intensity of OES increases with an increasing magnetic fields. It is possible that the nucleation and growth rate of the particles are also changing as seen by OES.

Acknowledgments

I wish to express my heartfelt gratitude to my parents, Mr. Dhannesswur and Mrs. Neya Ramkorun, whose unwavering support and encouragement have been instrumental throughout my academic journey. Despite limited financial resources, they stood by me during scholarship applications, and continued to support me as I pursued multiple degrees in the United States. I am also deeply thankful to my wife, Laura Elizabeth West, whose steadfast support helped me navigate the daily challenges of doctoral studies. Finally, I am grateful for the joy and motivation brought by our son, Joel Suresh Ramkorun, whose presence during my final semester made he completion of this journey more special.

I have had the privilege of working with several supervisors throughout my academic career, many of whom are acknowledged later in this dissertation. However, it is clear to me that Dr. Edward Thomas, Jr., my doctoral advisor, stands out as the most exceptional mentor I have ever had. His scientific insight is matched only by his kindness and compassion, qualities that have profoundly influenced both my professional and personal growth. The only regret I have about working in his laboratory is that I completed my Ph.D. so quickly (2021–2025), thereby depriving myself of additional years of learning under his guidance. The wisdom and mentorship I received from Dr. Thomas will continue to shape the course of my scientific journey for years to come.

My doctoral co-advisor, Dr. Ryan Comes, has also had a profoundly positive impact on my Ph.D. journey. I am deeply grateful for his ability to recognize and nurture the strengths in me, and for his consistent support throughout my doctoral work. The research involving titanium dioxide, which forms significant novelty in this dissertation, was inspired and guided by his expertise in the field.

As I proceed to acknowledge the members of my dissertation committee—who have continuously guided me since 2022—I would like to begin by expressing my sincere gratitude to Dr. Saikat C. Thakur. His consistent presence and mentorship in the laboratory directly contributed to several ideas and results presented in this dissertation. We are also very grateful to Dr. Minseo Park and Dr. Stuart Loch, whose thoughtful communication and guidance have been instrumental to this research. My discussions with Dr. Loch on optical emission spectroscopy and with Dr. Park on Raman spectroscopy significantly influenced their application in this project. I could not have asked for two more supportive and insightful committee members.

My day-to-day and often night-to-night research activities in the laboratory were greatly enriched by the presence of those around me, including Siddarth Bachoti, Dr. Saikat C. Thakur, Cameron Royer and Dr. Ellie Williamson. Taking breaks from research to engage with them was always a highlight, often leading to insightful and enjoyable conversations that reminded me of how much there is still to learn. For instance, I might never have appreciated Bayern Munich as a serious football club without Siddarth, learned how to assemble vacuum chambers without Cameron, understood the proper use of hiking sticks without Saikat, and developed a healthy skepticism toward computer modeling of argon plasmas without Ellie's thoughtful questioning. Their camaraderie and curiosity made the research laboratory a more vibrant and intellectually stimulating place.

Eating lunch at school became a daily highlight thanks to the regular company of either Dr. Nathan Davis, Dr. Guillaume Laurent, Paxton Wilson, Dane van Tol, Jody Davis, Scott Chumley, Jibril Ahammad, Swapneal Jain, and/or Courtney Wicklund. Without them, I might have spent all my lunch breaks alone in a quiet corner of the Leach Science Center. Instead, they made lunch more pleasant, filled with laughter, conversation, and the shared experience of trying a variety of foods, from tacos and curries to sandwiches and sushi. Their camaraderie brought warmth and community to my everyday routine.

I wish to express my sincere thanks to my master's thesis advisor, Dr. Shane Aaron Catledge, from the University of Alabama at Birmingham. Not only did he introduce me to the fields of plasma science and engineering and materials synthesis and characterization, but he also supported me selflessly when I chose to pursue my doctoral studies at a different institution. Many foundational aspects of my research, such as the use of advanced research equipment (X-ray diffraction, Raman Spectroscopy, Scanning Electron Microscope) were established in his laboratory during my time there from 2018 to 2020.

Finally, I would like to acknowledge Dr. Seth A. Smith from Vanderbilt University Medical Center, who graciously hosted me for a summer research experience in 2017, sponsored by the Berea College Office of Internships. This formative experience was pivotal in motivating my pursuit of doctoral studies. My time in his laboratory was remarkably productive; he allowed me to shadow himself, and students to collect data from magnetic resonance imaging, and entrusted me with data analysis responsibilities. During this period, I developed MATLAB scripts to process medical images and investigated trends in pediatric patients' data. Under Dr. Smith's mentorship, I co-authored my first peer-reviewed publication (Reynolds et al., 2019 [1]), contributed to several conference proceedings (SPIE Medical Imaging 2018 [2], Pediatric Blood and Cancer 2018 [3], and presented at the American Physical Society March Meeting (2018 [4]).

I gratefully acknowledge the financial support received from the National Science Foundation EPSCoR program (OIA-2148653), US Department of Energy - Plasma Science Facility (SC-0019176), and the NSF Major Research Instrumentation grant (NSF-DMR-2018794), which made this doctoral dissertation work possible.

Table of Contents

# List of Tables

# List of Figures

**Chapter 1**

**Background and Introduction**

## 1.1 Motivation

In physics and engineering, plasma are commonly used to grow materials. For example, the growth of nanomaterials from reactive chemical precursors in plasma is a common area of research. In the last three decades, nanometer to micrometer-sized particles have been grown from gases such as acetylene and silane. These particles levitate in the plasma due to a balance of several forces such as electric, gravitational, ion-drag and neutral-drag. They grow in size for an amount of time known as the growth cycle, after which the forces can no longer confine them. Then, they move away from the plasma, and a new generation of particles growth begins.

The particles and plasma can also be exposed to a background magnetic field during growth. In our laboratory, Couedel et al. (2019) grew nanoparticles in an argon (Ar)/acetylene ($C_2H_2$) plasma discharge under magnetic fields up to 2.5 Tesla (T), and showed that magnetic field strength significantly influenced nanoparticles' growth dynamics and morphology [5]. For example, at high magnetic fields, a cloud of particles were confined to the sheath above the grounded electrode, and large, porous spherical agglomerates were grown. The cloud effects were attributed to changes in the plasma electric field, as confirmed by particle-in-cell simulations. The porous nanostructures likely resulted from enhanced agglomeration driven by the modified plasma confinement due to the magnetic field.

At magnetic fields above 0.5 T, filaments formed in the plasma and aligned along the magnetic field between the electrodes. These are visible patterns that form in the plasma arising from a density imbalance between electrons and ions [6], for example, as seen in Fig. 1.1 [7].

Jaiswal et al. (2020) extended this work by examining the interaction between magnetic-field-induced filamentary structures and nanoparticle growth in a similar RF discharge [7].

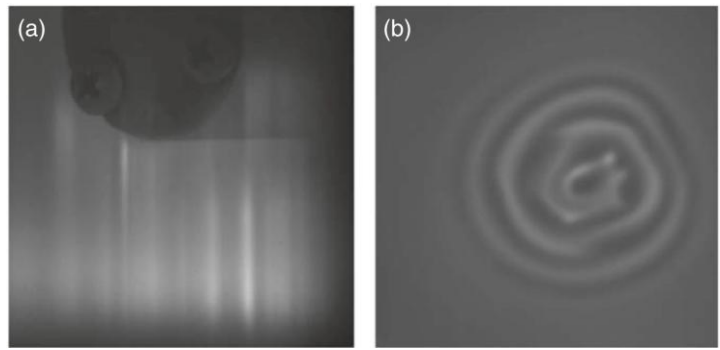

Figure 1.1: Filament structures in magnetized plasmas at Auburn University: (a) Side view in argon plasma at B = 1.76 T, and (b) Top view in argon plasma at B = 2.0 T (adapted from [7]).

They discovered that the particle growth cycle time decreases during the presence of the magnetic field. For example, Couedel et al. grew carbonaceous dust in $Ar/C_2H_2$ discharges and measured a cycle time of 40-60 seconds (s) without any magnetic fields, which was reduced to 15-20 s during fields of 0.032-0.1 T (320-1000 Gauss) [5]. The magnetic field did not seem to affect the material properties, as detected by Raman spectroscopy, with the samples consistently displaying a noncrystalline graphitic carbonaceous nature. These discoveries were also matched by other research, independent of ours, from other universities. For example, Carstensen, et al. (2010) showed that the growth cycle of carbonaceous dust in $Ar/C_2H_2$ discharges decreased from 180 s at 330 Gauss to 120 s at 670 Gauss.

The studies by Couedel and Jaiswal also revealed that while individual particles retained their spherical geometry, the particle growth dynamics and the cyclic nature of carbonaceous particle formation were significantly affected by even weak magnetic fields. This project builds on the original work of Couedel and Jaiswal, by asking **how does nanoparticle growth in a magnetized plasma modify their growth rate and confinement in the plasma?** Couedel's results suggest that for carbonaceous particles, the individual particles' shape were clearly affected. Jaiswal later showed that at magnetic fields greater than 0.5 T (5000 Gauss), the formation of filamentation in magnetized plasma appears to disrupt the particle growth cycle such that no cycle time is observed [7].

The primary goal of this work was to use an organometallic precursor—titanium iso-propoxide—to determine whether morphological differences exist in titanium dioxide nanoparticle formation compared to carbonaceous particles in both magnetized and unmagnetized plasmas. In this phase of the project, significant emphasis was placed on characterizing the material properties of the grown nanoparticles. The second part of this project focused on the interaction between the reactive carbonaceous plasma and magnetic fields varying in strength from 20 - 1000 Gauss, aiming to understand how this coupling alters the nanoparticle growth cycle.

## 1.2   Low Temperature Plasmas (LTPs)

In physics, plasma is generally known as the fourth state of matter, with the three most common states of matter being solid, liquid, and gas. Plasma is generally formed when enough energy is added to a gas to ionize a sufficient fraction (as little as $1 : 10^6$) of the atoms to allow long-range, collective effects to play a significant role in the system's dynamics. As a result, most plasmas contain a mixture of electrons, ions, and neutral atoms. Plasmas are typically characterized using the number density of electrons ($N_e$) and ions ($N_i$), the kinetic temperature of electrons ($T_e$) and ions ($T_i$), and the self-potential ($V_p = e(N_i - N_e)$), where $e$ is the electronic charge. A schematic representation of the four states of matter, i.e solid, liquid, gas, and plasma is shown in Fig. 1.2. Low Temperature Plasmas (LTPs) typically have $T_e \lesssim 10$ eV, while $T_i$ and $T_n$ are approximately equal to room temperature, $\left( \frac{1}{40} \text{ eV} \right)$. The ionization fraction is generally low, with $N_e/N_{gas} \sim 10^{-8}$ to $10^{-3}$, where $N_{gas}$ is the neutral density of the gas in the plasma.

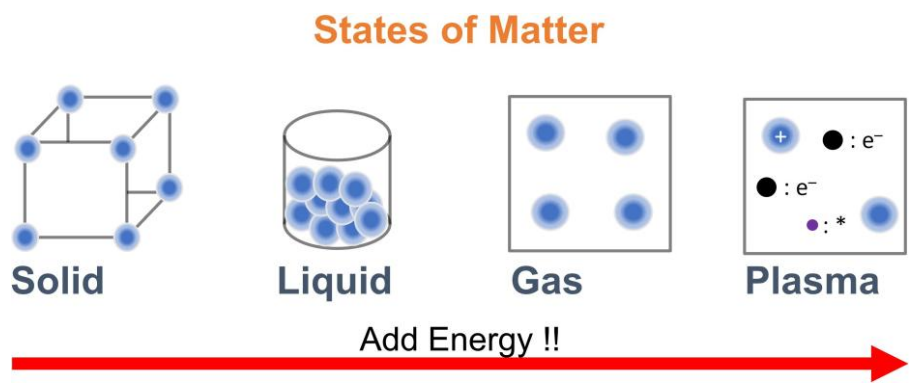

Figure 1.2: Adding energy changes solid into liquid into gas and into plasma.

When LTPs are composed of a mixture of noble and chemically active gases, the reactivity of these systems can be used to nucleate and grow materials from chemical precursors. Materials are formed through a process in which highly energetic electrons collide with gas molecules, leading to the dissociation, ionization, or excitation of the gas molecules and the production of highly reactive species. These plasmas operate far from equilibrium and provide a thermodynamic environment that favors nucleation and growth that would otherwise only occur at higher pressure and temperature regimes. **LTPs are commonly used in several areas of material growth, some of which are described below :**

1. **Plasma-enhanced chemical vapor deposition (PECVD)**

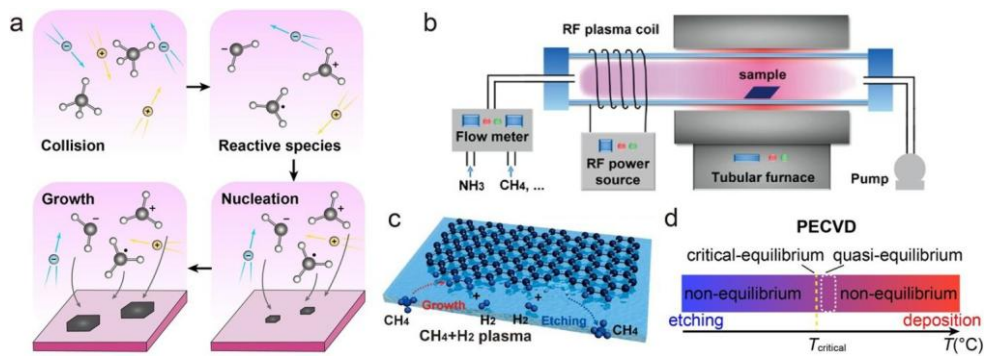

Figure 1.3: Growth of thin 2D materials via plasma enhanced chemical vapor deposition (PECVD): (a) growth mechanism; (b) PECVD system layout; (c) competing reactions in $CH_4/H_2$ plasma; (d) equilibrium states during deposition. (Adapted from [8])

PECVD is a well-known method used in industry for depositing thin films, as schematically depicted in Fig. 1.3. In some CVD processes, especially for materials like graphene, catalysts are needed to help generate enough reactive species and to keep the growth process under control [8]. PECVD works differently — it uses plasma instead of high temperature to break down the precursors. This means energy is transferred through charged particles, which allows the reaction to happen even at room temperature. Because of its efficiency, PECVD often does not need catalysts, making it easier to grow materials directly on different kinds of surfaces without needing complicated transfer steps. The electrons in the plasma can be used to bombard a substrate by negatively biasing the latter, in order to promote biased enhanced nucleation [9, 10].

2. **Pulsed Laser Deposition (PLD)**

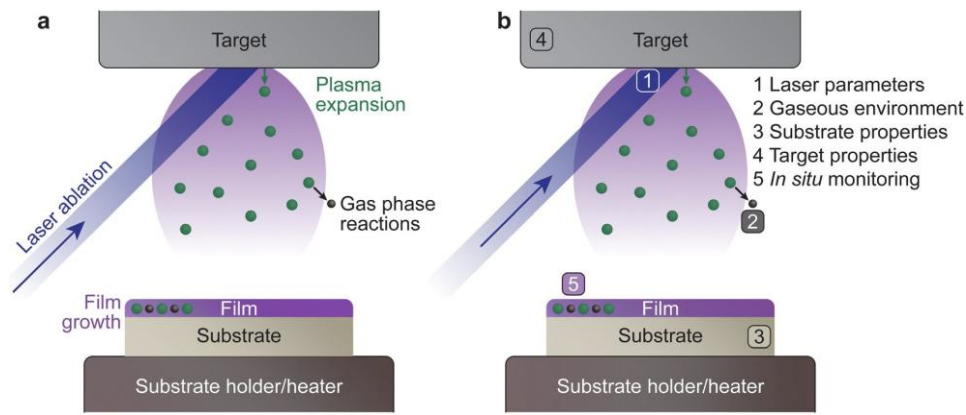

Figure 1.4: An overview of plasma in pulsed laser deposition (PLD). (a) A pulsed laser ablates the target, forming a plasma of energetic species that travel to the substrate, and (b) several experimental parameters can be adjusted to control film growth (adapted from [11]).

In PLD, a plasma forms when laser ablates a target, as shown in Fig. 1.4. This creates a vapor that absorbs energy through photoionization and inverse Bremsstrahlung. These processes generate free electrons, which increases the plasma ionization. As vapor density rises, collisions increases which expands the plasma, thus transitioning it from adiabatic expansion and free flight, depending on background pressure. During early expansion, lighter particles like electrons move ahead, while heavier ones lag, creating an electrostatic double layer. This layer accelerates ions and influences how many species reach the substrate. Upon arrival, these species can physisorb and later chemisorb onto the surface of the substrate to grow a thin film [11].

3. **Atomic Layer Deposition (ALD)**

In ALD, plasma forms when energetic electrons ionize gas molecules, creating reactive radicals and ions, several examples of which are shown in Fig 1.5. While radicals mainly drive surface reactions, ions gain energy in the plasma sheath due to electric fields caused by faster-moving electrons. This accelerates ions toward surfaces, enhancing reactions through bombardment. Ion energy depends on pressure and sheath thickness, with lower pressures enabling greater acceleration. Plasma in ALD offers three main benefits [13]:

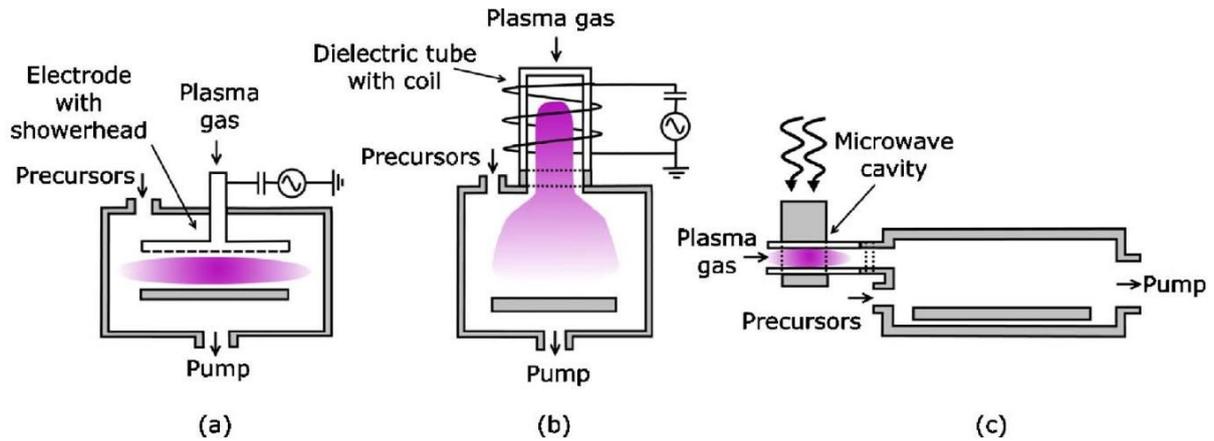

Figure 1.5: A schematic illustrates the three main types of plasma-assisted atomic layer deposition (ALD) (adapted from [12]): (a) direct plasma, (b) remote plasma, and (c) radical-enhanced. Each type involves distinct hardware setups and plasma sources.

(1) It generates highly reactive species in the gas phase, independent of substrate conditions, with tunable chemistry via gas, power, and pressure adjustments. (2) It delivers minimal heat to the surface, as only electrons are hot and exposure is brief. (3) Ion bombardment adds localized energy, enhancing reactions and diffusion, and can be controlled through plasma and substrate settings.

## 1.3 Dusty Plasma

A dusty plasma is a four component system consisting of electrons, ions, neutrals, and nanometer (nm) to micrometer (μm) sized charged solids, i.e., dust. All the charged species interact with each other and the background neutrals to form a complex, coupled system [14–16].

Since the charged dust particles are significantly more massive than the other species, these systems can exhibit a broad range of phenomena such as self-organization [17–20], clustering [21–23], and waves phenomena [24, 25]. Moreover, the large size of the dust particle enables direct visualization of these phenomena in the laboratory [26, 27], as illustrated in Fig. 1.6.

The study of dusty plasmas is important due to its wide range of scientific applications. Dusty plasmas can be formed in a variety of natural, and laboratory environments. They are commonly found in space environments like the lunar surface [28], planetary rings [29–31], and interstellar clouds [32]. Dusty plasmas also occur in industrial settings such as semiconductor manufacturing [33, 34] and fusion devices [35]. There is a need to understand the dynamics of the dust in the plasma, both in space and in the laboratory, to mitigate its devastating effects such as damage to astronauts' suits and lunar vehicles, and contamination in semiconductor processing and fusion experiments.

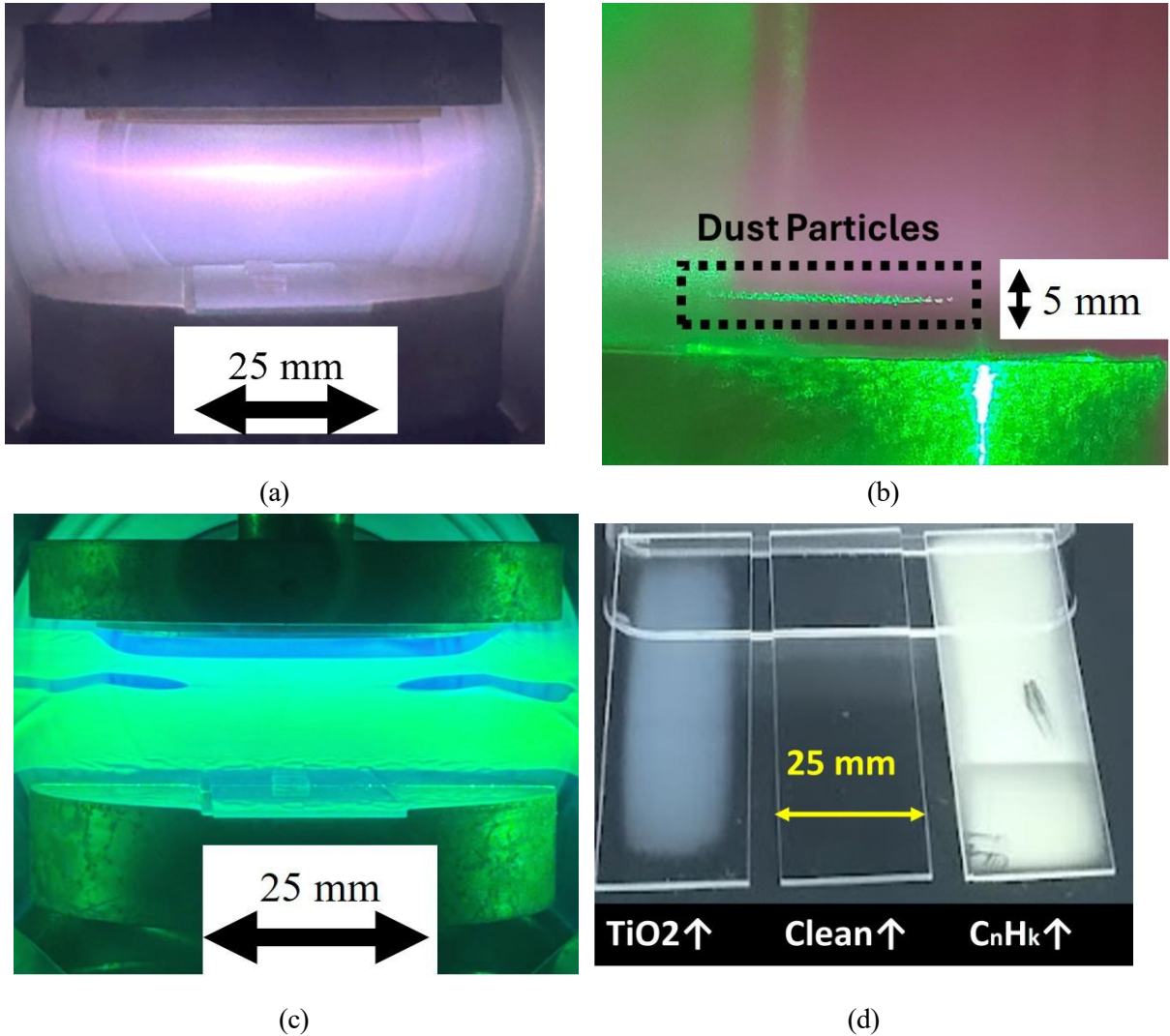

Figure 1.6: Example images of several laboratory plasma experiments to show (a) clean argon plasma, (b) a dust layer of micron-sized dust introduced in the plasma, (c) a cloud of nanometer-sized dust particles generated due to nucleation from reactive precursor gases, and (d) nano dust collected on fused silica for eventual material characterization.

Plasma processing is vital for material synthesis via some of the aforementioned techniques. To prevent wafer contamination, clean rooms and advanced gas handling systems are used, yet particle formation and trapping remain challenging. This issue, long recognized in PECVD for thin-film deposition, arises from the clustering of active species at high concentrations, which can lead to particle growth and electrostatic trapping. Excessive deposition rates can cause "snowing" in reactors, degrading film quality [36]. Particle contamination also affects other microelectronics processes like etching and sputtering. These particles may originate from process gases or electrode materials and are typically detected via light scattering, though only larger particles are visible this way. Smaller particles (10 nm) require ex situ methods like electron microscopy.

## 1.4 Nano Dusty Plasma

Many dusty plasma experiments are performed by directly introducing the dust particles into the plasma [37–39]. However, if reactive gas(es) is(are) used to generate the plasma, chemical processes can lead to the spontaneous nucleation and growth of dust particles directly from the gas phase. These dusts are monodispersed nanospheres that linearly increase in size while in the plasma. They usually range in radius from approximately 1 to 300 nm. They levitate and form a nano dusty plasma. The background plasma is usually ignited from an inert gas such as argon (Ar). Figure 1.6 shows several laboratory experiments at Auburn University in order to compare a clean argon plasma 1.6a, with dust introduced into the plasma 1.6b, a nano dusty plasma created from reactive precursor gases 1.6c, and the nano dust collected on fused silica slides after the experiment is turned off 1.6d.

Over the past three decades, different kinds of particle growth have been reported such as silicates [40], carbonaceous [41, 42], polyaniline [43] and organosilicon [44]. Extensive research has explored the spontaneous growth of particles from reactive gases in capacitively coupled plasmas (CCP), with a focus on generating either hydrocarbon or silicate dust, commonly from acetylene ($C_2H_2$) or silane ($SiH_4$) precursors, respectively [41, 42, 45–47]. In certain investigations, these particles have been collected and undergone detailed characterization using techniques such as deuteron-beam induced gamma-ray emission [42], near-edge x-ray absorption fine structure spectroscopy (NEXAFS) [45], and Raman spectroscopy [5]. Furthermore, scanning electron microscopy (SEM) [41] and transmission electron microscopy (TEM) [40] have consistently revealed that these particles exhibit spherical morphology and their radii exhibited linear growth characteristics over time. Dusty plasma growth processes offer precise control over particles' size and morphology [48–52].

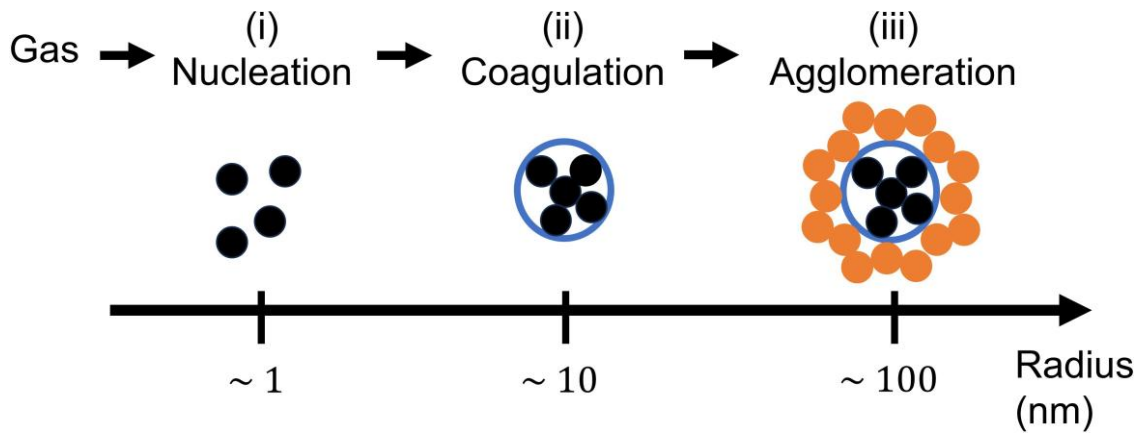

Figure 1.7: (Schematics not drawn to scale) The particle growth processes from the reactive gas: (i) Nucleation, (ii) Coagulation, and (iii) Agglomeration. After the first generation of growth ends, a new one begins. The particles continuously grow in cycles as long as the plasma is turned on [53].

Dusty plasma particle growth happens in three steps, (i) nucleation, (ii) coagulation, and (iii) agglomeration (accretion) [54–57]. The particles' size in the three steps are in the order of 1 nm, 10 nm and 500 nm, respectively. A general schematic figure summarizing the process is shown in Fig. 1.7. However, a more detailed description of each individual growth process is described below:

**Nucleation**

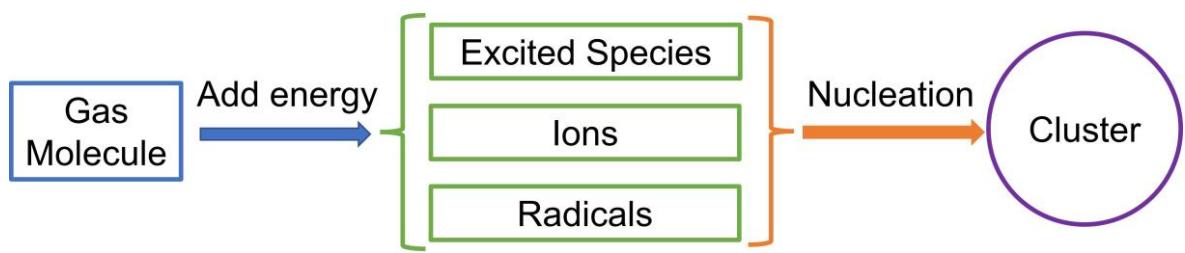

Figure 1.8: Schematics representation (not to scale) of the **nucleation** of clusters from from chemical reactions between neutrals, ions, radicals, and neutrals trapped in the plasma.

Energetic electrons from the background plasma constantly bombard the gas molecules to create charged and excited species. Charged species such as ions and radicals are formed by dissociative attachment of the gas in the plasma. Anions remain trapped in the bulk plasma whereas cations are attracted to the chamber walls and plasma sheaths. The negative ions react successively with molecules from the gas. Due to these ion-molecular reactions, clusters are nucleated [58]. This is schematically represented in Fig. 1.8. The nucleation process is thermodynamically favorable and occurs much more quickly than ion-ion recombination [54]. The clusters have diameters in the order of 1 nm. These clusters have stochastic charge fluctuations i.e, q = ±1 to 3e [59–61].

**Coagulation**

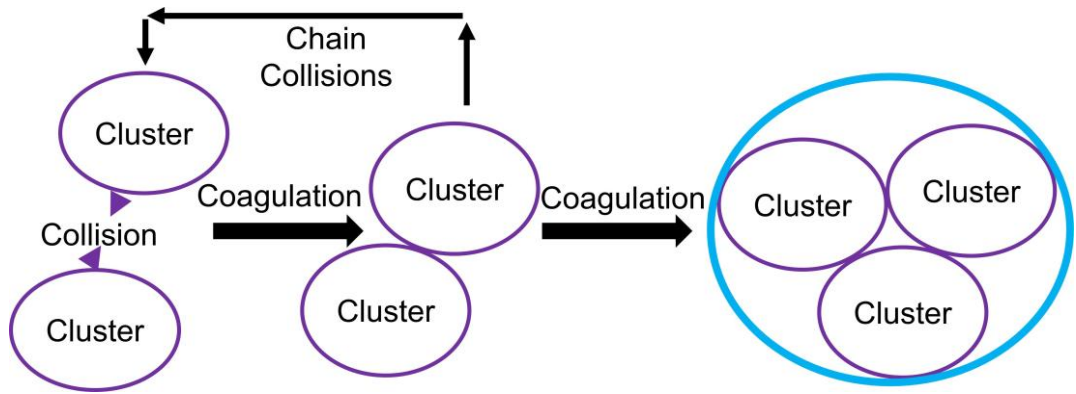

Figure 1.9: Schematics representation (not to scale) of the collision of clusters for **coagulation** into bigger clusters.

During coagulation, clusters and neutrals collide and react to form bigger particles [62]. This is schematically rerepsented in Fig. 1.9. Coagulation is dominated by chemisorption [63]. Chemisorption is the process by which strong chemical bonds are formed between the clusters as they collide with each other, resulting in the formation of the nanoparticle's core. As the concentration of clusters in the plasma increases, their mean free path decreases. This leads to cluster-cluster collision whereby some stick together. Coagulation occurs much faster than the reaction to grow new particles. Hence, new dust formation is suppressed. As the size of the particles increase, the rate of coagulation reaches a saturation [64]. At this stage, the particles have accumulated a net negative charge due to collisions with electrons and negative ions. Eventually, coagulation stops due to coulombic repulsion. The diameters are in the order of 10 nm.

**Agglomeration**

During agglomeration, also known as accretion or surface growth, new growth occurs on the surface of the negatively charged particles due to positive ions, radicals and neutrals colliding on its surface [65–67]. These clusters act like substrate onto which new surface growth happens. This is schematically represented in Fig. 1.10. Agglomeration is dominated by physisorption [63]. The concentration of particles in the plasma decreases. The clusters eventually agglomerate into larger dusty microparticles that have diameters around 500 nm. Eventually, the dust particles accumulate enough charge and mass so that they move away from the plasma.

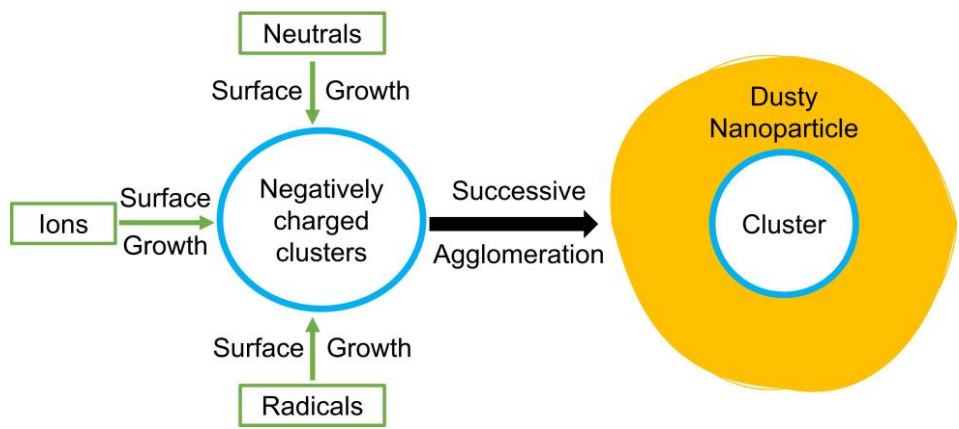

(a)

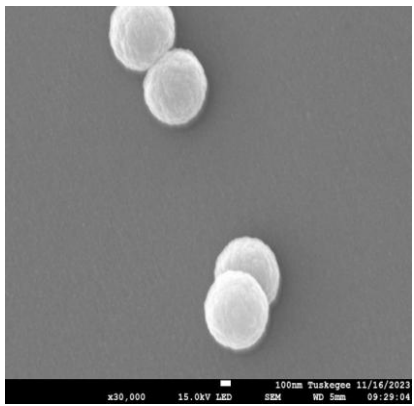

(b)

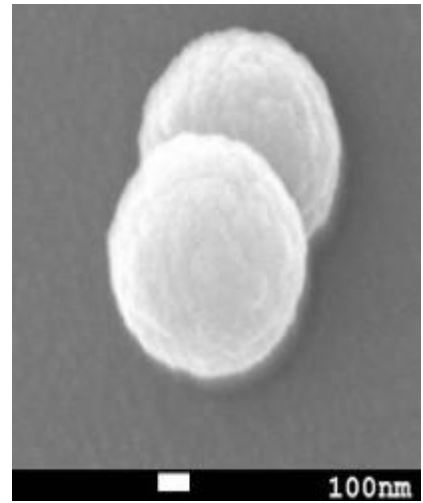

(c)

Figure 1.10: The agglomeration phase: (a) Schematic representation of **agglomeration**. Negatively charged clusters from coagulation repel positive ions and radicals due to electrostatic attraction. (b) SEM images of dusty nanoparticles after several seconds of agglomeration. (c) Magnified view of the particles revealing cauliflower shapes due to surface growth.

### 1.4.1 Growth Cycle

During growth, the particles form a dust cloud, as shown in Fig. 1.11a, that is levitated in the plasma due to a balance of several forces such as gravitational, electric, thermophoretic, ion drag, and neutral drag [68–71]. The forces' magnitudes depend on the radius of the particles, with gravitational being cubic, electric being linear, and the others being quadratic. Detailed description of these forces are found in section 3.3 of this dissertation. Initially, when the particles nucleate and start to grow, gravitational force is negligible and electric and ion drag forces are the dominant forces. As the particles further grow in size ($>100$ nm), the gravitational force can become dominant [71]. During the agglomeration phase, the particles accumulate enough mass and size so that gravitational and ion drag forces cause the particles to move away from the bulk plasma towards the chamber's wall, thus creating a "dust-free" region in the cloud, commonly referred to as a "void", as shown in Fig. 1.11b. The amount of time between nucleation and the void expanding is known as a growth cycle. A new cycle of growth immediately begins in the void, as shown in Fig. 1.11c, and similarly reported in other studies [7, 72, 73]. It is noted that gravity is not a requirement for the opening of the void and therefore for the cyclic particle growth, as cyclic particle growth has been observed even under microgravity conditions [74].

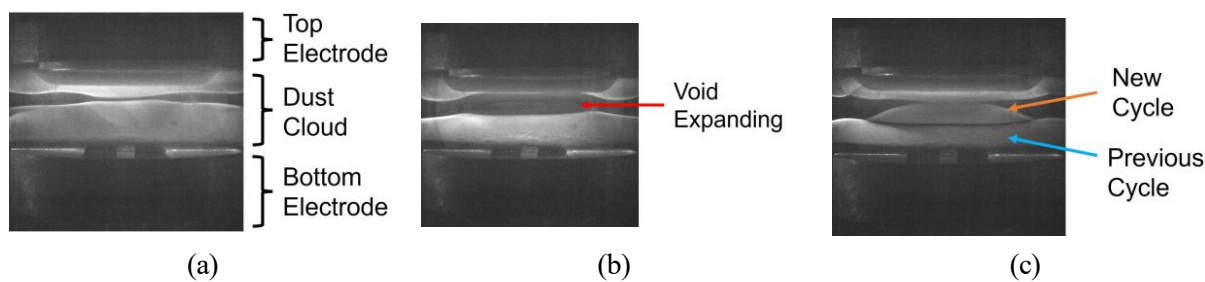

(a)  (b)  (c)

Figure 1.11: An example of the evolution of the dust cloud during titania particle growth in Ar/TTIP dusty plasma without magnetic field. (a) Dust cloud levitates in the plasma at 40 s. (b) Void starts to expand, signalling the end of a growth cycle at 71 s. (c) New growth cycle has started in the void. The previous cycle is gradually moving out of the plasma at 92 s. (Adapted from Ramkorun et al. [53].)

In capacitively coupled radiofrequency (rf) plasmas, particles have grown from different reactive gases, but the timing of their growth cycles have varied between experiments. The cycle time in plasma is influenced by various factors, including plasma conditions and the concentration of reactive gases [75]. Nonetheless, a consistent pattern emerges in each case: particles undergo a growth cycle, gradually increasing in size and mass before ultimately losing their electrostatic balance and becoming unconfined. A summary of several dusty plasma experiment, their cycle time and measurement method for determining cycle is summarized in Table 1.1

Table 1.1: Summary of growth cycles in various Ar-based dusty plasma experiments.

| Gas Mixture | Electrode Diameter | Electrode Spacing | RF Power (W) | Cycle Length | Measurement Method |
|---|---|---|---|---|---|
| Argon-Acetylene | 30 cm | 8 cm | 10–20 | 35 min | Rayleigh/Mie scattering of IR signals [41] |
| Argon-Acetylene | 75 mm | 25 mm | 5–30 | 55 s | CMOS camera imaging [5, 7] |
| Argon-Acetylene | 10 cm | 4 cm | 20 | 2 min | Scattered laser beam intensity [76] |
| Argon-Methane | 10 cm | 4 cm | 60 | 15 min | Scattered laser beam intensity [76] |
| Argon-Hexamethyl-disiloxane | 10 cm (top), 12 cm (bottom) | $3.5 - 4$ cm | 30 | 150 s | Mass spectrometry, OES, $T_e$, $N_e$ [77–79] |

### 1.4.2 Carbonaceous Nano Dusty Plasma

Carbonaceous nano dust has been grown and collected for material characterization in many previous studies. Prior to this dissertation, Couedel et. al studied Raman spectra of particles grown under magnetic fields of up to $\sim 2.5$ T [5]. The raw and samples' spectra are shown in Fig. 1.12. The raw data (a) shows a strong photoluminescence background and faint vibrational features. After background removal (b) two broad peaks, with the main G band between 1570 and 1600 $cm^{-1}$ and the D band around 1350 $cm^{-1}$. Graphitic carbon, of $sp^2$-hybridized carbon atoms, have two Raman active modes corresponding to $2E_{2g}$ arising from stretching of the C-C bond and their peak occur around 1570 $cm^{-1}$ [80]. D band arises from defects in the cabon samples [81]. For example, one such defect could be from the presence of diamond-like ($sp^3$-hybridized) carbon, which has also been detected by other groups studying carbonaceous dust grown in Ar/$C_2H_2$ plasma [45]. Additional peaks near 1200 cm$^{-1}$ and 1800-1900 cm$^{-1}$ may correspond to oxygen-containing bonds, though their origin is uncertain due to air exposure and substrate interference. The broad, overlapping nature of the bands makes precise peak identification difficult. To avoid model bias, spectra from reference materials like graphite (Gr.) and amorphous carbon (a-C:H) were used for comparison in (c), which also plots the G-band position against its FWHM, showing that the samples fall within the amorphous carbon region.

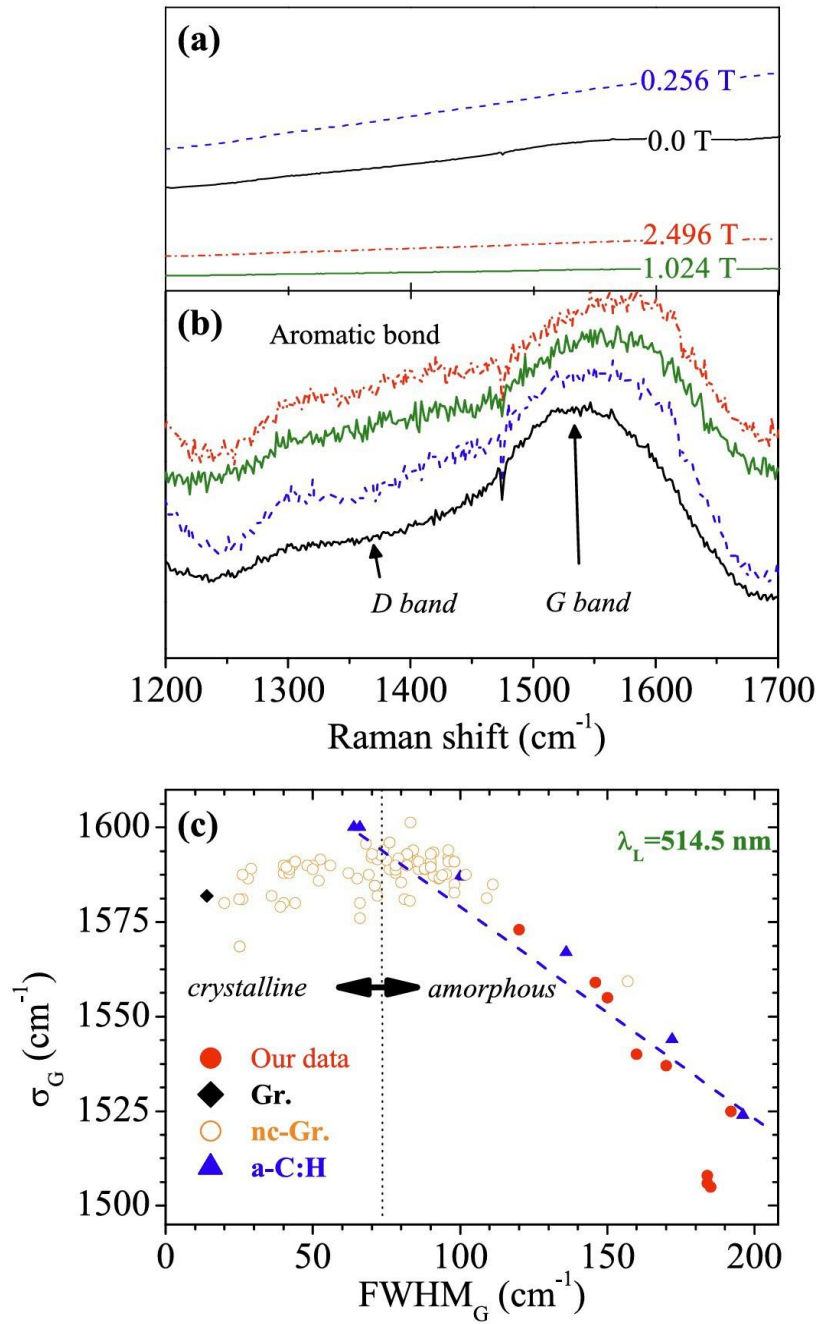

Figure 1.12: Raman spectra of carbonaceous dust particles (adapted from [5]). G band arises from graphitic ($sp^2$-hybridized) carbon, and D band arises from defects in the samples [80]. (a) Raw spectra (unprocessed). (b) Spectra after linear background subtraction and normalization to the G-band peak intensity. (c) Displays the G-band peak position ($\sigma_G$) plotted against its full width at half maximum (FWHM$_G$).

## 1.5   Overview of subsequent sections

This survey of previous studies shows that numerous questions remain about the formation of nanoparticles in reactive plasmas. The work presented in this dissertation will summarize three main experiments that were performed to investigate the growth of metallic and carbonaceous particles in plasmas with and without magnetic fields. This work is presented in this dissertation in the following order: **Chapter 2** describes the experimental setup used in this study. This includes detailed descriptions of the vacuum chamber, plasma source, plasma diagnostic tools, and materials characterization techniques. **Chapter 3** presents the theoretical background of dusty plasmas as it pertains to this experiment. It discusses the various forces necessary to levitate nanoparticles and introduces the electron Hall parameter.

**Chapter 4** will discuss the overall experimental results. First we show the growth of titania nano-dusty plasma from the organometallic precursor TTIP. The resulting particles are collected and characterized using scanning electron microscopy (SEM), X-ray diffraction (XRD), and Raman spectroscopy. The particle growth cycle is determined from intensity variations in camera images of the dusty plasma and from optical emission spectroscopy (OES) of the Argon I line at 763.5 nm. Second, we compare the growth of titania and carbonaceous dusty plasmas under the influence of a weak magnetic field of 500 Gauss. The magnetic field reduces the cycle time in both types of dusty plasma. The cycle time is determined from intensity variations in camera images and OES data. Particle size distributions from the first two growth cycles are analyzed using SEM. Third, we examines the changes in cycle time of carbonaceous dusty particles as the magnetic field increases from 18 Gauss to 1020 Gauss. It is observed that the cycle time reaches a minimum at the electron Hall parameter and remains at this minimum for higher magnetic fields. The nanoparticle growth rate, as determined from SEM images, also exhibits a minimum at the electron Hall parameter. These findings suggest that both the physical and chemical properties of the plasma evolve with increasing magnetic field strength.

Finally, **Chapter 5** summarizes the main findings of this dissertation and outlines potential directions for future research inspired by this work.

# Chapter 2

# Experimental Set Ups

This chapter discusses the primary experimental setup and diagnostics that are used throughout this work. Section 2.1 will begin with a discussion of the vacuum chamber / plasma source where experiments are performed. We will also discuss the chambers' accesories to measure pressure, pump down to vacuum and flow gases. Because the presence of a magnetic field is critical to this work, the mounting system used to place permanent magnets into the chamber is described in Sec. 2.2. Secs 2.3 to 2.6 will discuss the various diagnostics that are used for both non-invasive and in-situ measurements of the plasma as well as the material properties of the grown nanoparticles. Sec. 2.3. will discuss how laser light scattering measurements are used to detect the growing nanoparticles and identify the cyclical growth cycles that are the focus of much of this work. Sec. 2.5 will discuss the in-situ measurements made using Langmuir and emissive probes, respectively. Sec. 2.6 will discuss the materials characterization after growth and collection.

## 2.1   Vacuum Chamber Assembly

All the particle growth experiments in this study use a capacitively coupled plasma (CCP) driven by a radio frequency (RF) power supply to create a nonthermal plasma with electron temperature and density of approximately $2 - 5$ eV and $10^{15}$ m$^{-3}$, respectively. This kind of experiment and plasma is widely used for the large-scale synthesis of materials [82]. A general overview of the vacuum chamber for the purpose of dusty plasma experiment within is shown in Fig. 2.1. The vacuum chamber was a 6-way ISO-100 stainless steel cross vacuum chamber housing a pair of 75 mm diameter aluminum electrodes in a parallel plate configuration, separated by 25 mm to generate a capacitively coupled RF plasma. The base pressure of the chamber was maintained at 2–3 mTorr. An isolation valve was partially closed between the chamber and vacuum pump to limit the conductance of the pump and adjust the chamber pressure to 300 - 500 mTorr. This assembly is then slightly modified for particle growth experiment by adding different precursor gases, changing RF power, and addition of magnets to create varying magnetic fields. The electrodes' design are shown in Appendix A.

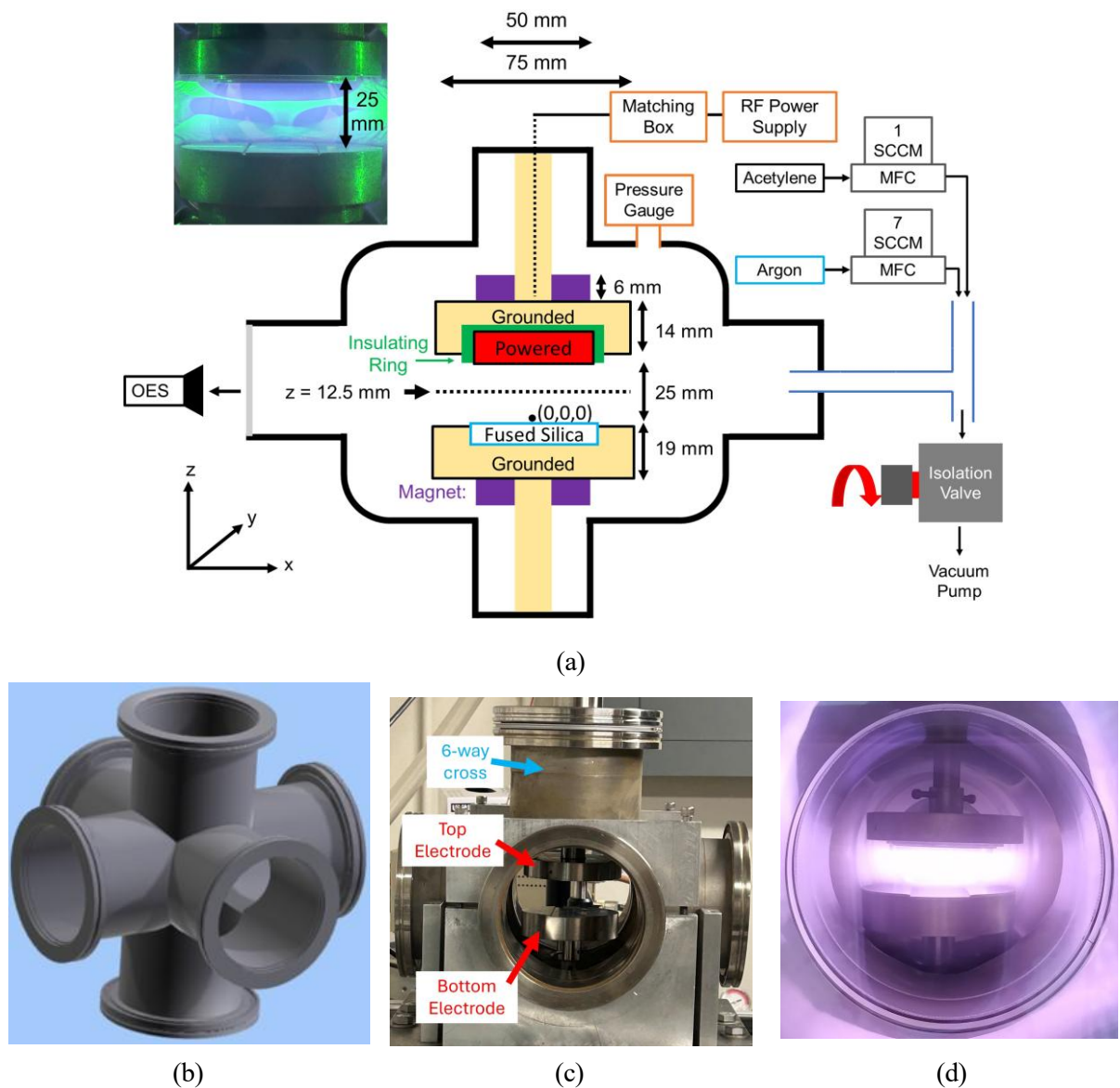

(a)

(b)                    (c)                    (d)

Figure 2.1: General assembly of the experimental chamber. (a) Schematic (not to scale) of the experimental layout (adapted from [83]), (b) Example 6-way cross from K.J. Lesker, (c) Electrodes fitted in our 6-way cross chamber, (d) A clean argon LTP between the electrodes.

Consider Fig. 2.2 to see a real image of a dusty plasma between the electrodes. Precursor reactive gases are used together with argon. Starting at a base pressure of 3 mTorr, for titania dust growth, TTIP vapor is introduced which raises the pressure to $35 \pm 3$ mTorr, while argon (Ar) gas was injected at 7 sccm, increasing the pressure to $45 \pm 3$ mTorr. The TTIP line gas line was heated to $75$ °C using a proportional–integral–derivative controller (PID) temperature controller (Inkbird) equipped with a Solid State Relay to maintain the temperature constant. For carbonaceous dust growth, the TTIP line was replaced with a $C_2H_2$ line regulated by a mass flow controller at 1 sccm, also achieving $35 \pm 1$ mTorr. Plasma ignition occurred at 60 W with minimal reflected power (1 W), and the configuration, including electrode placement and internal magnets, was optimized for dusty plasma formation and subsequent materials collection for characterization.

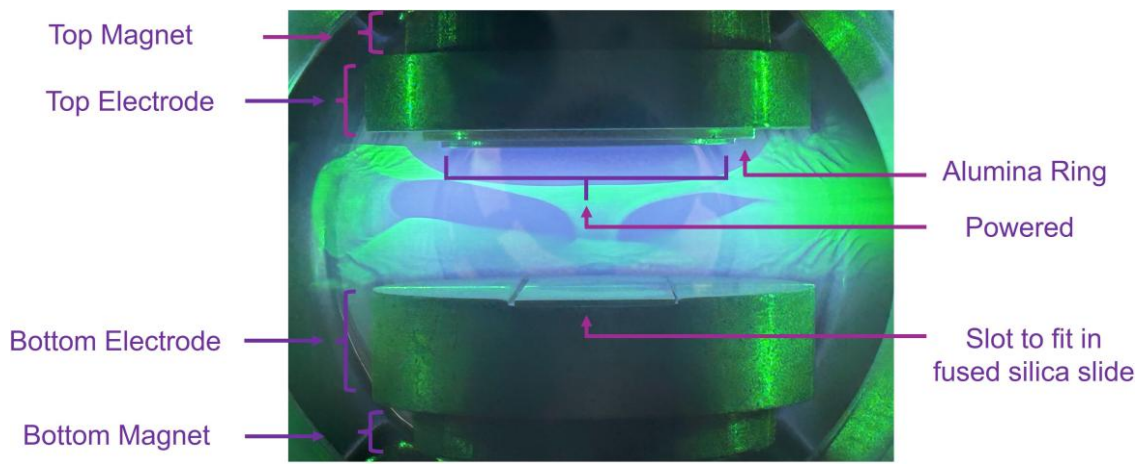

Figure 2.2: Side view of a dusty plasma between the two electrodes. Magnets are placed either above or below the bottom electrode to create a field strength of ∼ 500 Gauss in the middle of the plasma. The ground part of the top electrode was insulated from the powered by an alumina ring. The bottom electrode was always grounded and had a slot to fit in fused silica for dust collection.

More details about the electrodes figuration is shown in Fig. 2.3. The top electrode (Fig. 2.3a), was 14 mm thick, had a 50 mm powered region driven by a 300 W, 13.56 MHz RF generator (RF VII, Inc RF-3-XIII) and an auto-matching network (AIM-5/10), while the remaining portion was grounded and surrounded by a grounded counter ring (50 mm inner, 75 mm outer diameter). Power was supplied via a BNC feedthrough as indicated in Fig. 2.3b. The bottom electrode (Fig. 2.3c) was 19 mm thick and grounded, had a 1 mm deep slot to hold a $75\,mm \times 25\,mm$ fused silica slide for nanoparticle collection. The rod used to hold the bottom electrode inside the chamber was capped with a vacuum blank as shown in Fig. 2.3d.

In order to assist the experiment, the remaining four ports of the 6-way cross were used to (i) send gases in and out of the chamber, (ii) to send a laser sheet into the dust cloud, (iii) to image the dust cloud and (iv) to record optical emission spectroscopic data of the background plasma. These are all illustrated schematically in Fig. 2.4.

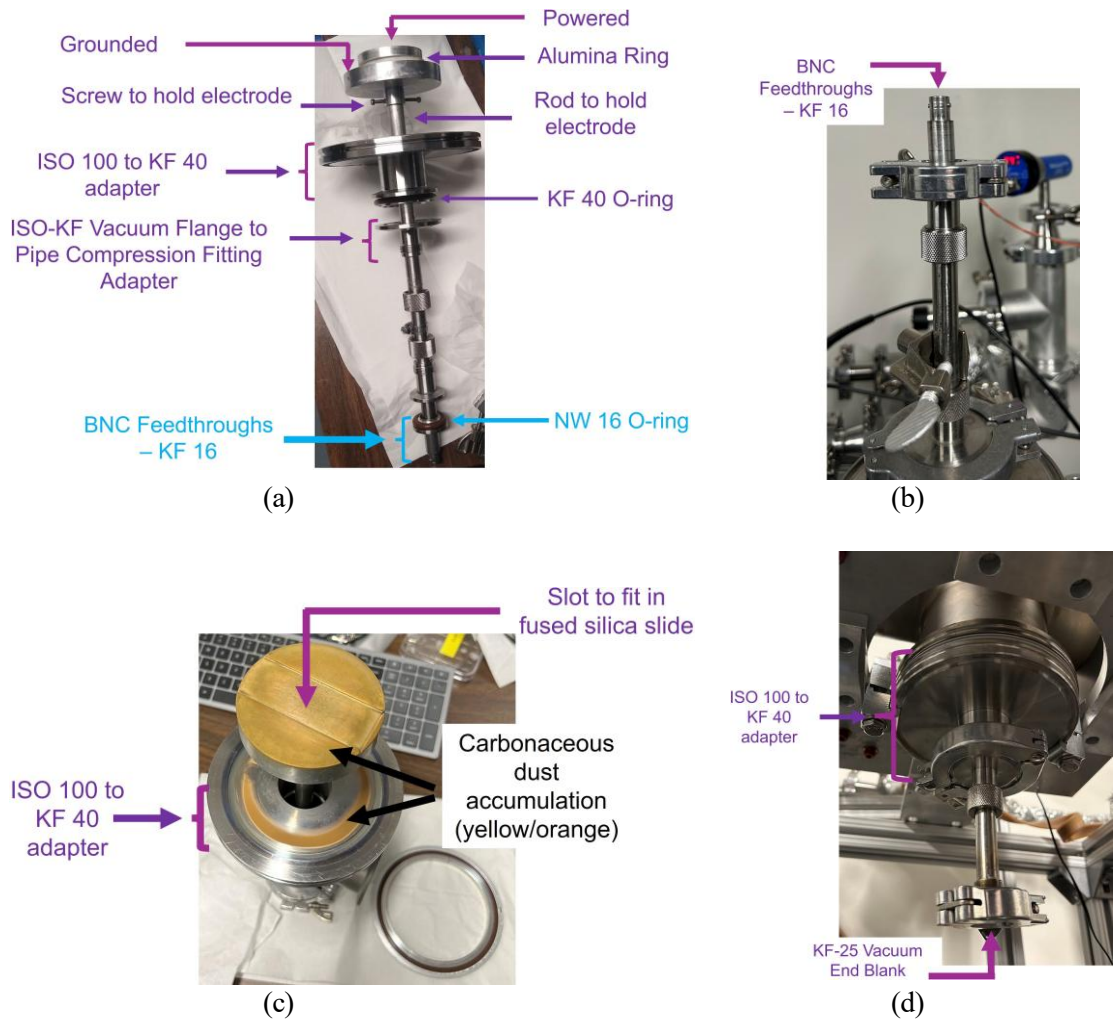

Figure 2.3: Detailed electrodes setup: (a) view of top electrode assembly, (b) BNC feedthrough at the top of the rod holding the elecrode, (c) view of the bottom electrode assembly. On the face of the electrode and on the flange, the brown color arises from carbonaceous nanoparticles that have been deposited on the plasma facing surfaces., and (e) vacuum blank at the bottom of the rod holding the electrode.

### 2.1.1 Pressure Gauge

A Kurt J. Lesker Company® (KJLC®) 275i Series convection vacuum gauge, as shown in Fig. 2.5a, is used to monitor the chamber's pressure. The pressure data can be sent to a computer via a Data Acquisition (DAQ) system, as shown in Fig. 2.5b. A LabVIEW code was written to collect the pressure measurement as shown in Appendix B.

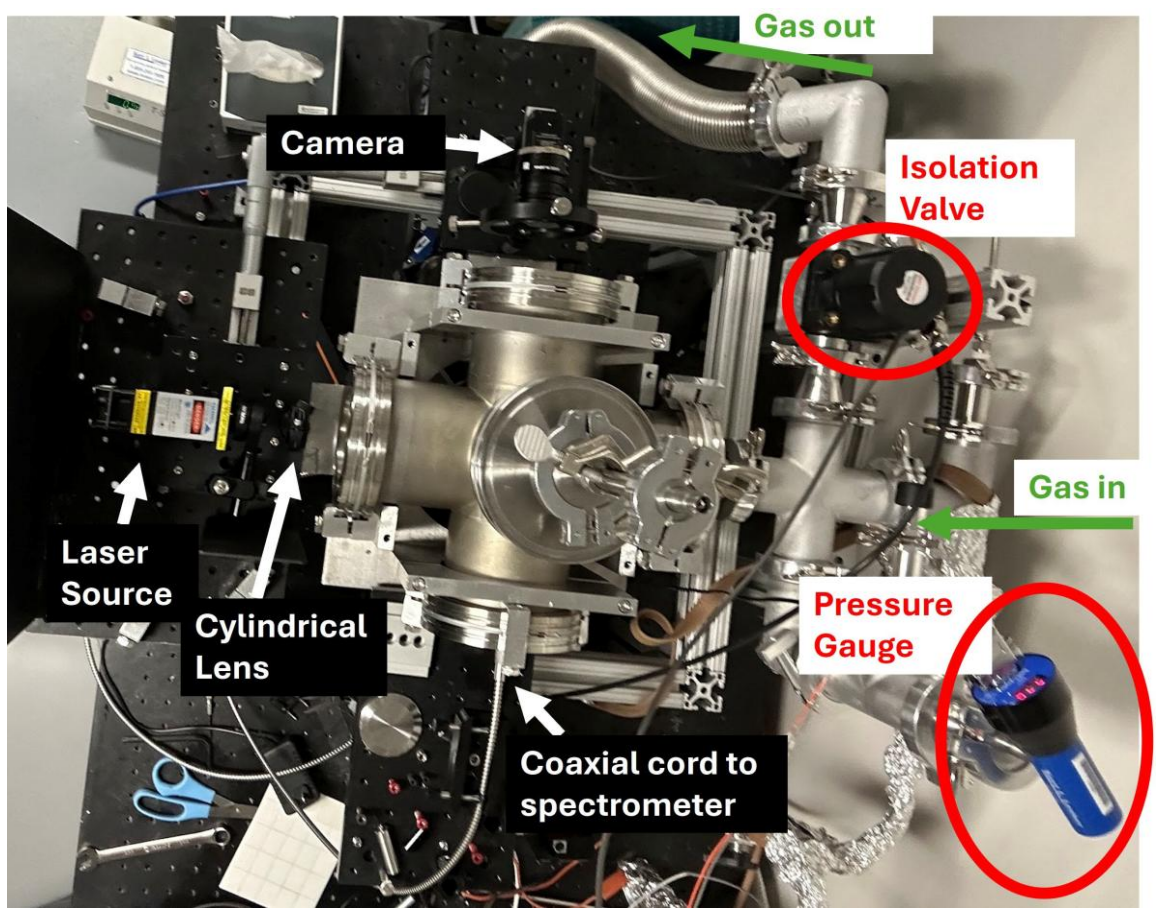

(a)

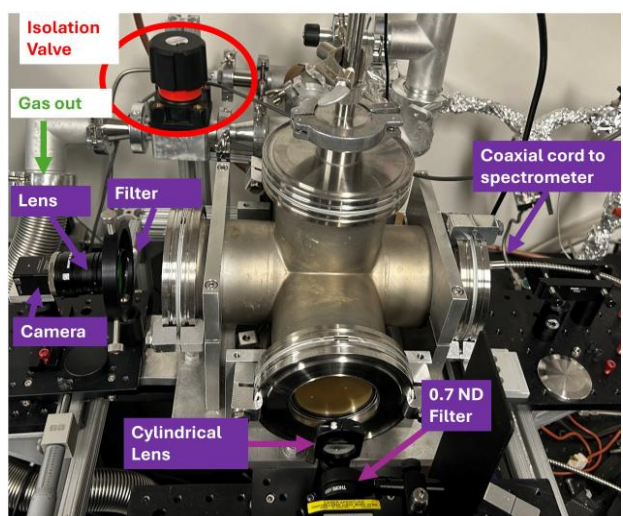

(b)

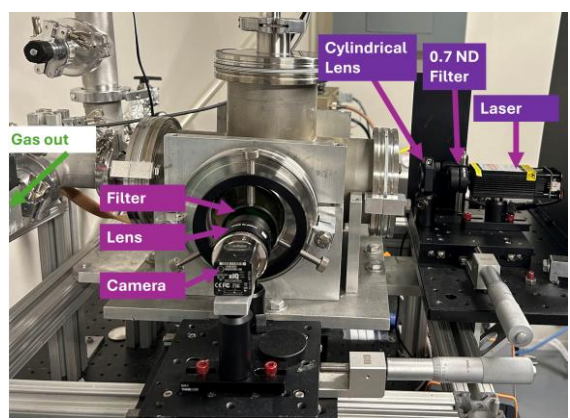

(c)

Figure 2.4: Detailed view of the experimental chamber assembly: (a) Top view showing camera, laser, cylindrical lens, pressure gauge, isolation valve; (b) and (c) are side views showing a 0.7 ND filter in front of the laser, the camera lens, and the camera filter.

## 2.1.2 Vacuum Pump

Three distinct vacuum pumps are employed throughout the experimental procedures to accommodate different operational requirements and plasma chemistries. Either a turbo pump or one

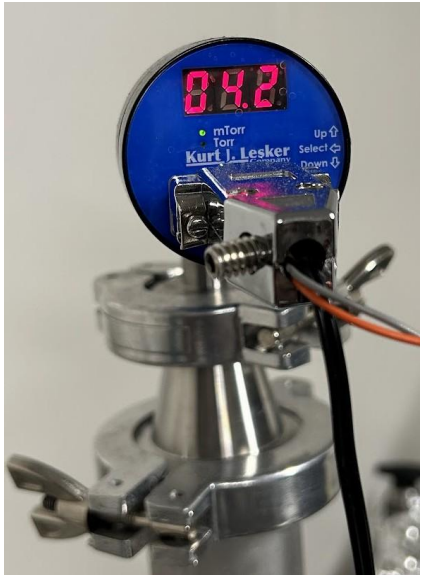
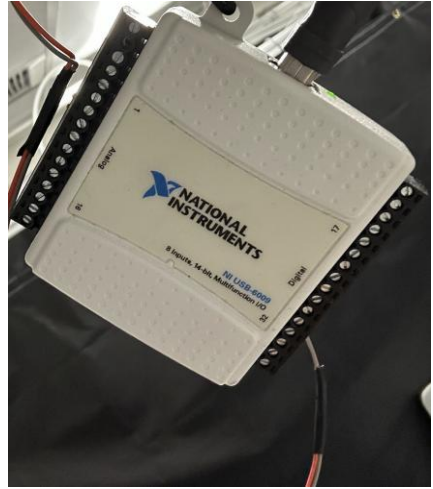

<div align="center">(a)                    (b)</div>

Figure 2.5: Equipments used to monitor chamber's pressure during experiment. (a) Pressure gauge and (b) data acquisition (DAQ) system.

of two rotary vane pumps are used, as shematically shown in Fig. 2.6. To maintain a base pressure below 1 mTorr, an Edwards model T-Station 75 integrated turbopumping stack, rated at 42 L/s, is used. This pump operates when no dusty plasma is required and creates a high-vacuum system (< 1 milliTorr) in the vacuum chamber, helping to minimize contamination from ambient gases and ensuring a clean environment for subsequent experiments. However, it is noted that over several weeks of dusty plasma operations, particles accumulate within the chamber, requiring a full disassembly and cleaning to maintain vacuum integrity and experimental consistency. After re-assembling the chamber, leak detection is performed by spraying methanol around vacuum seals and observing pressure fluctuations on the pressure gauge.

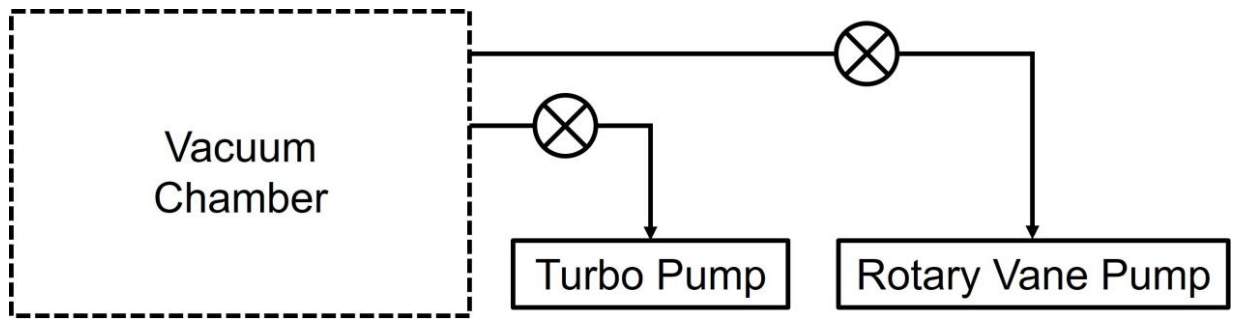

Figure 2.6: Schematic of vacuum chamber connection (not to scale)

During chemically active plasma experiments, the turbopump is closed off to prevent damage from reactive species and particulates. Instead, a rotary vane roughing pump is used to maintain operational pressure. For carbonaceous dusty plasma experiments, a KJLC RV206 rotary vane pump is sufficient. However, when working with organometallic precursors, such as in titania experiments, a more chemically robust solution is required. In these cases, an Edwards model RV8 rotary vane chemical pump is used to handle corrosive gases without compromising performance. Each pump is selected based on the specific plasma chemistry and cleanliness requirements.

### 2.1.3   Precursor Gases

#### Argon

Argon gas is used to ingite the plasma. A gas tank and regulator (Harris Model KH1139) is used to flow the gas into a mass flow controller (Alicat MCE Series Controller) and subsequently into the vacuum chamber.

#### Acetylene ($C_2H_2$)

Acetylene gas (see molecule in Fig. 2.7) is used to grow carbonaceous dusty nanoparticles in the plasma. A gas tank and regulator (Harris Model 9296NC) is used to flow the gas into a mass flow controller (Alicat MCE Series Controller) and subsequently into the vacuum chamber.

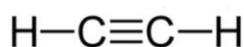

Figure 2.7: Acetylene molecule.

#### Titanium Isopropoxide (TTIP) ($Ti\{OCH(CH_3)_2\}_4$)

Liquid Titanium Isopropoxide (TTIP) (see molecule in Fig. 2.8a) is vaporized at 75 degrees Celsius (˚C) to flow TTIP vapor via a heated gas line and Swagelok VCR metering valve (SS-SVR4-VH) (as shown in Fig. 2.8b). The precursor liquid is bought prepacked in a cylinder (Fig. 2.8c) for plasma deposition.

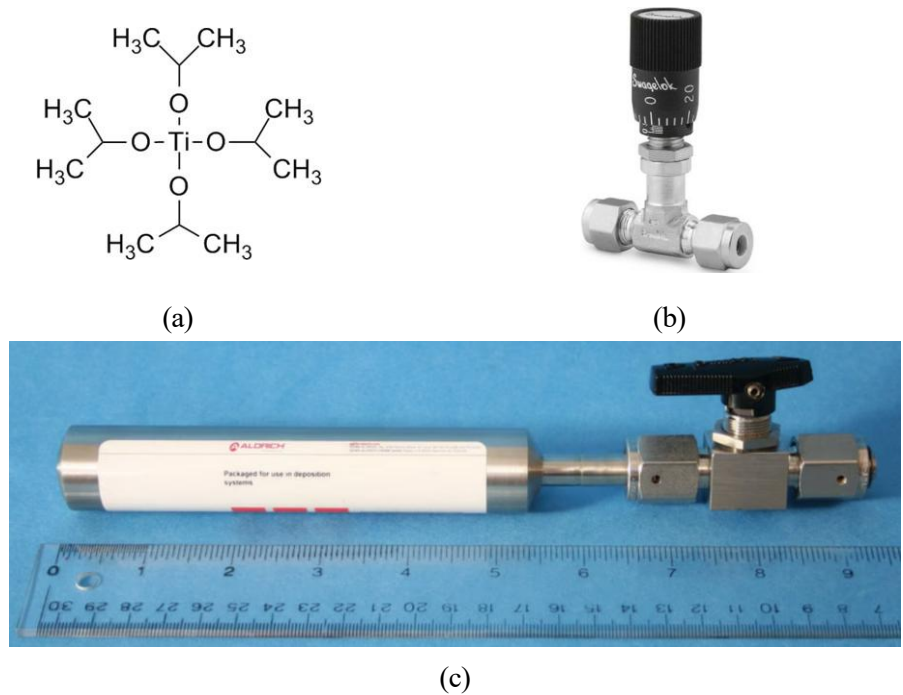

(a)                                        (b)

(c)

Figure 2.8: Titanium Isopropoxide (TTIP): (a) TTIP molecule (photo credit: Volochem), (b) Low flow metering valve (photo credit: Swagelok (SS-SS4-VH)) used to flow TTIP vapor into the chamber, and (c) TTIP cylinder (photo credit: Sigma Aldrich (687502)).

## 2.2   Permanent Magnets

Pairs of permanent magnets were used in this experiment to create magnetic fields between the electrodes. They were Neodymium Rare Earth Ring Magnet and their dimensions were 2" x 1" x 1/4" bought from (K&J Magnetics, Inc (RY0X04)). They were made from NdFeB, Grade N42 material. A picture of the magnet is seen in Figure 2.9. The magnet has axial magnetization (see Fig. 2.9b) placing the poles on the flat ends. It operates safely up to 176 ºF (80 ºC), with a maximum remnant magnetic field ($B_0$) of 13,200 Gauss on its surface.

3-D printed supports are used to place the magnets directly above the top electrode or the bottom electrode. For example, consider figure 2.10b, which show 1 magnet inside a 3-D printed part placed above the top electrode. Magnets are always used in pairs and similarly, one magnet would be placed directly below the bottom electrode, as schematically shown in Fig. 2.10a. This configuration produces a magnetic field strength of ∼ 500 Gauss at the center of the plasma. The dimensions of the magnets did not change, but several magnets were stacked together to increase the magnetic field strength. For example stacking 3 magnets above the top and below the bottom electrode created a maximum field strength of ∼ 1020 Gauss. A description of the magnetic field strength simulation in 2-D for the experiment be discussed in section 3.4.

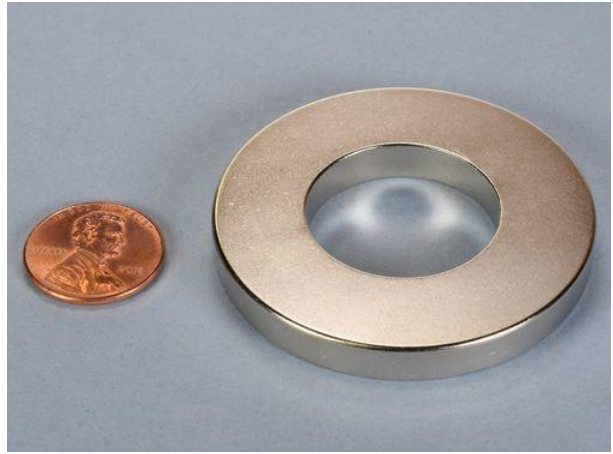

(a)

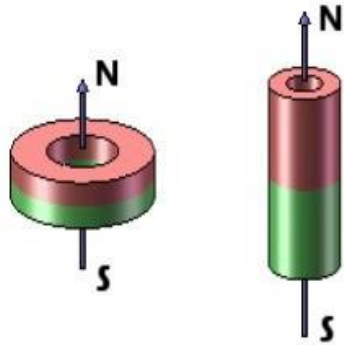

(b)

Figure 2.9: Permanent **axially magnetized** magnetic rings used in this dissertation: (a) actual image of the magnet (photo credit: K&J Magnetics, Inc (RY0X04)), and (b) schematic representation of axial and radial magnetization on rings (photo credit: KENENG hardware)

## 2.3 Dust Imaging from Laser Light Scattering

Laser light scattering is used in many dusty plasma experiments to directly visualize and measure the properties of the dust particles in the plasma. Shown in Fig. 2.4 are a "laser source". The laser power is initially 1 W, but it was brought down to 300 mW by using a 0.7 neutral density (ND) filter. The beam from the diode laser is expanded into a thin (~3 mm) wide, vertically oriented laser sheet using a "cylindrical lens" and injected into the plasma volume.

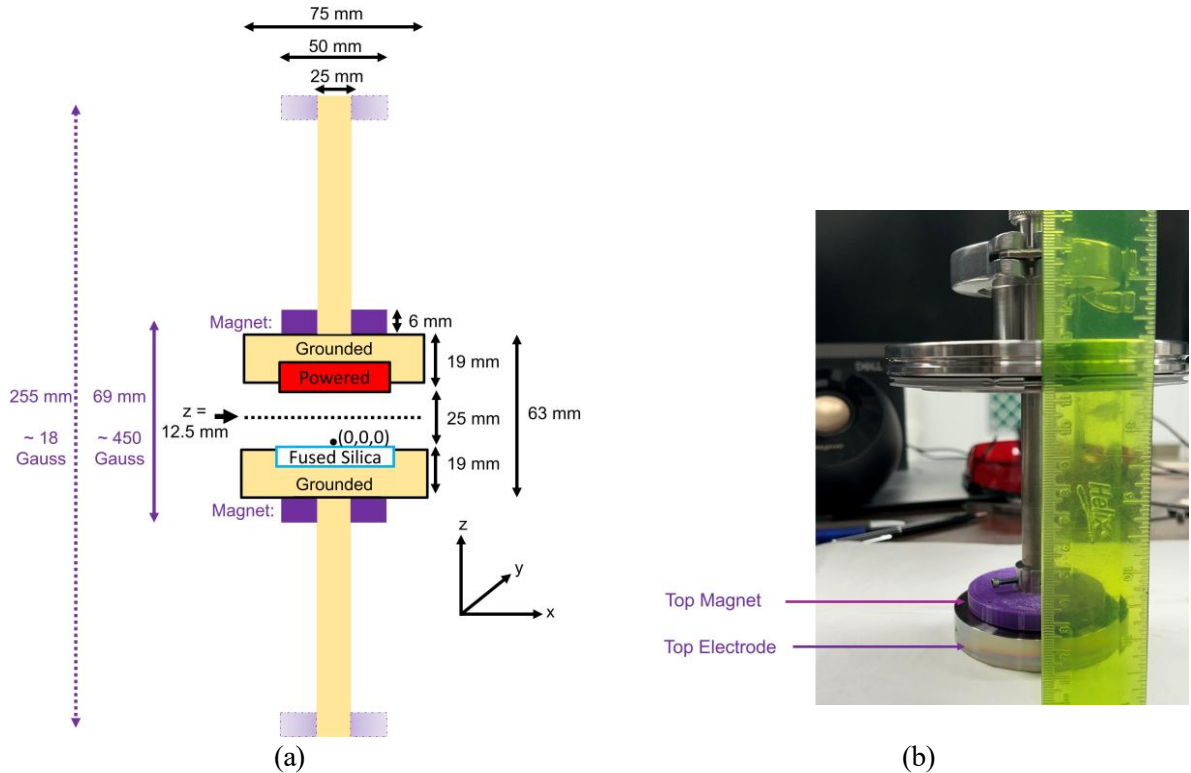

(a)                                    (b)

Figure 2.10: Configuration of magnets relative to the electrode setup. (a) One magnet placed directly above the top electrode, and (b) schematic representation of a pair of magnets directly above the top and below the bottom electrode.

This planar laser sheet illuminates the particles in the cloud as shown, for example, in Fig. 2.2. The light scattered from the illuminated nano dusty particle cloud is viewed at 90 degrees with a camera. The camera lens focused the imaging at the position of the laser sheet, and the filter in front of the camera maximized the green light from the laser to pass through, in order to see the dust cloud.

For the majority of these experiments, videos are recorded at 50 to 100 frames per second. The subsequent movies (or image sequences) are recorded from the moment the plasma is ignited and will continue to record for 100's of seconds through one or more dust growth cycles. An example of an image sequence of carbonaceous dust cloud is shown in Fig. 2.11. Fig. 2.11 (a) shows an image between the electrode when the plasma has not been turned on yet. Fig 2.11 (b) shows the first instance when the plasma is turned on which is also the start of the particle growth cycle. Between Fig. 2.11 (c) and (f), we show image of the dust cloud at increments of 1/8th of a growth cycle, until the end of a growth cycle.

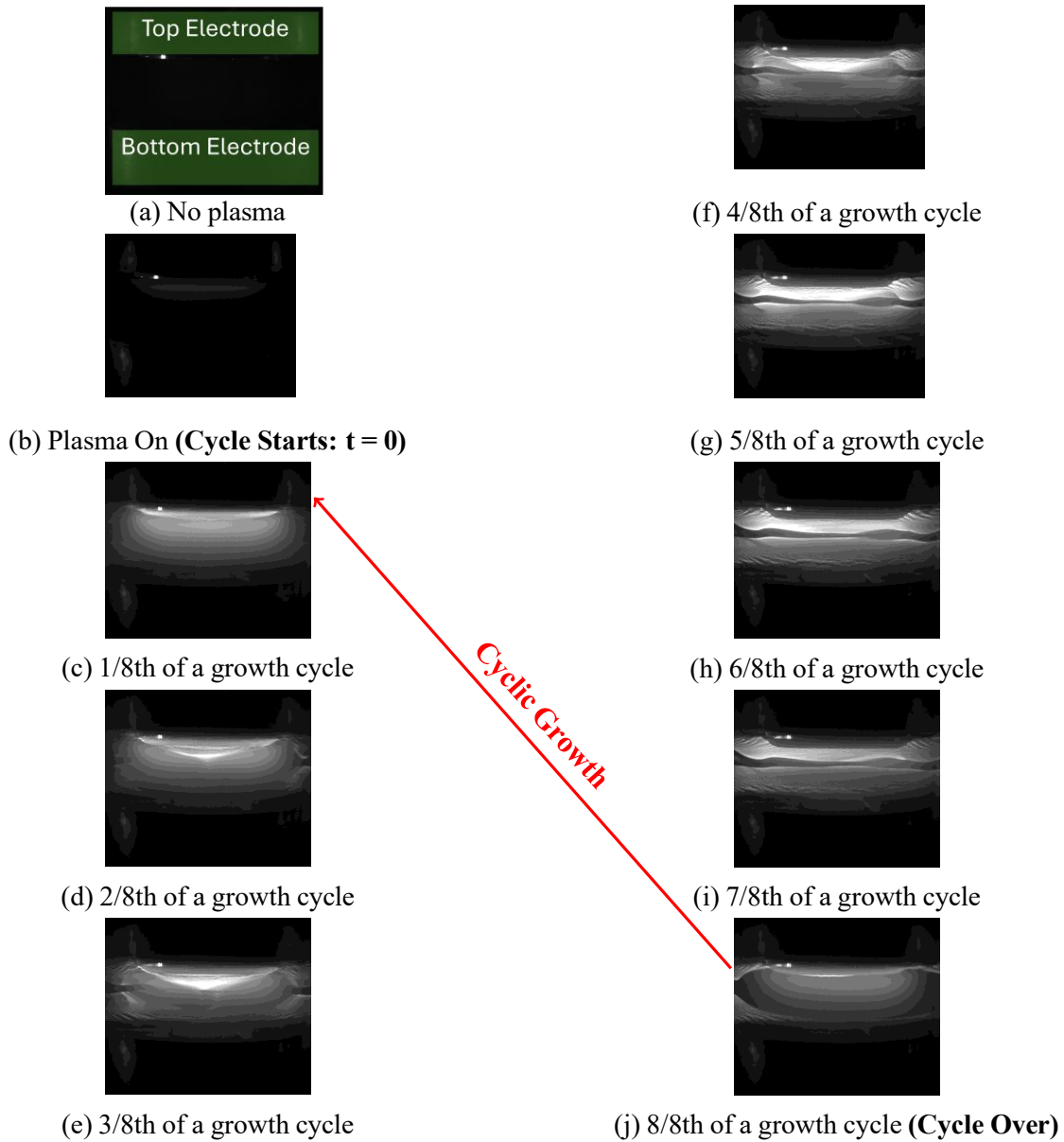

(a) No plasma

(b) Plasma On **(Cycle Starts: t = 0)**

(c) 1/8th of a growth cycle

(d) 2/8th of a growth cycle

(e) 3/8th of a growth cycle

(f) 4/8th of a growth cycle

(g) 5/8th of a growth cycle

(h) 6/8th of a growth cycle

(i) 7/8th of a growth cycle

(j) 8/8th of a growth cycle **(Cycle Over)**

Figure 2.11: Image sequence showing one cycle of carbonaceous dust particles in an argon-acetylene plasma at p = 300 mTorr without a magnetic field at RF power of 60 W. The images are recorded at 100 frames per second. The total time of this representative cycle is $\sim$117 s. (a) Image prior to the ignition of the plasma (green rectangles were overlaid to emphasize electrodes). (b) Image taken as the plasma is ignited, defining t = 0. (c - j) images at time intervals of $\triangledown$t = 1/8th of the growth cycle, i.e., every 1463 video frames. The increasing brightness of the images over time reflects both the increasing particle density and the increasing size of the particles. Note that it only takes about 1/8th of the growth cycle for the particles to grow large enough to be visible in the camera. By (e) 3/8th to (f) 4/8th of the cycle, particles are large enough to show collective dusty plasma effects in the form of dust density waves at the top, bottom, and sides of the particles cloud. Void structures are clearly visible by half-way (f) (4/8th) through the growth cycle. By the end of the cycle, a new generation of small particles are being formed in the void region as shown at (j) 8/8th. The red arrow indicates the cyclic growth from (c) to (j), as long as the plasma is on.

Fig. 2.11, has been recorded as part of a sequence of images over at least four growth cycles. Consider image (j) which is now shown in Fig. 2.12a. A purple box was drawn in the middle of the image, from which we obtain the average light scattering intensity from the laser, plotted as a function of time for the other images in the sequence in Fig. 2.12b. We see a cycle time of ∼ 117 s. Before the plasma is on, the light intensity is almost 0 arbitrary units (a.u.), but it quickly rises to almost 100 a.u. At the end of each cycle, there is a rapid drop in light intensity, but not as low as when the plasma is turned off and the particles fall.

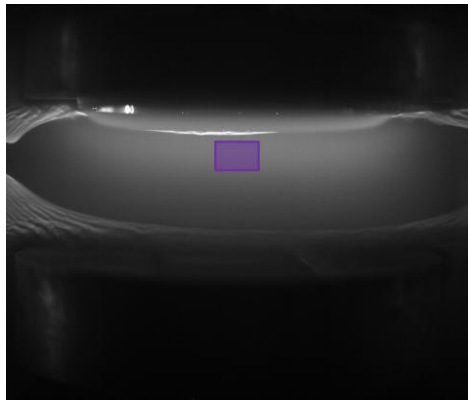

(a)

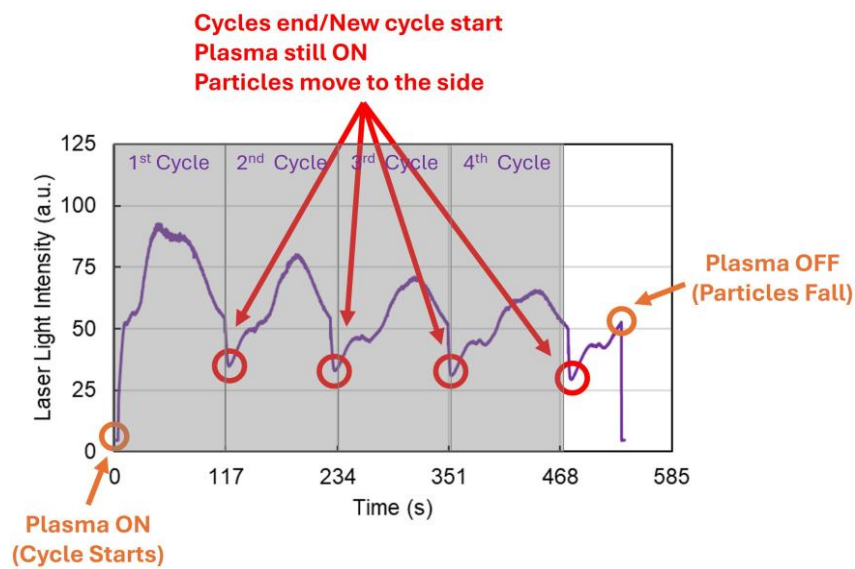

(b)

Figure 2.12: Extracting the cycle time ( ∼ 117 s) from (a) a small purple box in the dust cloud and (b) the full image sequence showing particle growth cycle phases.

Now, let us draw a 1-D red vertical line through the middle of the dust cloud, as shown in Fig. 2.13a, to analyze the dust cloud's pixel intensity at the end of the cycle. The line spans the electrode spacing, i.e. 25 mm, in the region of the plasma. By extracting the pixel intensity along this line and plotting it as a function of time for the images in the next 20 s, we get a temporal variation in pixel intensity, as seen in Fig. 2.13b. In Fig. 2.13c, we provide further details such as the sheath region and note that the dust indeed move down towards the bottom electrode when the cycle is over. However, the dust do not penetrate the sheath region above the bottom electrode. Instead they move sideways as will be discussed in Chapter 4.1.

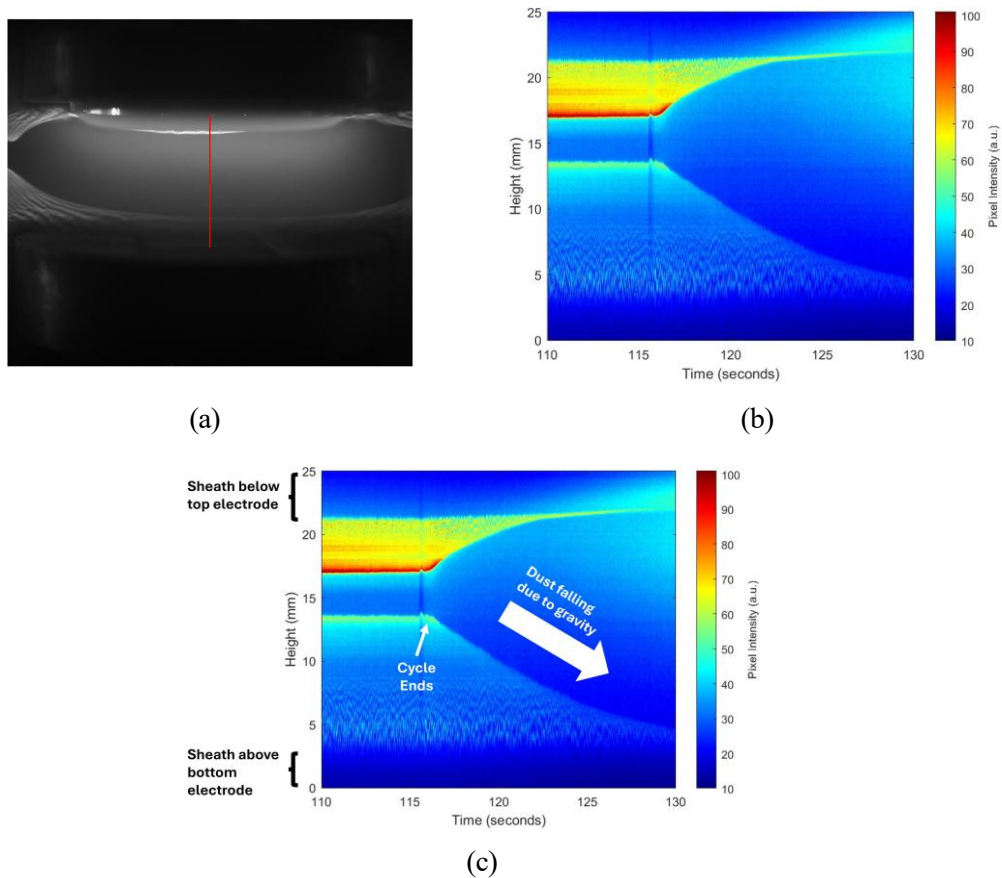

(a)

(b)

(c)

Figure 2.13: Analyzing a line through the middle of the carbonaceous dust cloud between the electrodes. (a) the red line drawn, (b) temporal variation of the pixel intensity from the line, (c) the dust fall after the cycle ends but do not penetrate the bottom electrodes' sheath.

Similar to Fig. 2.13, we draw a vertical line between the electrodes at middle of an Ar/T-TIP dusty plasma in order to extract. The result is shown in Fig. 2.14. After analyzing at least three growth cycles, we can determine a cycle time of ∼ 77 s.

We similarly draw a horizontal line that spans the diameter of the electrodes and at the middle of the plasma, i.e. z = 12.5 mm and compare the dust cloud for titania and carbonaceous dust, with and without magnetic fields in Chapter 4.2.

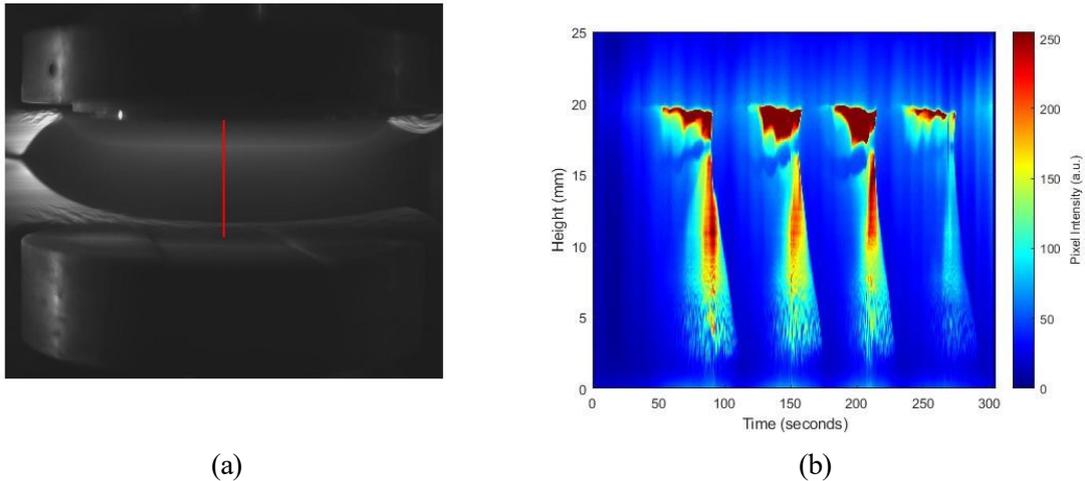

| (a) | (b) |

Figure 2.14: Analyzing a line through the middle of the titania dust cloud between the electrodes. (a) The red line drawn, (b) temporal variation of the pixel intensity from the line through at least 3 growth cycles.

## 2.4   Optical Emission Spectroscopy

The OES is obtained in our experiments using a broadband spectrometer. In Fig. 2.15 several Ar peaks are identified. We observed that all of the emission lines vary as a function of time during the growth process, but an examination of the (NIST) database reveals that the argon neutral (Ar I) line at 763.5 nm is a promising candidate for OES analysis. This line was chosen because it was the highest intensity line according to the survey scan without any other line blending. As seen in the inset in Fig. 2.15 - and confirmed by NIST database - this line, which was the electric dipole transition of Ar I from $3s^2 3p^5 \left( {}^2P^o_{3/2} \right) 4p$ to $3s^2 3p^5 \left( {}^2P^o_{3/2} \right) 4s$ was relatively isolated without any other line blending [84–87].

When the reactive gas was present, no new lines corresponding to emission of new species were seen. For example, it was anticipated to detect lines corresponding to either carbon and/or hydrogen arising from the reactive gas(es). The spectrometer may have not detected these lines either due its relatively low resolution and/or due to the higher flow rate of Ar increasing its intensity when compared to other species in the plasma [88]. An example of the cyclic variation of the OES intensity at 18 Gauss is shown in Figure 2.16a. The data were collected at time intervals of approximately 200 milliseconds (ms) and a spline interpolation was used to interpolate the data at every 100 ms in order to create a uniformly sampled dataset. From the interpolated data, a fast Fourier transform (FFT) was performed to determine the peak frequency of the cycles, as shown in Figure 2.16b. The average intensity was subtracted from each data point in order to extract any noise from the DC level. The FFT analysis code is provided in Appendex E.

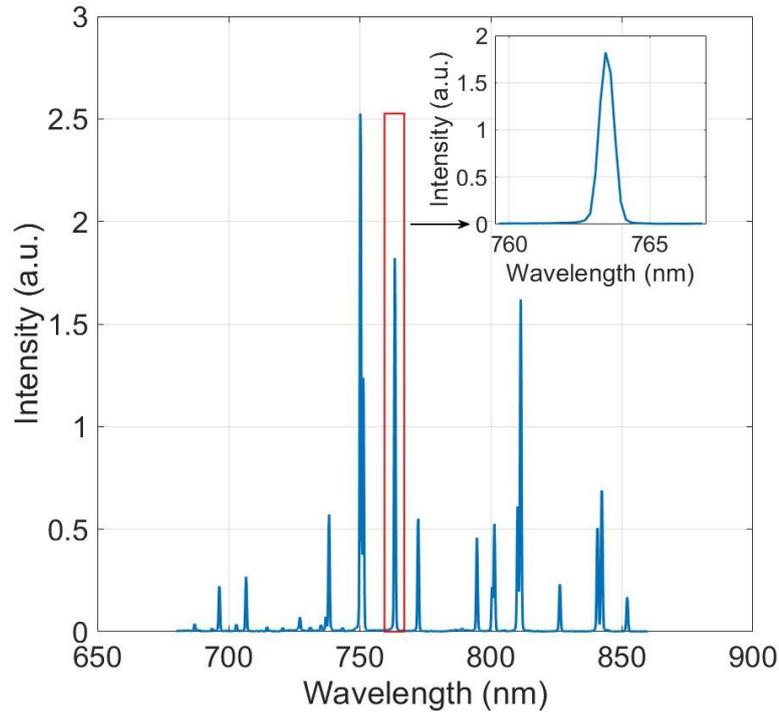

Figure 2.15: Survey scan between 680 and 860 nm revealing several Ar I peaks. No new lines were seen during the presence of either titanium isopropoxide or acetylene gases in the plasma. The line at 763.5 nm is an isolated peak (adapted from [53])

While OES can used effectively to monitor the cyclical growth process, the limited resolution (0.59 mm) and lack of an absolute intensity calibration of the spectrometer limits its use as an electron temperature diagnostic because of the possible blending of emission lines.

## 2.5   In-situ Plasma Diagnostics

Characterization of the plasma parameter is, of course, necessary to obtain a clear understanding of the nanoparticle growth process. This is complicated in multi-species plasmas because of the need to isolate the plasma parameters of each species, as well as the incorporating the influence of the growing nanoparticles. Therefore, care must be taken to make plasma measurements. For these studies, in-situ plasma measurements are made in argon-only plasmas using standard plasma probe techniques. A single-tipped Langmuir probe is used to measure the electron density and electron temperature. An emissive probe is used to measure the plasma potential. When nanoparticles are being formed, non-invasive measurement techniques are used. Camera images of laser light scattering is used to directly image the nanoparticle clouds and optical emission spectroscopy is used as a proxy for the plasma parameters. The remainder of this section will describe the various diagnostic techniques.

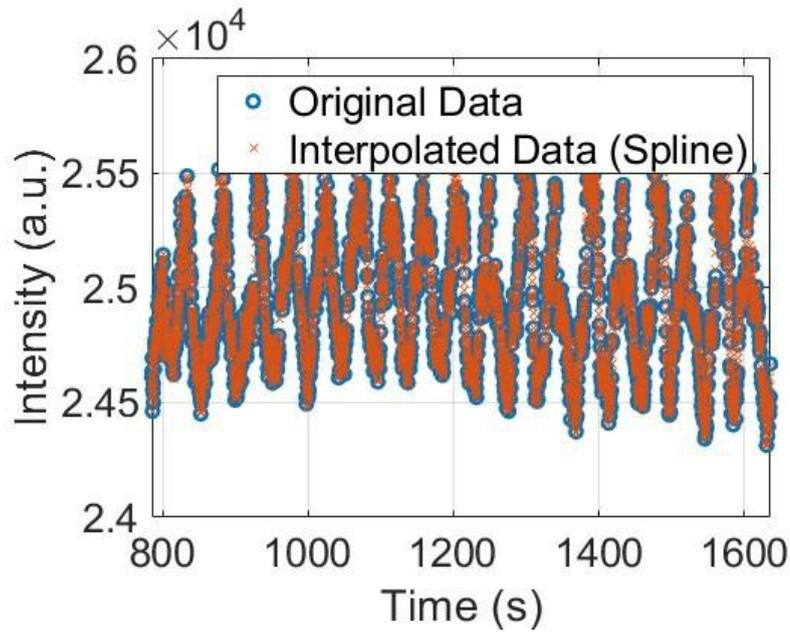

(a)

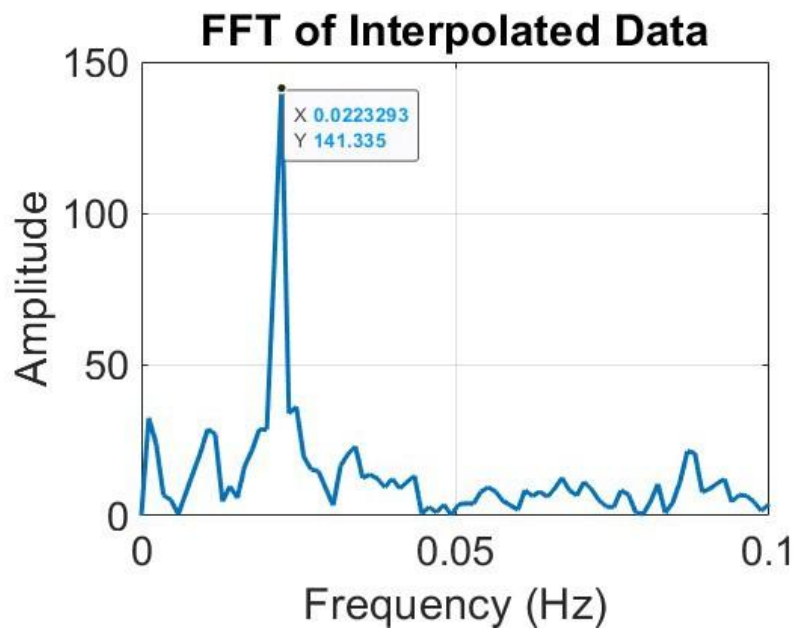

(b)

Figure 2.16: Optical emission spectroscopy analysis: (a) . (b) Temporal variation of 763.5 nm line intensity over time for carbonaceous dusty plasma at 1020 Gauss. (c) Corresponding FFT showing cycle frequency $\sim 0.022$ Hz (cycle time $\sim 45$ s).

### 2.5.1 Langmuir Probes

We used a single-tip Langmuir probe as a tool for measuring plasma properties, such as electron temperature, electron density, and floating potential. A Langmuir probe consists of a thin metallic wire (rod) that is inserted into the plasma and biased with respect to the plasma potential. We built a probe in-house with the wire's diameter and length being $\sim 0.5$ mm and $\sim 2.7$ mm, respectively, as shown in Fig. 2.17.

The theory behind a langmuir probe measurement which is operating near the plasma floating potential ($V_F$) where the current is small, is described by Eq. 2.1 [89]. Here I(V) is the total electrical current collected by the probe when a voltage V is used to bias the probe in the plasma. $N_e$ is the electron density, e is the electronic charge, $k_B$ is the boltzmann constant, $m_i$ is the mass of the ions, $m_e$ is the mass of the electrons, $A_p$ is the surface area of the probe, and $A_s$ is the area of the sheath surface around the probe, which, in this analysis, can be assumed to be approximately equal to $A_p$ [89].

$$I(V) = N_e e A_p \left(\frac{k_B T_e}{m_i}\right)^{1/2} \frac{1}{2} \left(\frac{2 m_i}{\pi m_e}\right)^{1/2} \left[\exp\left(\frac{eV}{k_B T_e}\right) - \frac{A_s}{A_p} \exp\left(-\frac{1}{2}\right)\right] \tag{2.1}$$

Eq. 2.1 needs to be slightly modified to accommodate a Langmuir probe in the presence of magnetic fields, because electrons and ions no longer move in a straight line [89]. In the experiments of this dissertation, we did not modify the Langmuir probe theory to accommodate our magnetic fields, because we worked with relatively weak fields of less than 1000 Gauss. The second term in Eq. 2.1 is the ion saturation current ($I_{sat}$), given by Eq. 2.2.

$$I_{sat} = e N_e \left(\frac{k_B T_e}{m_i}\right)^{1/2} A_p \exp\left(-\frac{1}{2}\right) \approx 0.61 e N_e A_p \left(\frac{k_B T_e}{m_i}\right)^{1/2} \tag{2.2}$$

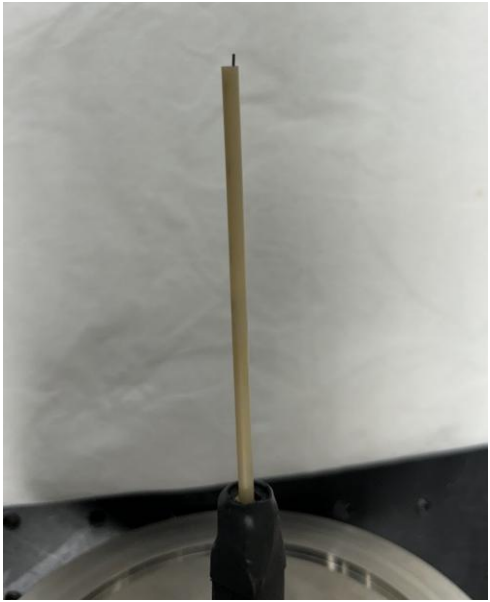

(a) Full Langmuir probe

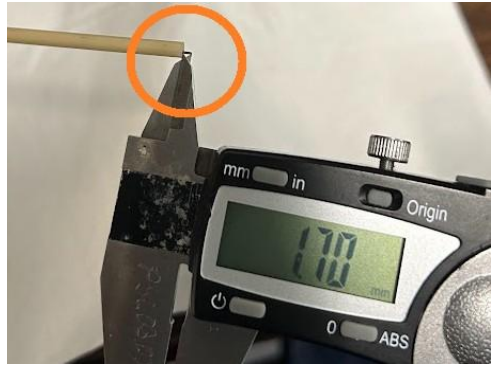

(b) Probe length view (inside orange circle)

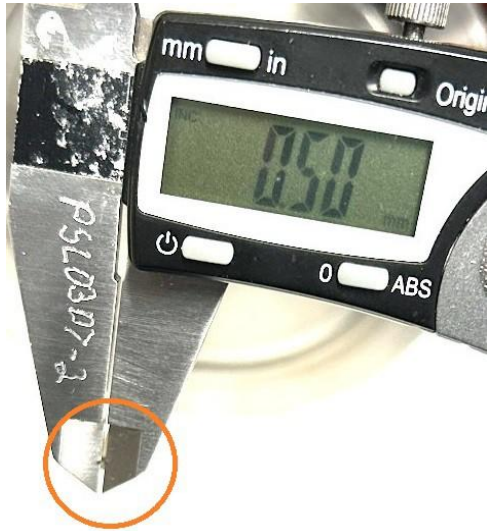

(c) Probe diameter view (inside orange circle)

Figure 2.17: Langmuir probe built in-house for several experiments, including length and diameter views

By taking the derivative of Eq. 2.1 with respect to V, we obtain approximately Eq. 2.3 for the electron temperature. Upon fitting a line to $\ln |I - I_{sat}|$ and plotting it against V, we can extract $T_e$ as inversely proportional to the slope, as seen in Eq. 2.4 [89].

$$T_e \approx \frac{1}{e/k_B(I - I_{sat})} / \frac{dI}{dV} \tag{2.3}$$

$$\ln |I - I_{sat}| = \frac{e}{k_B T_e} V + \text{constant} \tag{2.4}$$

Once $T_e$ has been determined, $N_e$ can be obtained by re-arranging Eq. 2.2 and making the latter as subject of the formula, given by Eq. 2.5, where we assume $A_s \approx A_p$.

$$N_e = \frac{I_{sat}}{eA_s \exp(-1/2)} \sqrt{\frac{m_i}{k_B T_e}} \tag{2.5}$$

A Keithley 2400 sourcemeter is a four-quadrant power supply capable of sourcing both current and voltage for test measurements. The specific instrument used for this project can source +/-200 V at +/- 1 A. Using the sourcemeter, a voltage sweep was done between $\sim$ - 40 V and 60 V, from which the current (I) experienced by the probe tip in the plasma was measured. Then, I− V was plotted on a graph (usually referred to as a Langmuir trace), from which several plasma parameters such as the plasma electron temperature ($T_e$), plasma electron density ($N_e$) and plasma floating voltage ($V_F$) can be obtained. A qualitative description of the Langmuir trace is provided, along with a description of how to get measurement of the aforementioned parameters. Consider the raw data collected without an external magnetic field in the center of our plasma (z = 12.5 mm and x = 0 mm), as seen in Fig. 2.18a. The floating potential ($V_F$) is defined as the voltage measured when the current collected by the probe tip is zero (circled in red in Fig. 2.18b). Here, $V_F \approx 8.55$V.

In order to determine $T_e$ and $N_e$, we need to work with the ion saturation current ($I_{sat}$). This is the current measured at low voltages of the Langmuir probe measurement, which is usually negative and approximately constant. In our I-V trace, $I_{sat}$ needed from a range of values in the green rectangle of Fig. 2.18c, for voltages below the predetermined $V_F$. However, since the value of current is changing in that region, we extrapolate a range of current from which a linear fit to determine $I_{sat}$. In order to extract the current, we plot the change of current against voltage, i.e. $dI/dV$ in Fig. 2.18d, to find a region of constant change. For example, we decide to take the current data between -20 V and 0 V. Below -20 V, there are several fluctuations in the data, which could arise from instrumental errors, and above 0V, the change in current with respect to voltage is starting to increase. In figure 2.18e, the blue line is the Raw Data as in Fig. 2.18a, but the data in red is the fit for the $I_{sat}$, and the data in black is the raw current minus the $I_{sat}$ plotted against the same voltage, i.e. $I - I_{sat}$. For a small range of $Ln[I - I_{sat}]$ above $V_F$, we can assume a linear fit from which we can determine electron temperature. In Fig. 2.18f, the linear fit is assumed for voltages between 2 and 9 V. We obtain $T_e \approx 2eV$ by taking inverse of the slope.

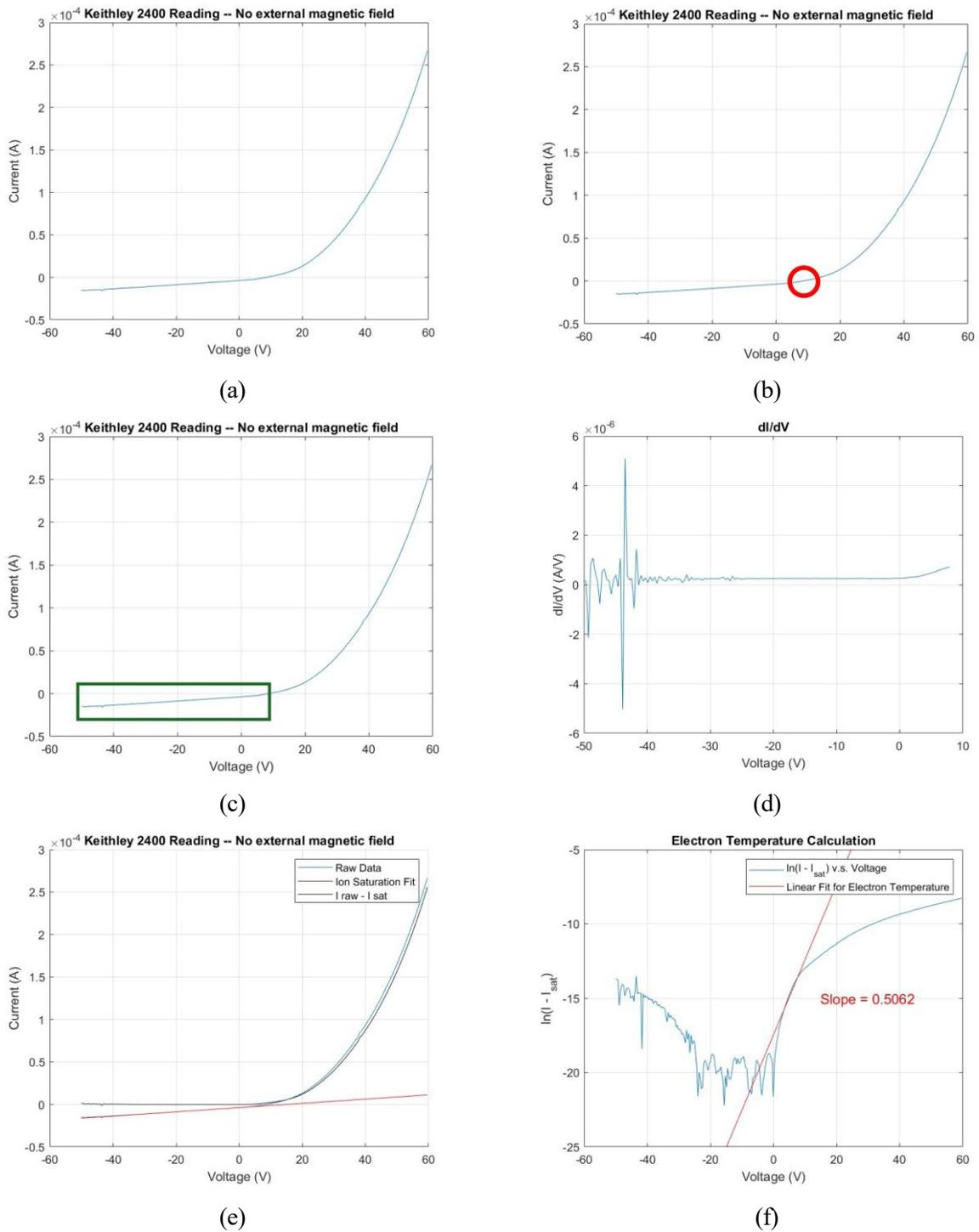

Figure 2.18: Qualitative description of a Langmuir probe trace in our plasma: (a) Raw data collected from a single tip Langmuir probe at z = 12.5 mm and x = 0 mm (center of the plasma). (b) Red circle highlighting the floating potential, which occurs when current is zero. (c) Green rectangle highlighting the region of ion saturation current at low voltage. (d) dI/dV of the raw data below the plasma floating potential. (e) Re-plot of the original data, ion saturation current fit, and subtracted data. (f) Linear fit to $\ln(I - I_{\mathrm{sat}})$ versus voltage, used to determine the electron temperature ($T_e \approx 2eV$).

## 2.5.2  Emissive Probes

An emissive probe was built in house for the experiment, as shown in Fig. 2.19a. Current was sent to the probe until it glows in the plasma, as shown in Fig. 2.19b. Then a Keithley 2400 was used to send voltage and measure current. From the I-V plot, the floating potential was extracted. The emissive probe operates by heating a filament that emits thermal electrons into the surrounding plasma. This emission reduces the voltage difference between the plasma and the filament, allowing the filament's voltage to serve as an estimate of the plasma potential ($V_P$).

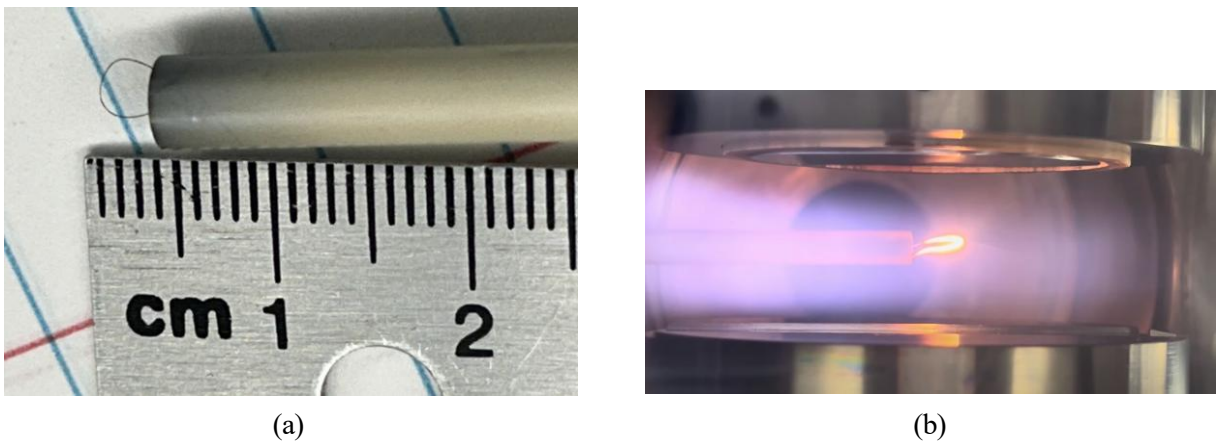

(a)                                    (b)

Figure 2.19: Two views of the emissive probe used in experiments. (a) Emissive probe built in house for several experiments. (b) Emissive probe glowing in the plasma.

There are several techniques for determining plasma potential using an emissive probe, two of which were considered for this dissertation, are described below, [90]:

1. **Inflection-Point Method:** By taking a langmuir trace, for example as in Fig. 2.20a there is supposed to be a rapid change in current seen by the probe due to the voltage supplied to it, close to $V_P$. Due to the rapid change, there will be a peak in the derivative of current to voltage, i.e. dI/dV, for example as in Fig. 2.20b. By taking langmuir trace for different heating currents, one can find the peak in dI/dV and interpolate $V_P$ at zero heating current [91, 92]. We attempted this method for this dissertation, as seen in Fig. 2.20c. Unfortunately, the data was too noisy, and we would get artificial peaks in dI/dV from the noise. It was impossible to determine the actual $V_P$. Therefore, we tried the next method.

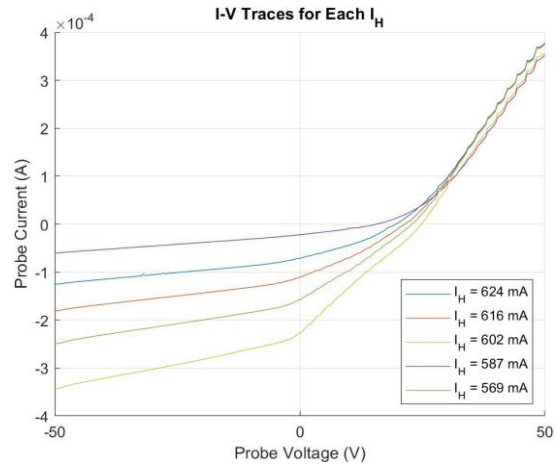

(a)

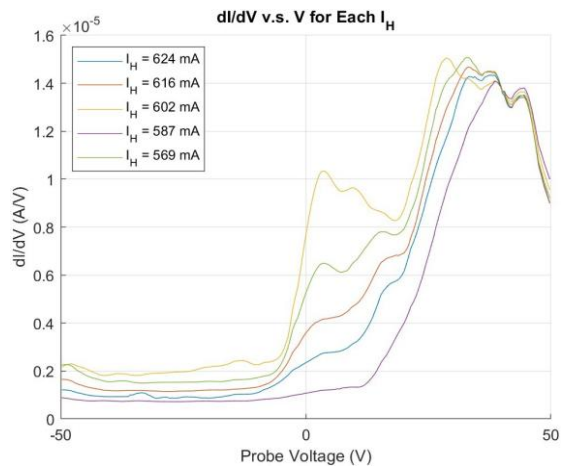

(b)

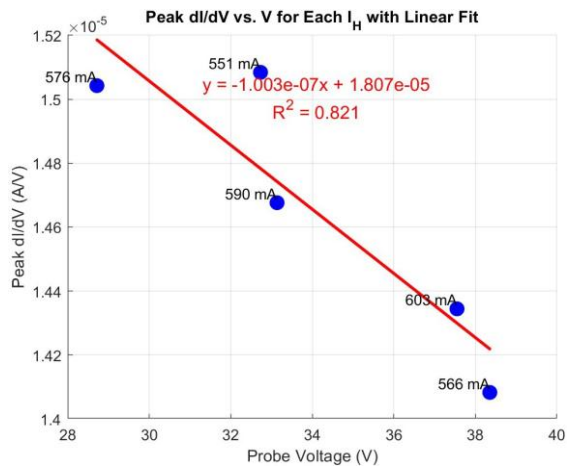

(c)

Figure 2.20: Inflection-Point method to determine the plasma potential: (a) I-V trace at different heating current ($I_H$) for the probe to glow in the plasma. (b) dI/dV as a function of V, and (c) interpolate the peak of (b) which was supposed to happen at the "knee" of (a).

2. **Floating Potential Method:** In this method, we assume that when an emissive probe is electrically floated and emitting electrons (glowing), $V_F \approx V_P$. The rationale is that emitted electrons balance the incoming flux, making $V_F$ a close estimate of $V_P$ [91]. In practice, a small, constant offset exists, but it cancels out when differentiating to obtain the electric field. A representative of $V_P$ measurement is outlined below. Current ($I_H$) was supplied using an Agilent E3630A, starting at 0 A and increased by 0.05 A every 20 minutes to condition the probe. For each increment, $V_F$ was extracted from I-V traces recorded with a Keithley 2400. Measurements at 0 Gauss are shown in Fig. 2.21a. Once $V_F$ saturates, $I_H$ is held constant, and $V_F$ is recorded across parameters such as height z (Fig. 2.21b). Each repetition begins with re-measuring $V_F$ vs. $I_H$, explaining the three experiments in Fig. 2.21a. After obtaining the new saturated $V_F$, the z-scan is repeated. Using a 2-D stage, $V_P$ is mapped over z and x for various magnetic fields. The experiment is conducted at 20 mTorr to reduce probe damage from collisions with background neutrals.

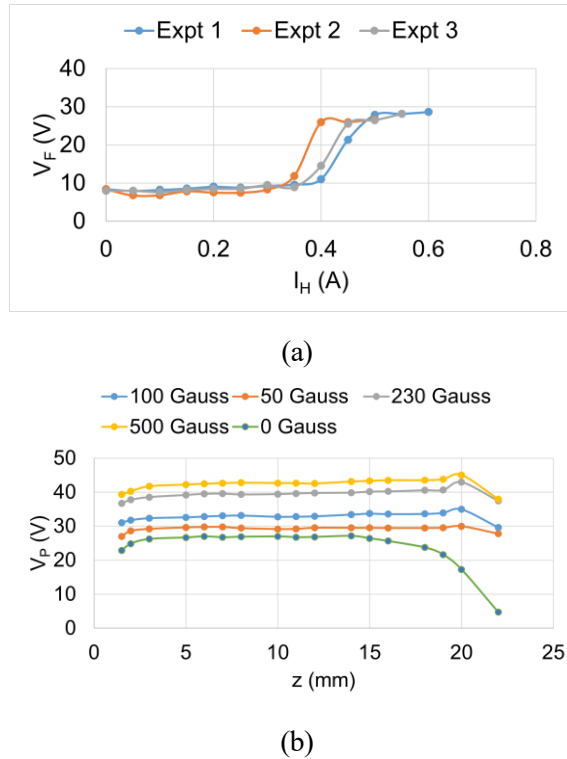

(a)

(b)

Figure 2.21: Emissive probe measurements: (a) saturation of floating potential with increasing electron emission, and (b) plasma potential variation with height under varying magnetic fields.

## 2.6 Material Characterization

### 2.6.1 X-ray Diffraction

Powder diffraction were performed on the titania samples using a Rigaku Smartlab. The $\theta - 2\theta$ scan ranged from $10°-80°$. The angle between the projected paths of the X-ray source and the detector is $2\theta$, which is why diffraction patterns obtained using this configuration are commonly referred to as $\theta$–$2\theta$ scans [93]. In this geometry, the X-ray source remains stationary while the detector rotates over a range of angles. The scan rate was $3°$ per minute. The as-grown titania particles were amorphous. However, when air annealed at $400°$ C and above, titania crystallized to anatase and rutile. The carbonaceous particles were always amorphous.

W. L. Bragg proposed a model for X-ray diffraction, where atomic planes in a crystal act like partially reflective mirrors. They produce constructive interference when the path difference between reflected rays equals an integer multiple of the wavelength, leading to the Bragg law, given by Eq. 2.6. Here, $\lambda$ represents the wavelength of the incident X-rays, $d$ is the distance between the crystal planes (also known as the interplanar spacing), $\theta$ is the incident angle, which is the angle between the incoming X-ray and the crystal plane, and $n$ is an integer known as the order of diffraction. This relationship is fundamental in determining the crystal structure and spacing in materials through XRD analysis.

$$n\lambda = 2d\sin\theta \qquad (2.6)$$

This predicts the observed diffraction patterns. The intensity of the diffracted beam arises from the coherent contribution of thousands of atomic planes, and while the Bragg law depends only on lattice periodicity, the atomic basis influences the relative intensities of diffraction peaks [94]. Two incident waves reflect from two adjacent crystal planes, with the difference in their path lengths represented by the dashed line, as seen in Fig. 2.22b [95].

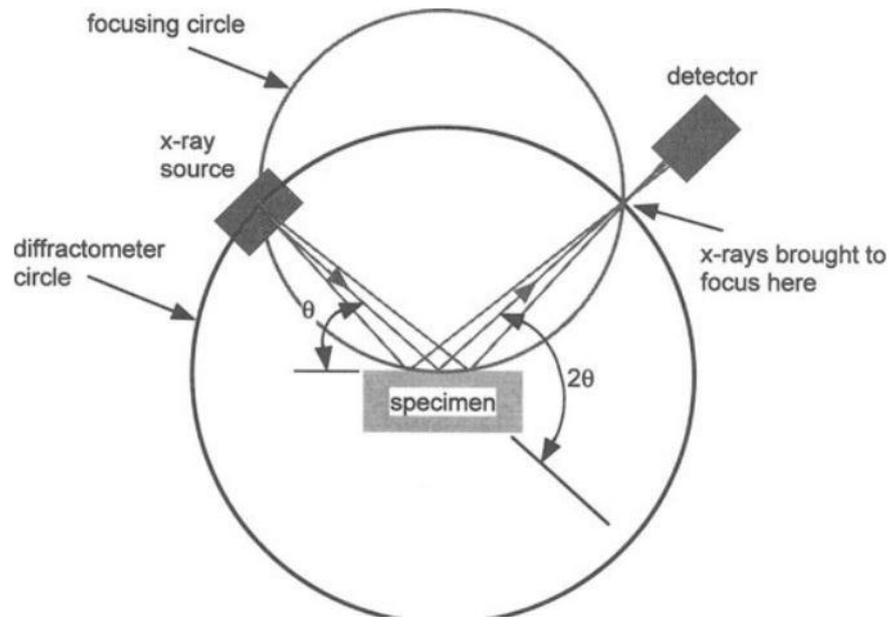

(a)

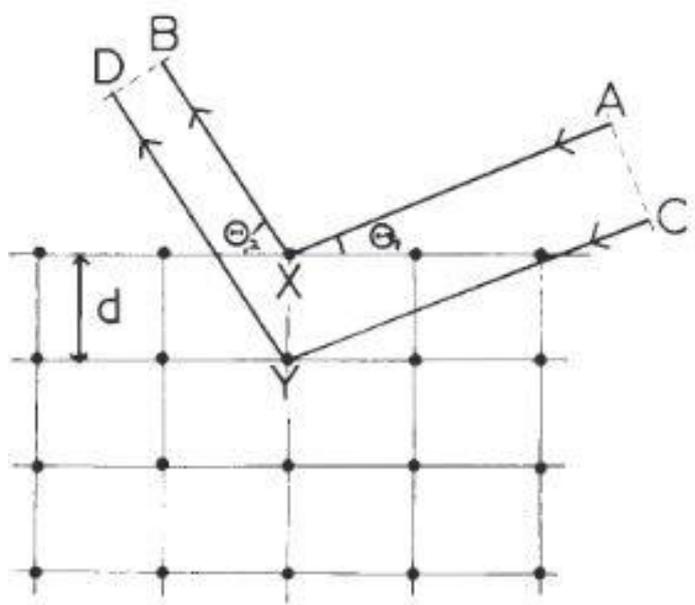

(b)

Figure 2.22: X-ray diffraction: (a) experimental setup schematic (adapted from [93]), and (b) Bragg's law for constructive interference (adapted from [95]).

Annealing to induce crystallization is a thermal process used to transform an amorphous or partially ordered material into a more crystalline structure. By heating the material and thus adding thermal energy (E) into the material to a specific temperature below its melting point, atoms gain enough mobility to rearrange into a more ordered and energetically favorable crystalline lattice. This process is commonly used in materials like thin films, glasses, and polymers to enhance properties such as electrical conductivity, optical clarity, or mechanical strength [96]. A schematic representation of annealing to make a sample more crystalline after starting from more disordered is shown in Fig. 2.23

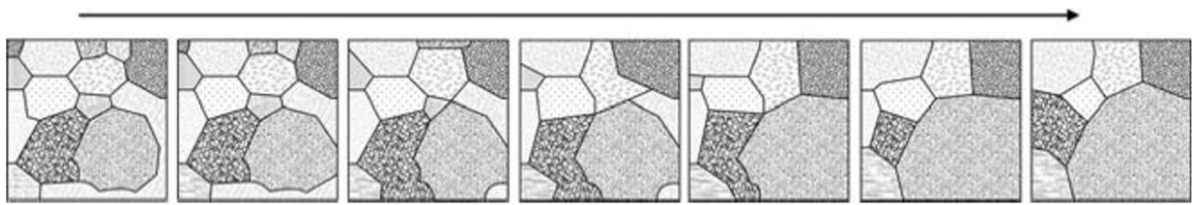

Figure 2.23: Schematic representation of annealing process: as time and temperature increases (arrow pointing to the right) samples become less disordered and more crystalline (adapted from [96]).

### 2.6.2    Scanning Electron Microscope (SEM)

Samples were deposited on to fused silica substates and were mounted directly on to aluminum support stubs with double stick carbon tape. All samples were sputter coated with about 25 nm of gold and then imaged at 10 - 20kv using a Zeiss EVO50 scanning electron microscope. We used secondary electron scanning to image the nanoparticles, which involves detecting low-energy electrons emitted from the surface of a sample when it is hit by the primary electron beam. These secondary electrons provide high-resolution images that reveal fine surface details and topography. Consider two examples of SEM images performed on carbonaceous nanoparticles in Fig. 2.24.

In Figure 2.24a, the original SEM image of carbonaceous nanoparticles displays several isolated particles as well as particles sticking together. However, in Figure 2.24b, only isolated nanoparticles, circled in teal, were chosen to determine the particle size distribution. This process was repeated for all imaged nanoparticles. When levitating in the plasma, nanoparticles are negatively charged and repel each other. However, once the plasma is turned off, they can experience charge fluctuations within milliseconds , including a complete reversal of charge from negative to positive during the afterglow period, that could allow particles to be attracted to each other and stick together while falling on the substrate [97]. These particles were excluded from the size distribution analysis because their shapes and edges could not be clearly delineated to determine radii.

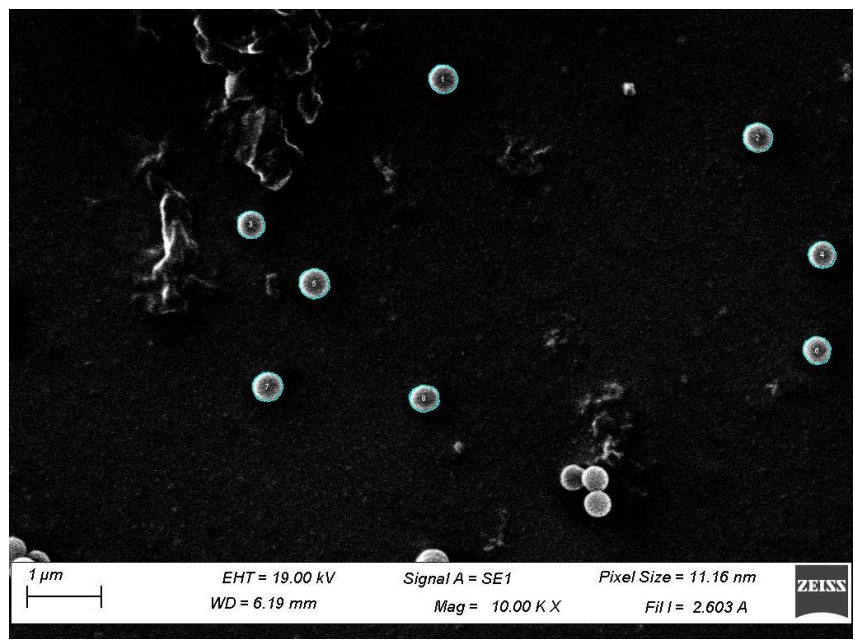

(a)

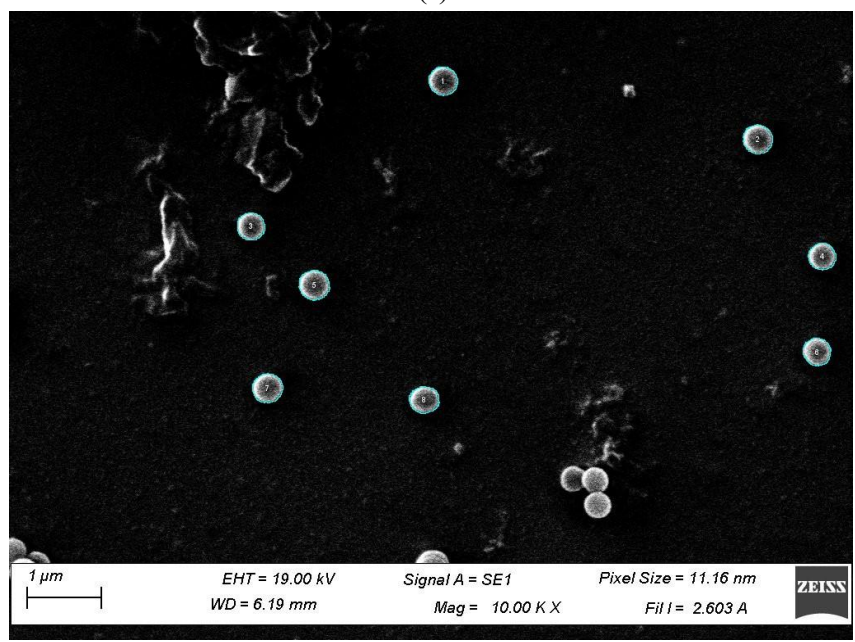

(b)

Figure 2.24: Scanning electron microscope images of carbonaceous nanoparticles grown for $T_c$ ⌣40 s at 790 Gauss: (a) Original SEM image. (b) The same image with isolated particles highlighted for size distribution analysis.)

Energy Dispersive X-ray Spectroscopy (EDS) in SEM is a technique used to analyze the elemental composition of a sample by detecting X-rays emitted when the sample is bombarded with the electron beam. Each element emits X-rays at characteristic energies, allowing for identification and quantification. EDS is used to characterize the titania nanoparticles in order to see the effects of annealing on the samples' composition. EDS can not detect hydrogen due to its low atomic number and mass. Thus it was not used to characterize the carbonaceous samples, because we would not be able to determine the ratio between carbon and hydrogen in the samples.

### 2.6.3   Raman Spectroscopy

Raman spectroscopy measurements were performed on samples in Fig. 4.8b, 2.26b and 2.27b using a custom-made system with a Horiba HR spectrometer equipped with a 1200 grooves/mm grating and a 532 nm excitation laser source. The measurements were taken using a 10× microscope objective lens. Laser power was kept at 25 mW, and accumulation times ranged from 1 to 5 s. An example raman spectra of carbonaceous samples is shown in Fig. 2.25. This is showing a peak around 1400 - 1500 cm$^{-1}$, similar to Couedel et. al.'s raman spectra in Fig. 1.12 [5].

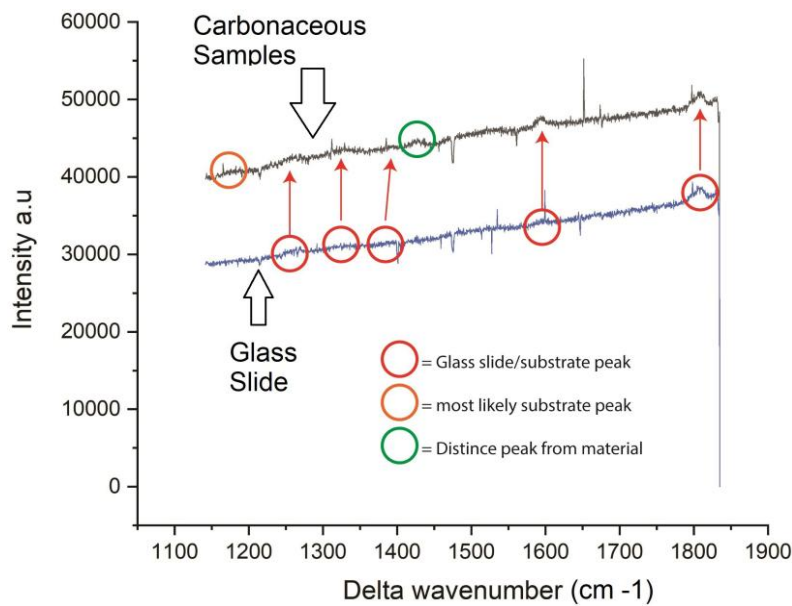

Figure 2.25: An example of raman spectroscopy performed on carbonaceous materials. The background glass slide spectra is also plotted.

Raman active modes for rutile and anatase will be explored from group theory. We can determine Raman active modes by considering the symmetry properties of titania crystals using its space group. First, the symmetry of the system is identified, and the total number of vibrational modes is calculated based on the number of atoms. These vibrational modes are then classified into irreducible representations using the character table of the point group. A vibrational mode is Raman active if its symmetry corresponds to that of a component of the polarizability tensor, which transforms like quadratic functions such as $x^2$, xy, or $z^2$ [98]. By matching the symmetry of each vibrational mode to the symmetries listed under the quadratic terms in the character table, one can predict the number of Raman active modes.

**Rutile**

Consider that tetragonal rutile has the space group $D_{4h}^{14}$ ($P4_2/mnm$), 6 atoms (N) in the unit cell, and 15 vibrational modes ($3N - 3$). Then, its character table is given in Fig. 2.26a [98]. Then its irreducible representation (IRREP) is $A_1^g + B_1^g + B_2^g + E_g + 2B_1^u + A_2^g + A_2^u + 3E_u$. The first four are Raman active, whereas the last two are infrared (IR) active. Literature suggest that these corresponds to peaks in energy shifts at $\sim 144$, 448, 612, and 827 $cm^{-1}$, as seen in Fig. 2.26b. However, peaks at 235 and $320 - 360$ $cm^{-1}$ may be observed either due to disorder induced in the material or second order scattering effects [99].

**Anatase**

Consider the space group of tetragonal anatase $D_{4h}^{19}$ ($I4_1/amd$), with 6 atoms in the unit cell and 15 vibrational modes ($3N - 3$). Its character table is given in Fig. 2.27a [98]. It has 6 Raman active modes: $A_1^g + 2B_1^g + 3E_g$, 3 IR active modes: $A_2^u + 2E_u$, and 1 inactive mode: $B_2^u$. Literature suggest that these corresponds to peaks in energy shifts at $\sim 147$, 198, 398, 515, 640 and 796 $cm^{-1}$, as seen in Fig. 2.27b. However, peak at $\sim 398$ $cm^{-1}$ may arise due to a first overtone of $B_{1g}$, at $320 - 360$ $cm^{-1}$ either due to disorder induced in the material or second order scattering effects, or even at 448 $cm^{-1}$ due to early formation of rutile [99].

| $D_{4h}$ | E | $2C_4$ (z) | $C_2$ | $2C'_2$ | $2C''_2$ | i | $2S_4$ | $\sigma_h$ | $2\sigma_v$ | $2\sigma_d$ | linear functions, rotations | quadratic functions | cubic functions |
|---|---|---|---|---|---|---|---|---|---|---|---|---|---|
| $A_{1g}$ | +1 | +1 | +1 | +1 | +1 | +1 | +1 | +1 | +1 | +1 | - | $x^2+y^2$, $z^2$ | - |
| $A_{2g}$ | +1 | +1 | +1 | -1 | -1 | +1 | +1 | +1 | -1 | -1 | $R_z$ | - | - |
| $B_{1g}$ | +1 | -1 | +1 | +1 | -1 | +1 | -1 | +1 | +1 | -1 | - | $x^2$-$y^2$ | - |
| $B_{2g}$ | +1 | -1 | +1 | -1 | +1 | +1 | -1 | +1 | -1 | +1 | - | xy | - |
| $E_g$ | +2 | 0 | -2 | 0 | 0 | +2 | 0 | -2 | 0 | 0 | $(R_x, R_y)$ | (xz, yz) | - |
| $A_{1u}$ | +1 | +1 | +1 | +1 | +1 | -1 | -1 | -1 | -1 | -1 | - | - | - |
| $A_{2u}$ | +1 | +1 | +1 | -1 | -1 | -1 | -1 | -1 | +1 | +1 | z | - | $z^3$, $z(x^2+y^2)$ |
| $B_{1u}$ | +1 | -1 | +1 | +1 | -1 | -1 | +1 | -1 | +1 | -1 | - | - | xyz |
| $B_{2u}$ | +1 | -1 | +1 | -1 | +1 | -1 | +1 | -1 | +1 | -1 | - | - | $z(x^2-y^2)$ |
| $E_u$ | +2 | 0 | -2 | 0 | 0 | -2 | 0 | +2 | 0 | 0 | (x, y) | - | $(xz^2, yz^2)$ $(xy^2, x^2y)$, $(x^3, y^3)$ |

(a)

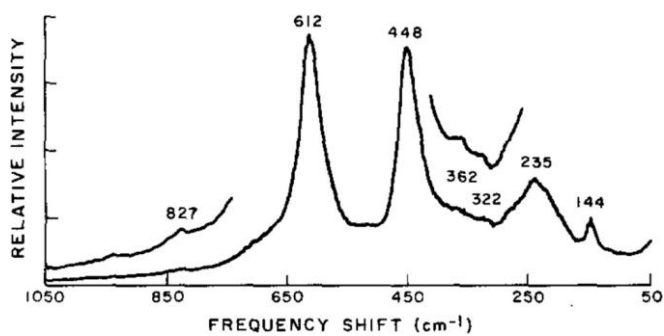

(b)

Figure 2.26: Raman active modes of rutile: (a) Consider the character table for rutile $D_{4h}^{14}$ ($P4_2/mnm$, adapted from [98]), where $A_g + B_g + B_g + E_g$ are Raman active; (b) these correspond to peaks at ~144, 448, 612, and 827 cm$^{-1}$, adapted from [99].

| $D_{4h}$ | E | $2C_4 (z)$ | $C_2$ | $2C'_2$ | $2C''_2$ | i | $2S_4$ | $\sigma_h$ | $2\sigma_v$ | $2\sigma_d$ | linear functions, rotations | quadratic functions | cubic functions |
|---|---|---|---|---|---|---|---|---|---|---|---|---|---|
| $A_{1g}$ | +1 | +1 | +1 | +1 | +1 | +1 | +1 | +1 | +1 | +1 | - | $x^2+y^2, z^2$ | - |
| $A_{2g}$ | +1 | +1 | +1 | -1 | -1 | +1 | +1 | +1 | -1 | -1 | $R_z$ | - | - |
| $B_{1g}$ | +1 | -1 | +1 | +1 | -1 | +1 | -1 | +1 | +1 | -1 | - | $x^2-y^2$ | - |
| $B_{2g}$ | +1 | -1 | +1 | -1 | +1 | +1 | -1 | +1 | -1 | +1 | - | $xy$ | - |
| $E_g$ | +2 | 0 | -2 | 0 | 0 | +2 | 0 | -2 | 0 | 0 | $(R_x, R_y)$ | $(xz, yz)$ | - |
| $A_{1u}$ | +1 | +1 | +1 | +1 | +1 | -1 | -1 | -1 | -1 | -1 | - | - | - |
| $A_{2u}$ | +1 | +1 | +1 | -1 | -1 | -1 | -1 | -1 | +1 | +1 | $z$ | - | $z^3, z(x^2+y^2)$ |
| $B_{1u}$ | +1 | -1 | +1 | +1 | -1 | -1 | +1 | -1 | -1 | +1 | - | - | $xyz$ |
| $B_{2u}$ | +1 | -1 | +1 | -1 | +1 | -1 | +1 | -1 | +1 | -1 | - | - | $z(x^2-y^2)$ |
| $E_u$ | +2 | 0 | -2 | 0 | 0 | -2 | 0 | +2 | 0 | 0 | $(x, y)$ | - | $(xz^2, yz^2) (xy^2, x^2y), (x^3, y^3)$ |

(a)

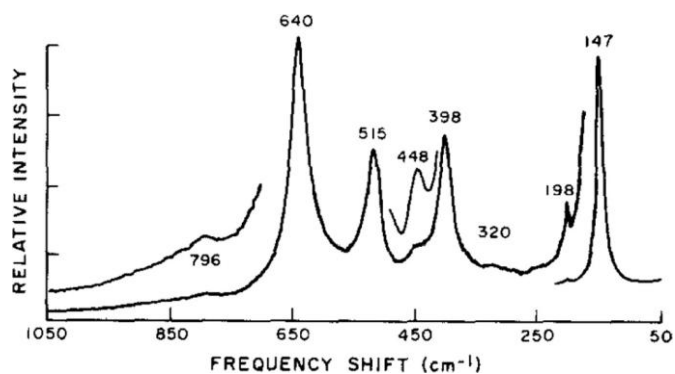

(b)

Figure 2.27: Raman active modes of anatase: (a) Consider the character table for anatase $D_{4h}^{19}$ ($I4_1/amd$), adapted from [98], where $A_{1g} + 2B_{1g} + 3E_g$ are Raman active; (b) these correspond to peaks at ~144, 448, 612, and 827 cm$^{-1}$, adapted from [99].

**Chapter 3**

**Theoretical Background**

This chapter presents relevant calculations and estimates that are important for characterizing the plasma, the forces that influence the dynamics of the various charged species in the plasma, and the magnetic field in the plasma. Several calculations are presented in the sections that follow.

In Sec. 3.1, a discussion of the relevant Debye lengths for the plasma and dusty plasma is given. This is important to understand the spatial scales over the various charged species may be impacted by static electric fields applied at plasma boundaries.

Sec. 3.2 will introduce the concept of dust charging. The fact that the grains are charged is critical because that is how the dust dynamics become coupled to the surrounding electrons and ions in the plasma. That is, the dust particles are an integral part of the plasma and not just tracers of plasma flows. Additionally, many of the forces that govern the dust grain dynamics in the plasma are determined by the grain charge. Sec. 3.3 will provide a summary and estimates of the different forces that will govern the dynamics of the dust particles in the plasma.

In Secs. 3.4 the modeling of magnetic fields produced by permanent magnets is presented followed by a discussion of charged particle dynamics in the magnetic fields as characterized by the hall parameter in Sec. 3.5.

## 3.1   Deybye Length

The debye length is a characteristic length over which charged species in a plasma are shielded. The shielding arise due to the species own charge causing them to either repel electrons and attract ions, or vice-versa, which produces a polarization charge over the characteristic length [100]. In this dissertation, the debye length will be considered when estimating the electric force in Eq. 3.9. The dust's Debye length ($\lambda_D$) is:

$$\lambda_D = \sqrt{\frac{\lambda_{De}\lambda_{Di}}{\lambda_{De}^2 + \lambda_{Di}^2}}$$

(3.1)

where $\lambda_{De}$ and $\lambda_{Di}$ are the electron and ion Debye length, respectively:

$$\lambda_{De} = \sqrt{\frac{\epsilon_o k_B T_e}{e^2 n_e}}$$

(3.2)

$$\lambda_{Di} = \sqrt{\frac{\epsilon_o k_B T_i}{e^2 n_i}}$$

(3.3)

## 3.2   Dust Charging

Since the indiviual dust grains in the plasma are typically approximated as spheres, then a model for estimating the charge of the grain would be that of a spherical capacitor [68].Under the spherical capacitor assumption, equations 3.4 and 3.5 give the relationship between the dust grain charge ($Q_d$), radius (r), and its surface potential ($\phi_s$).

$$C = 4\pi\epsilon_o r,$$

(3.4)

$$Q_d = C\phi_s$$

(3.5)

Generally, $\phi_s$ is negative because electrons are more mobile than ions, because the lower mass of the electrons (compared to the ions), make them the more mobile species in the plasmas [69]. As a result, when an uncharged object is placed in a plasma, the initial flux of electrons to the surface is greater than that of the ions. The surface will acquire a net negative charge until the flux of electrons and ions is balanced. Under this condition, the dust grain will have approximately no net current reaching the surface and achieve an "electrically floating" condition.

To determine the dust grain charge, it is necessary to have an estimate of the grain surface potential. This surface potential is defined by the aforementioned electron and ions fluxes. Numerous authors in the dusty plasma community have developed models to solve for the potential [101–106]. Eq. 3.6, below, uses the equation used by Thomas, et al., in Ref. [105], that makes an estimate of the grain surface potential, $\phi_s$, that incorporates the quasineutrality condition for the dusty plasma: $q \times N_i = q \times N_e + e \times Z_d \times N_d$. Most importantly, $\phi_s$ depends on $r$, and the equation can be numerically solved as a function of $r$ if we know the other parameters.

$$\left(1 + \frac{4\pi\varepsilon_0 r N_d \phi_s}{e N_e}\right) \left(\frac{m_i T_e}{m_e T_i}\right)^{1/2} \exp\left(\frac{e\phi_s}{k_B T_e}\right) = 1 - \left(\frac{e\phi_s}{k_B T_i}\right) \quad (3.6)$$

Most of the parameters in Eq. 3.6 are well known, for example $T_e \approx 3$eV and $T_i \approx \frac{1}{40}$eV, but $N_d$ can be estimated from measuring dust density waves (DDW) in the plasma and solving for its dispersion relation [105, 107]. We do not solve any DDW in this dissertation, but we do see them on the dust clouds, as seen in Fig. 3.1 for a carbonaceous dusty plasma without magnetic field. In general, when dust is introduced into the plasma to create a dusty plasma, $n_d \approx 10^9 \text{m}^{-3}$ [105], however, when dust grows in the plasma from gaseous chemical precursors, $N_d \approx 10^{13} \text{m}^{-3}$ [107]. Since $N_i \approx N_e \approx 10^{15} \text{m}^{-3}$, when $N_d$ increases from $\sim 10^9$ to $10^{13}$, $Z$ decreases from $\sim 10^6$ to $10^2$. After numerically solving Eq. 3.6 in Fig. 3.2, we see that $\phi_s$ decreases from about $\sim$ -7 V to almost $\sim$ -0.48 V for a dust of radius 300 nm, which is the typical maximum radius of dust in this dissertation.

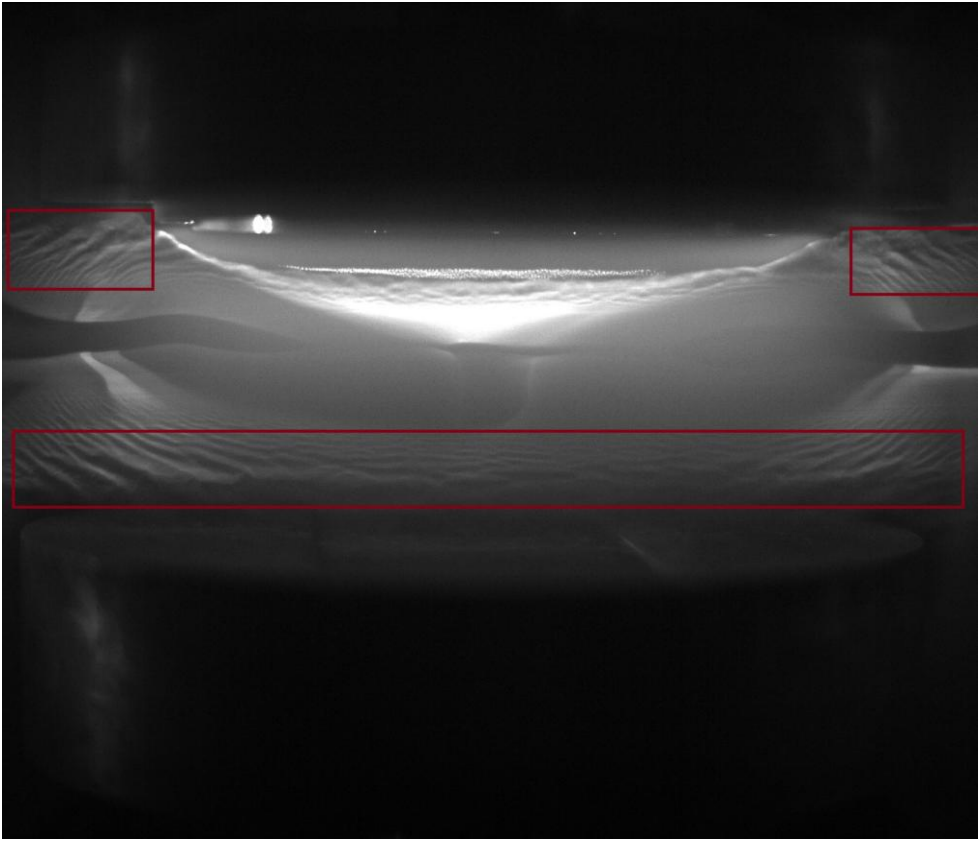

Figure 3.1: Dust density waves seen inside the red rectangles in a carbonaceous dusty plasma

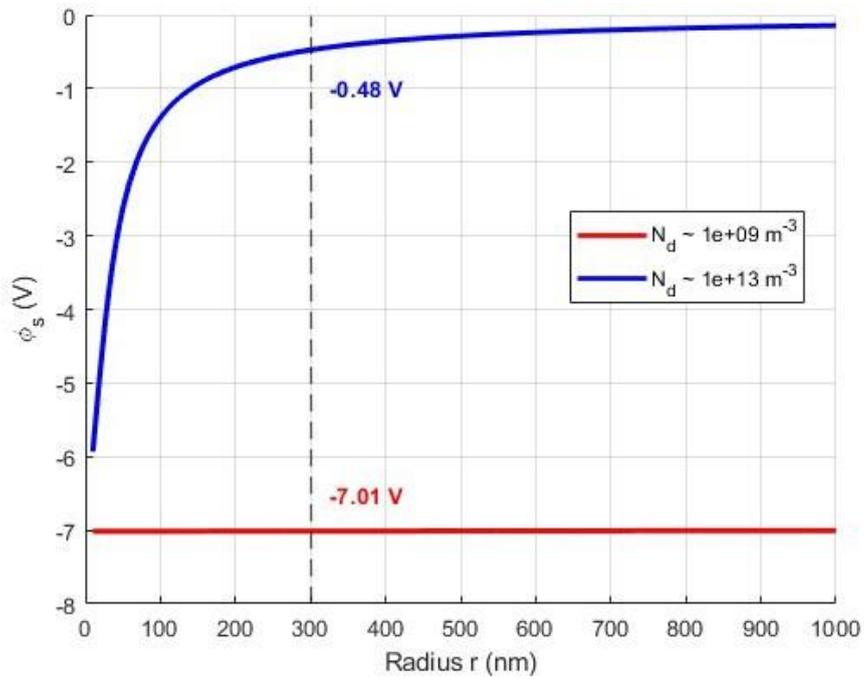

Figure 3.2: Numerical solution for the dust surface potential relative to the background plasma ($\phi_s$) as a function of dust radius r, for dust number density ($N_d$) of either $10^9 \text{m}^{-3}$ or $10^{13} \text{m}^{-3}$

## 3.3 Forces in Dusty Plasma

The dusty plasma system behaves collectively through several forces, including gravitational force ($\vec{F}_g$), electric force ($\vec{F}_e$), neutral-drag force ($\vec{F}_n$), and ion-drag force ($\vec{F}_i$). The dust particles form a cloud suspended between the electrodes by a balance of these forces [69,71]. There exist other forces in a dusty plasma, but their magnitude may be much less than the magnitude of the four mentioned forces, and hence their contribution is minimal [68]. The magnitude of these forces depends on the particle radius, with $\vec{F}_g$ scaling cubically, $\vec{F}_e$ scaling linearly, and the other forces ($\vec{F}_i$, $\vec{F}_n$) scaling quadratically. In our experiments, the particle growth cycle concludes when $\vec{F}_g$ surpasses the other forces, causing the particles to acquire enough mass to charge ratio, so that it can not be balanced, and they transport away from the central region of the plasma. We measure all the cycle time of the dusty plasma using optical emission spectroscopy from the middle of the plasma, that is in the center between the electrodes, as shown in Fig. 3.3. In this region, we assume that the longitudinal electric field $E_z \gg E_r$, the transverse electric field. Hence, we ignore $E_r$ when considering the cycle time. However, we know that $E_r$ is relevant at the edges of the plasma, as we earlier showed dust density waves forming in those locations in Fig. 3.1.

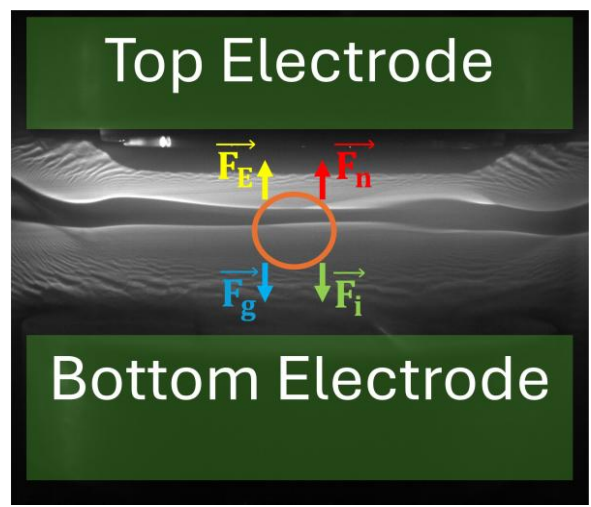

Figure 3.3: Forces in the middle of the plasma where optical emission spectroscopy records the cycle time (orange circle). electric $\left(\vec{F}_e\right)$ and neutral drag force $\left(\vec{F}_n\right)$ mostly point upwards and balance the gravitational $\left(\vec{F}_g\right)$ and ion-drag force $\left(\vec{F}_i\right)$.

### 3.3.1 Gravitational Force

As dust particles grow in size in the plasma, they become increasingly influenced by the gravitational force. As given in Eq. 3.7, the force is proportional to the mass density of the dust ($\rho_d$), the radius radius of the dust (r), and the acceleration due to gravity ($\vec{g}$).

$$\vec{F}_g = \frac{4}{3}\pi\rho_d\, \vec{g}\, r^3 \tag{3.7}$$

The gravitational force will pull the dust downward, towards the bottom electrode; however, other forces in the plasma can either balance or outweigh it, so that dust is levitated in the plasma.

### 3.3.2 Electric Force

Electric force arises from charged species interacting with the background electric field due to a potential difference between the powered electrode and grounded electrode. The electric force is given by Eq. 3.8, where $\vec{E}_{eff}$ is the effective electric field acting on each particle, given by Eq. 3.9 [68]. The effective electric force is used to accurately represent any polarization that occurs around the dust particle due to its debye length, given in Eq. 3.1, [108].

$$\vec{F}_e = Q_d\vec{E}_{eff} = 4\pi\epsilon_0\phi_s\vec{E}_{eff}r \tag{3.8}$$

$$\vec{E}_{eff} = \vec{E}\left[1 + \frac{(r/\lambda_D)}{3(1 + r/\lambda_D)}\right] \tag{3.9}$$

### 3.3.3 Neutral-Drag Force

Neutral-drag force arises from dust collisions with the neutrals atoms in the plasma, as given by Eq. 3.10 [71]. $m_n$ is the mass of the neutrals, and $\vec{u}_{dn}$ is the average velocity difference between the dust particles and neutral gas atoms [69].

$$\vec{F}_n = \frac{4}{3}\pi m_n n_n\sqrt{\frac{8k_BT_n}{\pi m_n}}\vec{u}_{dn}r^2 \tag{3.10}$$

### 3.3.4 Ion-Drag Force

Ion-drag force arises from momentum transfer between dust particles and ions which interact with the background electric field of the plasma [109]. When the electron debye length (Eq.3.2) is $\gg$ than the dust radius r, the ion drag force becomes a Coulomb collision problem [110]In our experiment, by assuming a maximum $r = 300\text{nm}$, the ratio of $\lambda_{De} : r \rightarrow 4 \times 10^{-4} : 300 \times 10^{-9} \sim 1000$. Hence, the Coulomb collision assumption is valid. Barnes et. al (1992) broke the force into two components: $F_c$ and $F_o$. The former arises from ions that gets collected on the dust surface due to Coulomb attraction, hence the term $F_c$, and the latter arises from ions that orbits around the dust, without getting attached to the dust, hence the term $F_o$. $F_c$ is given by Eq. 3.11, where $\chi \equiv -\frac{Ze\phi_s}{T_i}$ is the normalized potential, $v_{ti} = \sqrt{\frac{2T_i}{m_i}}$ is the ion thermal speed, $u = v_f/v_{ti}$, and $v_f$ is the ion flow velocity [110].

$$F_c = n_i r^2 m_i v_{ti} \sqrt{\frac{\pi}{2}} u(2u^2 + 1 + 2\chi)e^{-u^2}$$
$$+ \left[ 4u^4 + 4u^2 - 1 - 2(1-2u^2)\chi \right] \sqrt{\frac{\pi}{2}} \text{erf}(u) /u^2 \qquad (3.11)$$

$F_o$ is given by Eq. 3.12, where $G(u) \equiv \left[ \text{erf}(u) - \frac{2ue^{-u^2}}{\sqrt{\pi}} \right] /2u^2$, is the Chandrashekhar function [110]. $\ln \Lambda$ was initially introduced by Barnes at. al (1992) [109], but was later improved by Khrapak et. al (2002) [111].

$$F_o = n_i \frac{q^2\phi^2}{m_i v_{ti}^2} r^2 8\pi \, G(u) \ln \Lambda \qquad (3.12)$$

### 3.3.5 Other forces

There can be other forces in the dusty plasma, but it is generally accepted that their magnitude are smaller than the four aforementioned forces. For example, in our experiment, there can be a **thermophoretic force** ($F_{th}$) which arise from a temperature gradient between the eletrodes and a **radiative pressure force** ($F_{rad}$) which arise from the laser sheet that cuts through the dusty plasma in order or it to be visualized.

### 3.3.6 Forces estimation

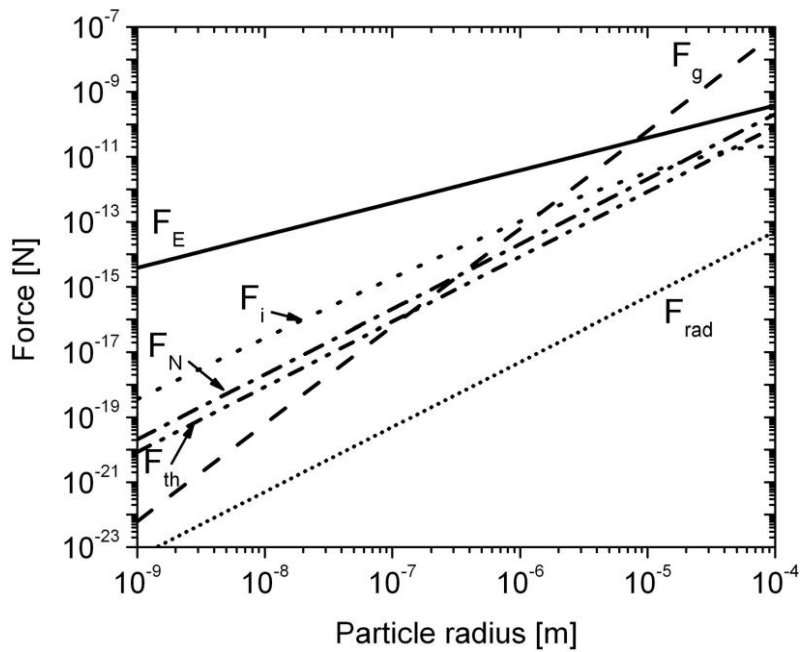

Figure 3.4: Magnitude of several forces as a function of dust radius in a dusty plasma (adapted from [70]). Thermophoretic force ($F_{th}$) and radiative pressure force $F_{rad}$ are the two weakest. However, towards the end of the growth cycle, gravitational force ($F_g$) and electric force ($F_e$) are the strongest.

It is generally agreed that the two dominant forces acting on the dust particles are the gravitational and electric force. When the particles' radii are small enough, electric force confines the dust between the electrodes, but as the particles grow in size, gravitational force cause them to no longer be balanced between the electrodes. However, the ion drag force contributes to move the dust particle sideways, away from the bulk plasma and towards the chambers's wall. This allows space for a new generation of particle growth. Consider, for example, a plot of the magnitude of forces as function of dust radius adapted from [70] in Fig.3.4. For $F_{rad}$, the laser power is assumed to be 300 milliWatts (similar to ours), and the area of the dusty plasma is assumed to be 13 cm$^2$, hence, we expect our $F_{rad}$ to be less than in this plot. Furthermore, $F_{th}$ assumes a temperature gradient of 200K/m. We have never measured the temperature gradient between our electrodes, but we do not expect much deviation from this value in our experiments.

We estimate the two strongest force at the end of the growth cycle for our carbonaceous and titania dusty plasma using several parameters listed in Table 3.1, with their source arising either from our study in this dissertation or from recent literature. For example, we have approximated the longitudinal electric field $\vec{E}$ in our bulk plasma (ignoring the sheath) to be about 120 V/m from 1-D COMSOL Multiphysics simulation ran at 300 milliTorr and 60 Watts, however we can only estimate the dust surface potential $\phi_s$ as function of radius from numerical calculation that assumes $n_d \approx 10^{13} \mathrm{m}^{-3}$.

Table 3.1: Estimation of parameters for some forces in our dusty plasma.

| Parameter | Description | Value | Source |
|---|---|---|---|
| $T_i/T_n$ | Ion/Neutral temperature | 300 K | Room Temperature |
| $T_e$ | Electron temperature | 3 eV | Langmuir Probe |
| $n_e$ | electron density | $10^{15}\,\mathrm{m}^{-3}$ | Langmuir Probe |
| $n_i$ | ion density | $10^{15}\,\mathrm{m}^{-3}$ | quasi-neutrality |
| $m_n$ | Mass of neutral gas | $6.6 \times 10^{-26}\,\mathrm{kg}$ | Argon |
| $m_i$ | Mass of ionized gas | $6.6 \times 10^{-26}\,\mathrm{kg}$ | Argon |
| $\rho_c$ | Mass density of carbon dust | $1300\,\mathrm{kgm}^{-3}$ | [112] |
| $\rho_c$ | Mass density of titania dust | $3700\,\mathrm{kgm}^{-3}$ | [113] |
| $g$ | Acceleration due to gravity | $9.8\,\mathrm{ms}^{-2}$ | Gravity |
| $E$ | Electric field | $20\,\mathrm{Vm}^{-1}$ | Maximum particle size |
| $e$ | Electronic charge | $10^{-19}\,\mathrm{C}$ | Constant |

We use the parameters from Table 3.1 to plot a graph of the magnitude of gravitational and electric force as a function of radius for carbonaceous and titania dust, from which we can predict the former to grow bigger than the latter. hence, carbonaceous dust should have a longer growth cycle than titania, because the latter will lose its force balance faster. The code to plot the graph is given in Appendix F. This is illustrated in Fig.3.5.

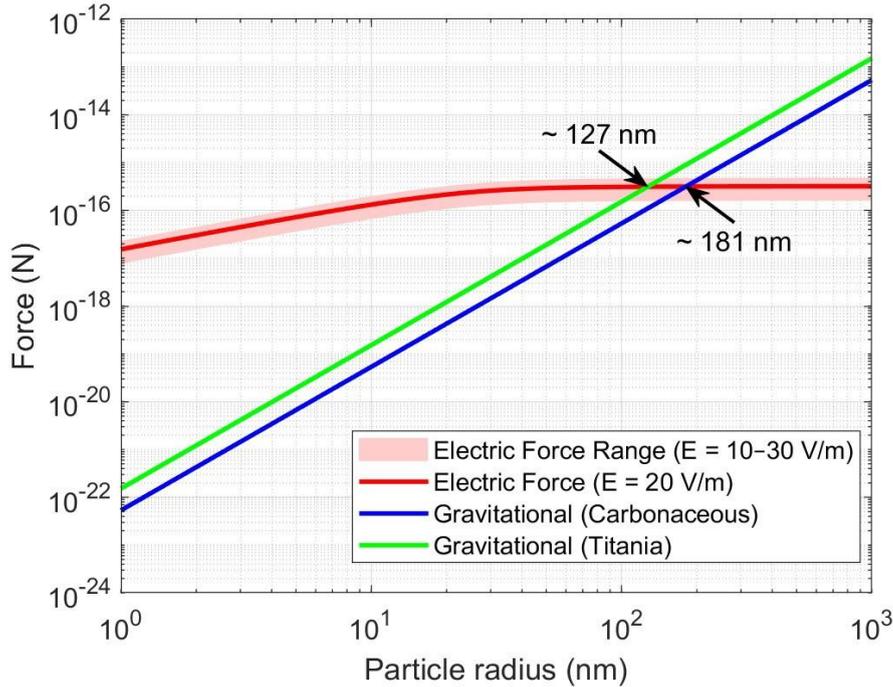

Figure 3.5: Forces as a function of radius. The electric force is calculated with the $\phi_S$ as a function of radius and assuming the magnitude of E to be in a range of 10 to 30 V/m. The mass density of carbonaceous and titania nanoparticles are estimated to be 1300 kg/m$^3$ and 3700 kg/m$^3$, respectively. According to this plot, the maximum radius of carbonaceous dust is slightly bigger than titania (pointed by the black arrow), due to the former's smaller mass density.

## 3.4 Magnetic Fields from Axially Magnetized Cylinders

In this dissertation, we use permanent cylindrical magnets in pairs to create a region of magnetic fields between the electrodes. For example, we show a picture of the magnet holder assembly in Fig. 2.10 in section 2.2. Magpylib is used to simulate the magnetic field strength created by several combinations of magnets. Maypylib is a Python package which can be imported to calculate magnetic fields in 3-D for several geometries [114]. The code used in this dissertation is provided in appendix D. The documentation for the package can be found on this website: `https://magpylib.readthedocs.io/en/stable/`

The equations used for the field computation are provided in this section in order to describe the magnetic field simulation of permanent magnet rings used in this dissertation. First of all, it is assumed that a permanent magnetic cylinder can be approximated an ideal solenoid [116]. An ideal solenoid is modeled as a thin azimuthal sheet of infinite number of wires, carrying current $I_{total}$, tightly wrapped around a cylinder. The magnitude of magnetization is $|\vec{M}| = \mu_0 = nI$, where n is the number of turns per unit length and I is the current per turn. Consider then a cylinder of length L = 2b and radius R = a, wrapped with an arbitratry magnetization $\vec{M}$. A schematic representation of the cylinder and the field lines around it is shown in Fig. 3.6a [115] and 3.6b [116], respectively. Since $\vec{M}$ has a longitudinal (axial) ($\hat{z}$) and transverse ($\hat{\rho}$) component in cylindrical coordinates, then the magnetic field $\vec{B}$ can also be decomposed into a radial ($B_\rho$) and axial ($B_z$) direction. An angular direction ($\hat{\phi}$) is absent due to radial symmetry. Derby et. al (2009) has solved this problem (provided in Eq. 3.13 and 3.14), which is the physics used by Magpylib for the simulation of permanent cylindrical magnets [116]. Eq.3.16 is a generalized elliptic integral. The expressions for the magnetic field components are:

$$B_\rho = B_0 \left[ \alpha_+ C(k_+, 1, 1, -1) - \alpha_- C(k_-, 1, 1, -1) \right] \qquad (3.13)$$

$$B_z = \frac{B_O a}{a + \rho} \Big[ \beta_+ C(k_+, \gamma^2, 1, \gamma) - \beta_- C(k_-, \gamma^2, 1, \gamma) \Big] \qquad (3.14)$$

with the following definitions:

$$B_O = \frac{\mu_O}{\pi} nI \qquad (3.15)$$

where $B_O$ is the net remnant magnetic field on the surface of the cylinder, a value which is provided by the magnets supplier (13200 Gauss for our magnets).

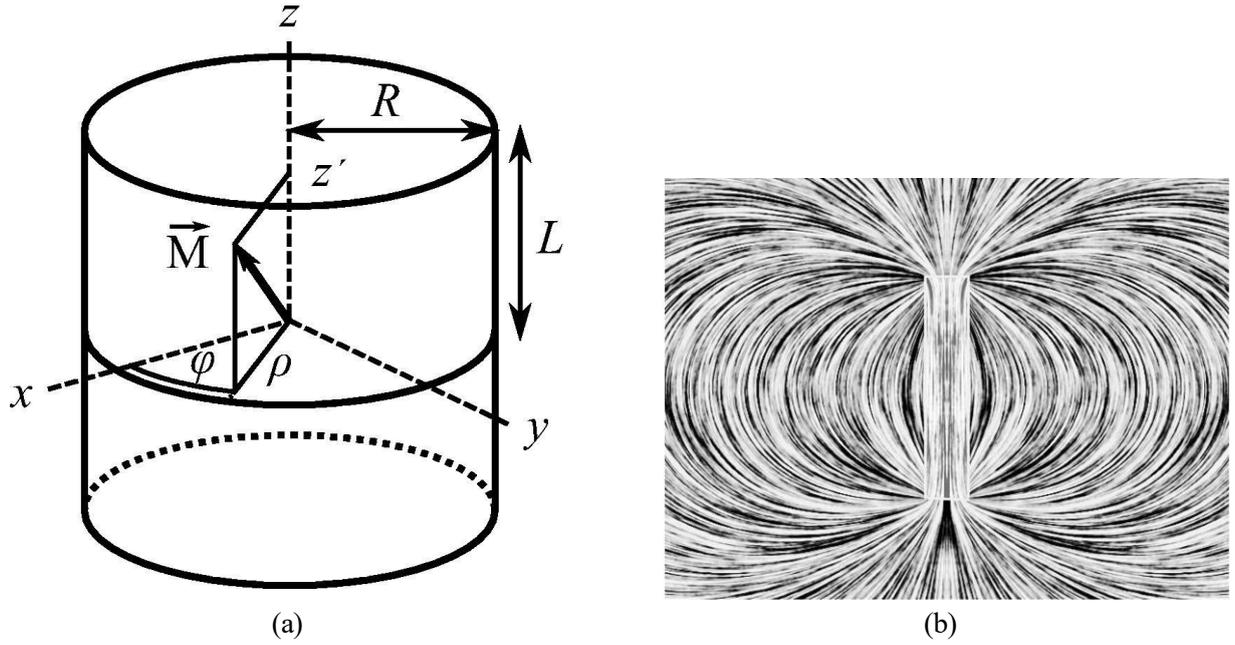

(a)                                    (b)

Figure 3.6: Cylinder magnets approximated by an ideal solenoid. (a) Schematic representation of a cylinder of length L = 2b and radius R = a wrapped tightly with a sheet of magnetization $\vec{M}$ (Adapted from [115]). The geometry is defined in both cylindrical ($\rho, \phi, z$) and Cartesian (x, y, z) coordinates. (b) Schematic representation of 2-D field lines around a cylinder (perfect solenoid) where length is 5 times the diameter (L = 5 × 2R) (adapted from [116]).

$$C(k_c, p, c, s) = \frac{\pi}{2} \int_0^{\frac{\pi}{2}} \frac{c \cos^2 \varphi + s \sin^2 \varphi}{(\cos^2 \varphi + p \sin^2 \varphi) \sqrt{\cos^2 \varphi + k_c^2 \sin^2 \varphi}} d\varphi \qquad (3.16)$$

where $C$ is a generalized complete elliptic integral which can be solved numerically both inside and outside the solenoid.

The remaining terms below are all shorthand notations that feed into Eq. 3.13 and 3.14.

$$z_\pm = z \pm b \qquad (3.17)$$

$$\alpha_\pm = \frac{a}{\sqrt{z_\pm^2 + (\rho + a)^2}} \qquad (3.18)$$

$$\beta_\pm = \frac{z_\pm}{\sqrt{z_\pm^2 + (\rho + a)^2}} \qquad (3.19)$$

$$\gamma = \frac{a - \rho}{a + \rho} \qquad (3.20)$$

$$k_\pm = \sqrt{\frac{z_\pm^2 + (a - \rho)^2}{z_\pm^2 + (a + \rho)^2}} \qquad (3.21)$$

In general, Eqs. 3.13 and 3.14 can be solved using the elliptic integral $C$ at various geometrical limits such as when $\rho = 0$, when $b = 0$, and even at distances $\gg (a, b)$. However, $C$ does not behave well on the edges of the cylinders, that is at $\rho = a$ and $z = \pm b$, where division by zero can arise. In our experiments, we will focus in the region between two permanent magnets and not worry about the location at edges of the magnets. Hence, we do not expect to be impacted by the limitations of the numerical solution.

### 3.4.1 Magnetic Rings

We will use the numerical solution for magnetic fields around permanent cylinders in order to build the fields around a magnetic ring, which is simply the difference between a cylinder of a larger radius ($R_o$) and a cylinder of a smaller radius ($R_i$), as schematically shown in Fig. 3.7. $R_o$ and $R_i$ are commonly referred to as the "outer radius (OD)" and "inner radius (ID)", respectively. Eq. 3.13 and 3.14 are still valid when computing the magnetic field from the principle of superposition taking the difference between two cylinders.

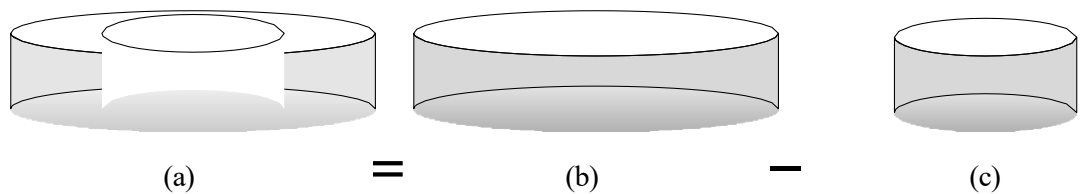

(a) $\qquad$ = $\qquad$ (b) $\qquad$ − $\qquad$ (c)

Figure 3.7: Schematic representation of (a) a magnetic ring, which is the difference between (b) a cylinder of larger radius $R_o$ and (c) one of smaller radius $R_i$, both of equal height.

These parameters were input in Magpylib to simulate the magnetic field: the remnant magnetic field $B_o = 1.32$ T, the length of the cylinder $2b = 6.35$ mm, the outer radius $R_o = 50.8$mm, the inner radius $R_i = 25.4$mm and the position(s) of the center along the Z-axis $(0, 0, Z_i)$. For every magnets used $(0, 0, Z_i)$ was specified, for example, see $(0, 0, Z_1), (0, 0, Z_2)$, and $(0, 0, Z_3)$ for three magnets stacked in Fig. 3.8. In our experiments, magnets were always used in pairs. For example, if 3 magnets were stacked above the top electrode, then 3 magnets were stacked below the bottom electrode, equidistant from the center of the electrode spacing.

### 3.4.2 Examples of magpylib simulations in 2-D

### ∼ 500 Gauss

One pair of magnet was used and each magnet was placed above and below the plasma electrodes as shown in Fig. 2.10a. This produced a nominal magnetic field at the mid-point between the electrodes of $\approx 500$ Gauss. The magnetic field lines distribution in a two-dimensional plane at the center of the electrode is shown in Fig. 3.9a. Moreover, Fig. 3.9b shows the radial profile of $B_z$ at the middle of the two electrodes, i.e. $z = 12.5$ mm, and Fig. 3.9c shows the profile of $B_z$ as a function of z at x = 0 mm. The red values are measured using F.W. Bell Gauss/Teslameter Model 5080 and blue lines are calculated using magpylib.

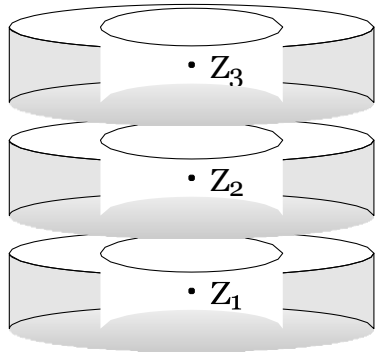

Figure 3.8: Schematic representation of three stacked magnetic rings labeled at their centers as $(0, 0, Z_1)$, $(0, 0, Z_2)$, and $(0, 0, Z_3)$. There is a gap of approximately 3 mm between each magnet because a "spacer" is used between the magnets.

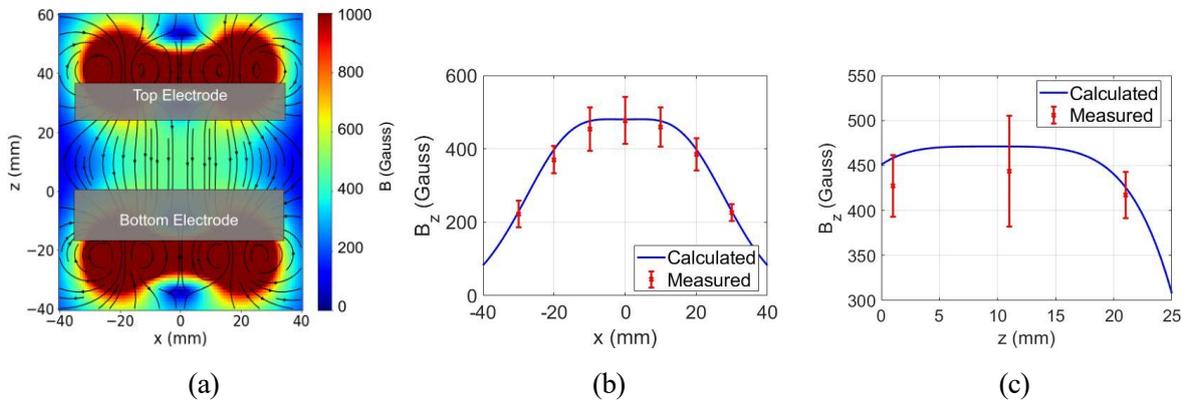

Figure 3.9: Magnetic field lines between the two permanent magnet rings in the plasma region. (a) Two-dimensional plane at the center of the electrodes. The arrows and lines represent the net magnetic field direction and the colors represent the magnetic field net as labeled by the color bar. (b) Radial variation of $B_z$ vs x at $z = 12.5$ mm. (c) Transverse variation of $B_z$ vs z at x = 0 mm. (Adapted from [53].)

### ~ 18 - 1020 Gauss

To produce a range of magnetic fields from ~ 18 - 1020 Gauss, either 1 or 2 or 3 pairs of permanent magnet rings were used. These magnets were placed symmetrically at z = 12.5 mm from the center of the electrode spacing to maximize the magnetic field at the center of the plasma as measured at the geometric center between the top and bottom electrodes at x = 0 mm and z = 12.5 mm. The maximum separation between a pair of magnets (255 mm) created a field strength of approximately 18 Gauss. The spacing between the magnets was gradually decreased in each experiment to produce various field strengths. The minimum separation between a pair of magnets (64 mm) created a field strength of approximately 460 Gauss. At this separation, either one or two additional magnets were stacked on the original pair to increase the magnetic field strength to either 790 or 1020 Gauss, respectively. Two examples of magnetic field strength are shown in Fig. 3.10 for 18 and 1020 Gauss. 2-D plots of magnetic field strength that span the region between the electrodes are shown, along with a line plot of magnetic field strength along the middle at z = 12.5 mm. A summary of the different magnetic field strengths, the number of pairs of magnets, and their $Z_i$ locations at the top and bottom is provided in Table 3.2.

Table 3.2: Magnetic field strengths and corresponding locations $Z_i$ of magnet pairs.

| Field Strength (Gauss) | Pairs of Magnets | Bottom $Z_i$ (mm) | Top $Z_i$ (mm) |
|---|---|---|---|
| 18 | 1 | -115 | 140 |
| 26 | 1 | -100 | 125 |
| 73 | 1 | -65 | 90 |
| 87 | 1 | -60 | 85 |
| 120 | 1 | -52 | 77 |
| 195 | 1 | -39 | 64 |
| 233 | 1 | -35 | 60 |
| 335 | 2 | -39 | 64 |
| | | -47 | 72 |
| 460 | 1 | -22 | 42 |
| 790 | 2 | -22 | 42 |
| | | -30 | 50 |
| 1020 | 3 | -22 | 42 |
| | | -30 | 50 |
| | | -38 | 68 |

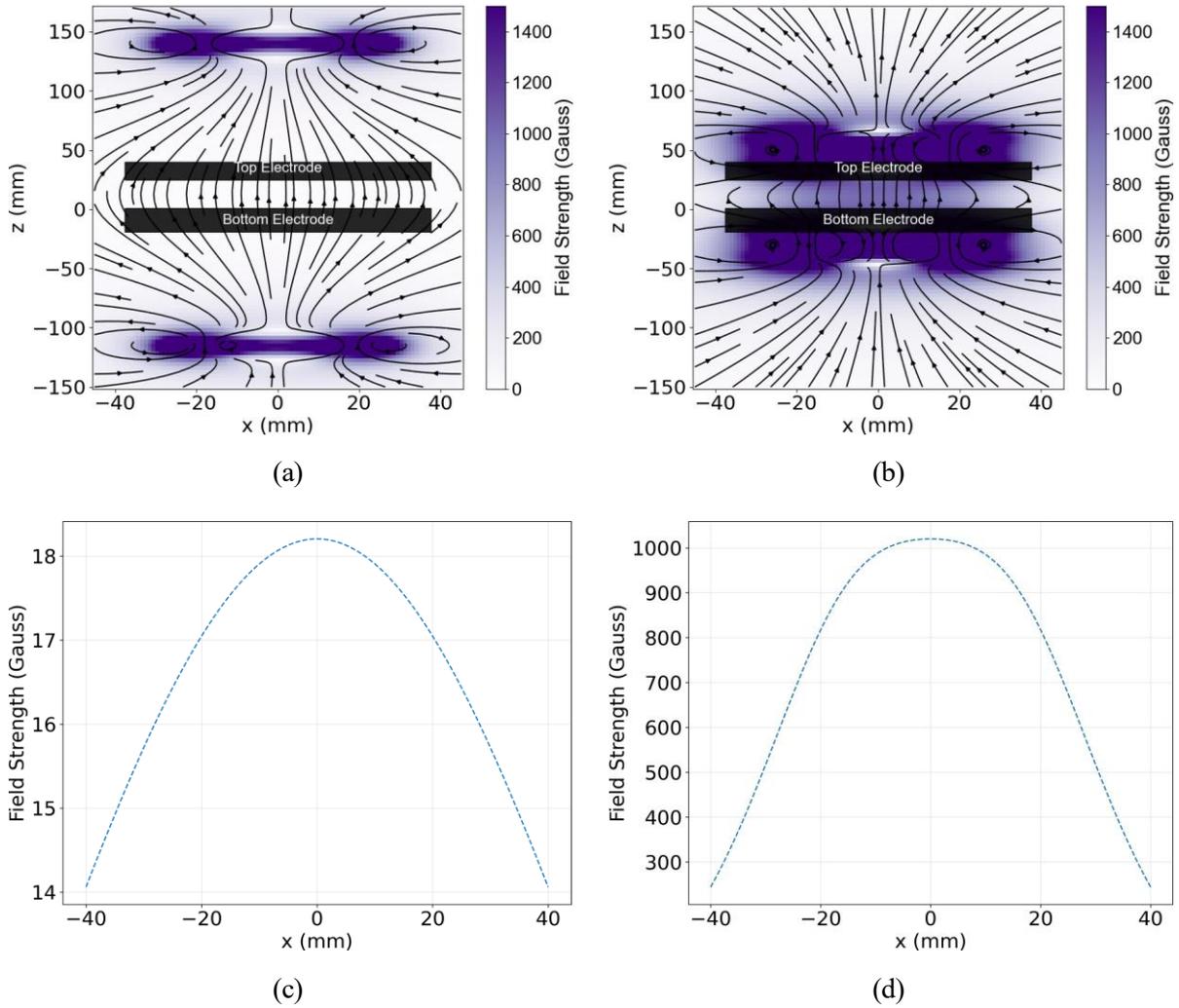

Figure 3.10: Example of the 2-D magnetic field between the electrodes. (a) At the largest magnet separation (255 mm), the field reaches a maximum of approximately 18 Gauss at the center of the electrodes (x = 0 mm, z = 12.5 mm). (b) At the closest separation (66 mm), two additional pairs of magnets are stacked, producing a maximum field of approximately 1020 Gauss at the same location. (c) and (d) show the magnetic field strength along x at z = 12.5 mm for 18 Gauss and 1020 Gauss, respectively. (Adapted from [83].)

## 3.5  Modified Electron Hall Parameter $H_e{}'$

The Hall parameter has been used by a number of dusty plasma research groups to quantify the degree of magnetization of charged species in plasmas [117, 118]. In general, it is defined as a ratio of the gyrofrequency ($\omega_c$) to neutral collision frequency ($\nu_n$), that is ($H = \omega_c/\nu_n$), where if $H \gg 1$, then a charged particles is considered to be magnetized. As noted in Table 3.3, only the electrons satisfy that criterion. Moreover, within our group, the definition of the Hall parameter has been modified to provide a more physical interpretation of this dimensionless parameter [119, 120]. Since $\omega_c$ is an angular frequency (rad/s) and $\nu_n$ is a linear frequency (Hz), a normalization factor of $2\pi$ is introduced in the denominator, yielding the modified Hall parameter $H_e{}'$ as shown in Eq. 3.22. The remainder of this section focuses on the definition of the modified electron Hall parameter ($H_e{}'$) that will be used in this work.

$$H_e{}' = \frac{\omega_{c,e}}{2\pi\nu_{n,e}} \tag{3.22}$$

### 3.5.1 Collision Frequency, $\nu_n$

As discussed in the introduction, LTPs such as those in described in this work, are weakly ionized. For typical conditions in these experiments, at a neutral pressure of p = 300 mTorr ($N_n \sim 1 \times 10^{22} \mathrm{m}^{-3}$) and electron/ion densities $N_e \approx N_i \approx 10^{15}\mathrm{m}^{-3}$, then the ionization fraction is $\sim 10^{-7}$.

Consequently, the role of charged species - neutral collisions must be taken into account when trying to understand the dynamics of the plasma. For the electrons and ions, the collision frequency will be defined by three terms, the number density of neutrals in the plasma ($N_n = k_B T/p$), a characteristic collision cross-section ($\sigma$), and the thermal velocity of the charged species ($v_{th}$). The collision frequency ($\nu_{n,e}$) between electrons and neutrals is given by Eq. 3.23,where $N_n$ is the neutral density. $V_{th,e}$ is the thermal speed of electrons, given by Eq. 3.24.

$$\nu_{n,e} = N_n V_{th,e}\sigma_{n,e} \tag{3.23}$$

$$v_{th,e} = \sqrt{\frac{8k_B T_e}{\pi m_e}} \tag{3.24}$$

$\sigma_{n,e}$ is the collision cross section between neutrals and electrons, which we assume to be $\sim 5 \times 10^{-20}\,\mathrm{m}^2$ based on recent literature [121].

We can also calculate the collision frequency for ions ($\nu_{n,i}$) as given by Eq. 3.25. In the literature, $\sigma_{n,i} \approx 3.27 \times 10^{-18} m^2$, a value that previous students have used in our laboratory [118] based on Khrapak et. al's (2013) calculations [122]. Furthermore, the ion thermal velocity, $V_{th,i}$, given in Eq. 3.26, can be calculated by assuming $T_i = 300K$.

$$\nu_{n,i} = N_n V_{th,i} \sigma_{n,i} \tag{3.25}$$

$$v_{th,i} = \sqrt{\frac{8k_B T_i}{\pi m_i}} \tag{3.26}$$

The collision cross section calculation for dust particles is slightly modified when compared to electrons and ions given the much more massive size of the dust. Ions and electrons are basically point particles whereas dust are macroscopic particles. When dust move through a column of background neutrals, we need a $\pi r^2$ term to account for the dust object moving through a column of gas. We also need to account for momentum transfer with neutrals collision, hence the presence of $m_n V_{tn} N_n$ terms where $m_n$ is the mass of the neutrals, $V_{tn}$ is the thermal speed of the neutrals (given by Eq. 3.27) and $N_n$ is the neutral density. The dust colision frequency is then given by Eq. 3.28, where $m_d$ is the mass o the dust [123].

$$v_{th,n} = \sqrt{\frac{8k_B T_n}{\pi m_n}} \tag{3.27}$$

$$v_{n,d} = \frac{4\pi r^2 m_n v_{tn} N_n}{3m_d} \tag{3.28}$$

### 3.5.2 Cyclotron Motion – Gyrofrequency

Charged species experience circular motion around magnetic field lines due to the Lorentz force as described in Eq. 3.29 [124]. $m$ is the mass of the species, $q$ is the charge of the species, $\vec{E}$ is the background electric field experienced by the plasma, $\vec{v}$ is the velocity of the species and $\vec{B}$ is the background magnetic field.

$$m\frac{d\vec{v}}{dt} = q\vec{E} + q\left(\vec{v} \times \vec{B}\right) \tag{3.29}$$

The permanent magnets are aligned so that the dominant field lines are in the $\hat{z}$ direction. Generally, there is a small field in in the $\hat{\rho}$ direction, but, at the midpoint between the electrodes where the studies of this dissertation are focused, $B_\rho = 0$, thus $\vec{B} = (0, 0, B_z)$. Similarly, $\vec{E} = (0, 0, E_z)$ in further treatment of Eq. 3.29 and decompose it into:

$$m\dot{v}_x = qBv_y \tag{3.30}$$

$$m\dot{v}_y = -qBv_x \tag{3.31}$$

$$m\dot{v}_z = qE_z \tag{3.32}$$

The presence of the vertical electric field will appear as a drift parallel to the magnetic field. For the components of particle motion transverse to B, equations 3.30 and 3.31 are differentiated with time to obtain:

$$m\ddot{v}_x = qB\dot{v}_y = qB\left(-\frac{qB}{m}v_x\right) = -\frac{q^2B^2}{m}v_x \tag{3.33}$$

$$m\ddot{v}_y = -qB\dot{v}_x = -qB\left(\frac{qB}{m}v_y\right) = -\frac{q^2B^2}{m}v_y \tag{3.34}$$

This can be considered as:

$$\ddot{v}_x = -\omega_c^2 v_x, \qquad \ddot{v}_y = -\omega_c^2 v_y \tag{3.35}$$

Eq. 3.35 shows that the motion of a charged particle in a magnetic field will be in the form of periodic motion (e.g., harmonic oscillator-like) in the direction transverse to the magnetic field. The characteristic frequency of this motion, the cyclotron frequency, $\omega_c$, is applicable to each charged species (i.e., electrons, ions, and charged dust). Each species will have a characteristic scale because of the differences in charge and mass. Table 3.3 compares the characteristic gyromotion for the electrons, ions, and dust particles in this work, based on a maximum magnetic field strength of B = 1000 Gauss = 0.01 T, which is the maximum magnetic field achievable. Circular motion around magnetic field lines behaves like a simple harmonic oscillator with cyclotron frequency of $\omega_c$:

$$\omega_c = \frac{|q|B}{m} \tag{3.36}$$

Table 3.3: Cyclotron frequencies ($\omega_c$), collision frequencies ($\nu$), and Hall parameters (H) of charged species in a magnetic field of B = 0.01 T. We assumed the dust had a radius of 300 nm and mass density of 1300 kg/m$^3$ (carbonaceous dust). The electron and ion temperature were 3 eV and 1/40 eV, respectively.

| Species | Charge $|q|$ [C] | Mass m [kg] | $\omega_c$ [rad/s] | $\nu$ [s$^{-1}$] | H |
|---|---|---|---|---|---|
| Electron | $1.6 \times 10^{-19}$ | $9.1 \times 10^{-31}$ | $1.76 \times 10^{9}$ | $6 \times 10^{7}$ | 29 |
| Argon ion (Ar$^+$) | $1.6 \times 10^{-19}$ | $6.6 \times 10^{-26}$ | $2.42 \times 10^{3}$ | $1 \times 10^{7}$ | $2 \times 10^{-4}$ |
| Dust particle | $1.6 \times 10^{-17}$ | $1.5 \times 10^{-16}$ | $1.07 \times 10^{-3}$ | 214 | $5 \times 10^{-6}$ |

Table 3.3 gives estimates of the neutral collision frequencies of the electrons, ions, and dust for the conditions that are used in these experiments. It is shown that for the ions and the dust grains, the neutral collision frequencies substantially exceed their gryofrequencies, while the opposite is true for the electrons. This fact motivates the consideration of a normalized quantity known as the Hall parameter as a measure of the "degree of magnetization" of charged particles in a plasma.

### 3.5.3 Plots of $H_e{}'$

The modified Hall parameter is a function of magnetic field strength, B, and pressure P, since $N_n = P/k_B T_n$, where $T_n$ is neutral temperature. Therefore $H_e{}'$ scales as Eq. 3.37.

$$H_e{}' = \frac{\frac{|q_e|B}{m_e}}{n_n V_{th,e} \sigma_{n,e}} \rightarrow \frac{B}{P_n} \tag{3.37}$$

This means $H_e{}'$ is directly proportional to B and inversely proportional to pressure. We plot $H_e{}'$ as a function magnetic field strength for two operating chamber pressures in this dissertation (300 mTorr and 500 mTorr), and as a function of pressure for two constant magnetic field strengths (110, and 335 Gauss ) in Fig. 3.11 and 3.12, respectively.

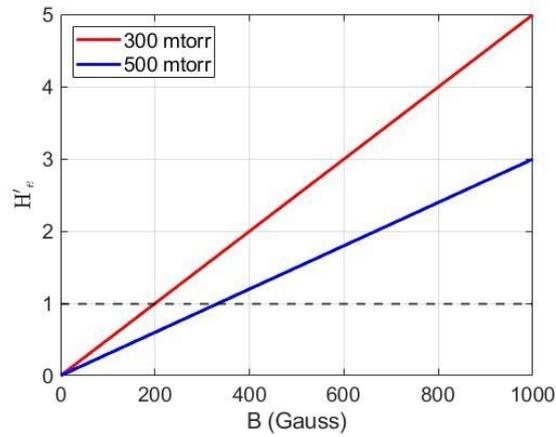

Figure 3.11: Modified electron Hall parameter ($H_e^{'}$) as a function of magnetic field strength for two experimental pressures in this dissertation.

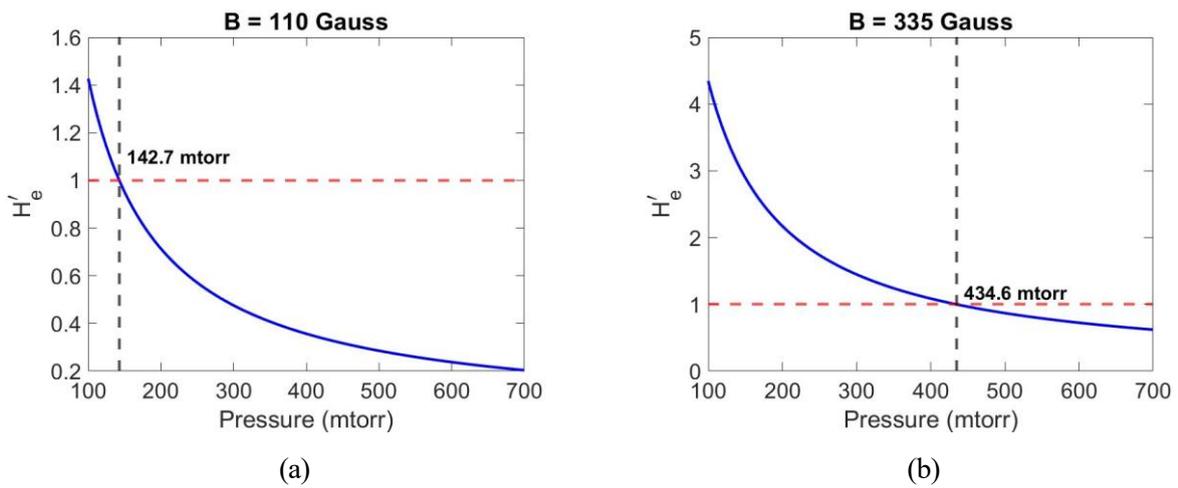

(a)                                    (b)

Figure 3.12: Modified electron Hall parameter ($H_e^{'}$) as a function of pressure for two constant magnetic field strengths: (a) 110 Gauss, (b) 335 Gauss. This shows the inverse relationship between $H_e^{'}$ and pressure.

# Chapter 4

# Experimental Results

For the experiments performed for this dissertation, there were three major tasks. The first task was to demonstrate if the TTIP precursor could successfully form titania nanoparticles in the plasma. Section 4.1 discusses the experimental procedure, the characterization of the grown nanoparticles, and describes some differences in the growth cycle of titania nanoparticles that distinguish them from earlier studies at higher pressure.

These differences motivated the second task, which is a comparison of the growth processes of the titania nanoparticles vs. carbonaceous nanoparticles under conditions with and without magnetic fields as discussed in section 4.2. Here, it is shown that at magnetic field strengths $\sim 500$ G (0.05 T), where the electrons are magnetized, but the ions and dust particles are not, structural differences in the formations of the two types of particle clouds arise and the particle growth cycle time decreases.

This then motivated the third and final set of experiments, in section 4.3, which focused on a detailed investigation of the growth of carbonaceous nanoparticles as a function of a range of magnetic field strength. This work will show that there is evidence that the onset of electron magnetization may be an important threshold condition for the evolution of nanoparticle growth.

Together, this series of experimental studies have led to one of the first demonstrations of titania (TiO2) nanoparticle growth in a laboratory dusty plasma and gives new insights into the influence of magnetic fields on reactive plasma processes.

## 4.1    Growth and characterization of monodispersed titania nanospheres as dusty plasma

The results presented in this section were previously published in *B. Ramkorun, et al, Appl. Phys. Lett. 124, 144102 (2024)* [72]. This chapter will present that work with some additional text to explain its connection to the other studies conducted as part of this dissertation.

### 4.1.1    Introducing Titanium Dioxide (Titania)

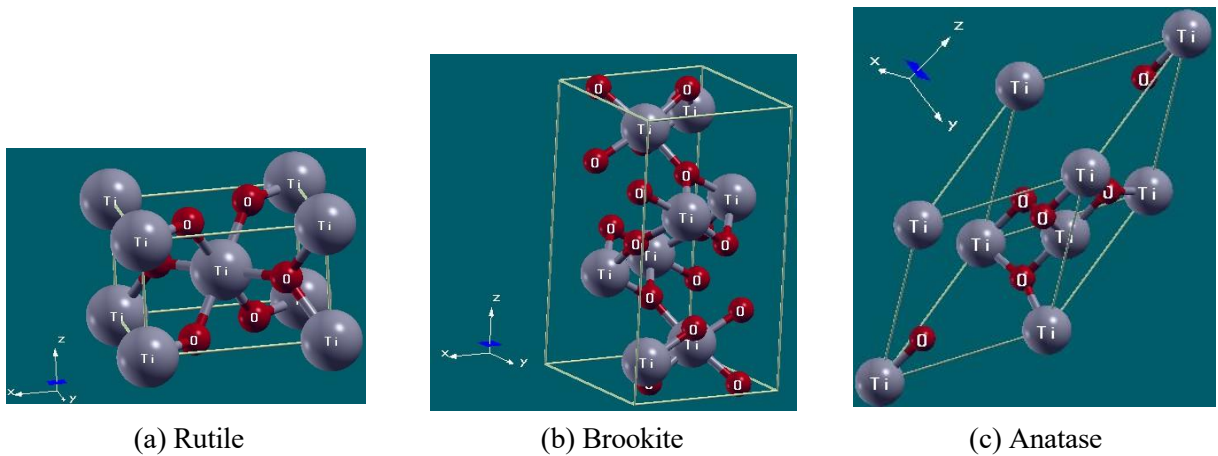

(a) Rutile                    (b) Brookite                    (c) Anatase

Figure 4.1: Crystal structures of (a) rutile, (b) brookite, and (c) anatase.

Titanium dioxide ($TiO_2$) (a.k.a Titania) is a transition metal oxide which exists in multiple polymorphs, most commonly rutile, brookite, and anatase. Rutile is naturally abundant as a mineral and is the most common polymorph. Brookite and anatase are other naturally occuring polymorphs which are rarer than rutile. Studies related to rutile, anatase and brookite are commonly found in literature with their bandgap reported as 2.98 eV, 3.05 eV and 3.26 eV respectively [125–127].

Rutile is tetragonal with space group $P4_2/mnm$. Its crystal stucture is shown in Fig. 4.1a. The cell parameters are a = b = 4.673 Å and c = 3.023 Å. The nearest neighbor distance is 1.96 Å. Brookite is orthorhombic with space group $Pbca$. Its crystal stucture is shown in Fig. 4.1b. The cell parameters are a = 5.267 Å, b = 5.543 Å and c = 9.352 Å. The nearest neighbor distance is 2.02 Å. Anatase is body centered tetragonal with space group $I4_1/amd$. Its crystal stucture is shown in Fig. 4.1c. The cell parameters are a = b = 3.87 Å and c = 9.69 Å. The nearest neighbor distance is 2.00 Å. Rutile and anatase have 6 atoms in the unit cell, in contrast to brookite which has 24. All three polymorph have one Ti octahedrally coordinated to 6 O.

The growth process was initiated using the chemical precursor titanium (IV) isopropoxide (TTIP) ($Ti(OC_3H_7)_4$) as the reactive precursor. X-ray diffraction (XRD) and Raman spectroscopy were employed to measure the crystal phases. SEM was used to image and subsequently measure the sizes of the particles grown as a function of time. SEM energy-dispersive X-ray spectroscopy (EDS) was used to ascertain the particles' chemical composition. Nucleation of nanoparticles and the formation of spherical aggregates are essential prerequisites for material growth. This phenomenon has been well-documented not only for $TiO_2$ thin films [128, 129], but also for various films containing carbon [10, 130–132]. Dusty plasma has already exhibited rapid nucleation of nm-sized carbon particles within ms, which can swiftly grow to hundreds of nanometers within seconds [59, 133]. Here we demonstrate that dusty plasma processes can grow $TiO_2$ nanospheres within 10 s. These spheres continue to grow linearly until the end of a growth cycle. Although the as-grown particles collected after 70 s were initially amorphous, they crystallized into either anatase or rutile upon annealing at different temperatures.

The experimental steps of this chapter follows closely the steps provied in the flowchart of Fig. 4.2, except that we do not grow any carbonaceous dust here. A schematic of the experimental set up is shown in Fig. 4.3a, which shows the PID temperature control to flow TTIP vapor into the chamber. The plasma was energized at 60 Watts (W) during the initial 10 s and eventually at 30 W for the remainder of the experiments. This was to promote nucleation for 10 s, followed by continued growth at a reduced power [134].

Throughout this chapter, several terms will be used that are described here. *Energize* refers to the application of RF power to the top electrode for ionizing gases and igniting plasma. *De-energize* means the removal of RF power, causing the plasma to revert back to neutral gases. *Transport* occurs when the charged dust particles, having accumulated sufficient mass, leave the central plasma but are deflected by electric forces in the plasma to the sides of experimental volume, as depicted in, as depicted in Fig. 4.4a. *Deposition* occurs when the plasma is de-energized, and dust particles experience gravitational and neutral drag forces, leading them to fall on the substrate, as depicted in Fig. 4.4b.

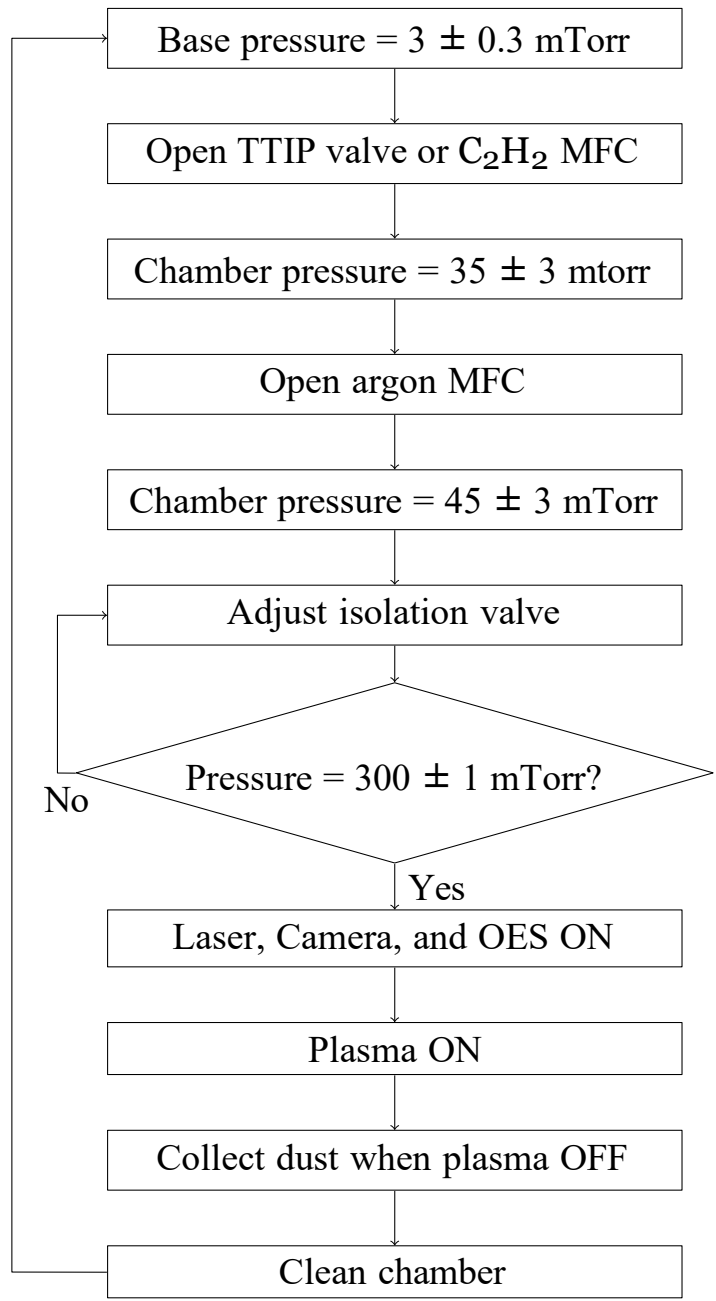

Figure 4.2: Flowchart to summarize the steps of the experiments. (Adapted from [53].)

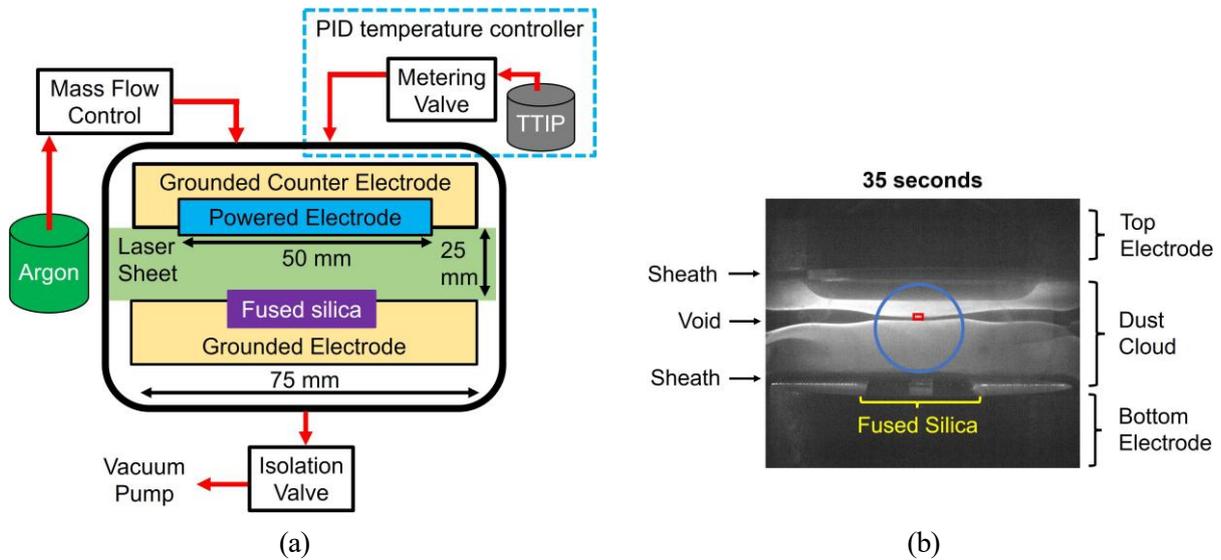

(a)                                                  (b)

Figure 4.3: (a) Schematic view of the modified experimental setup (not to scale). (b) Photograph of dusty plasma at 35 seconds showing sheath, void, dust, electrodes, and fused silica (adapted from [72]).

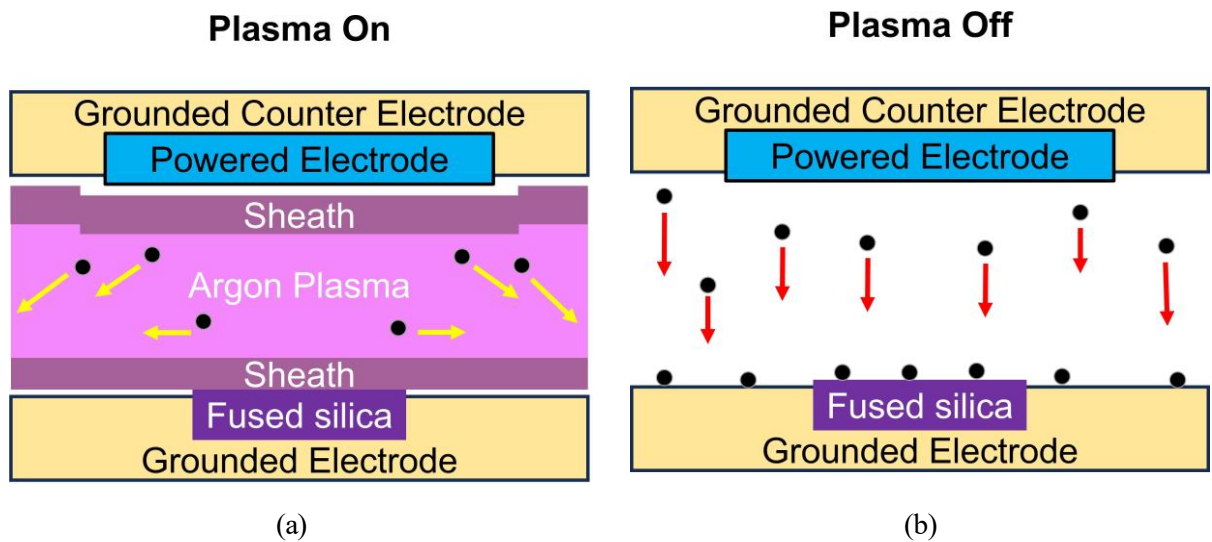

(a)                                                  (b)

Figure 4.4: (a) Dust (black) transports (yellow arrow) out of the plasma at the end of a cycle with plasma energized. (b) Dust (black) deposits (red arrow) on the bottom electrode, including fused silica, when plasma is de-energized. (Schematics not to scale) (Adapted from [72].)

Three types of experiments were conducted to describe this dusty plasma. In *experiment A*, the upper electrode was energized for 260 s, leading to consecutive growth cycles of dusty plasma between the electrodes. The objective was to determine the cycle time of dust particle growth. In *experiment B*, the upper electrode was energized for a specific duration, and upon de-energization, dust particles deposited on the substrate. The objective was to gather particles at different times, analyze their radii, and establish correlations with the growth cycle. In *experiment C*, the upper electrode was energized for 70 s followed by a 45 s de-energization. The procedure was repeated twenty times, without breaking vacuum, to increase the density of dust particles collected for material characterization. During each plasma, the forward power was 60 W initially for 10 s and maintained at 30 W for the remainder 60 s.

### 4.1.2 Experiment A: Growth cycle

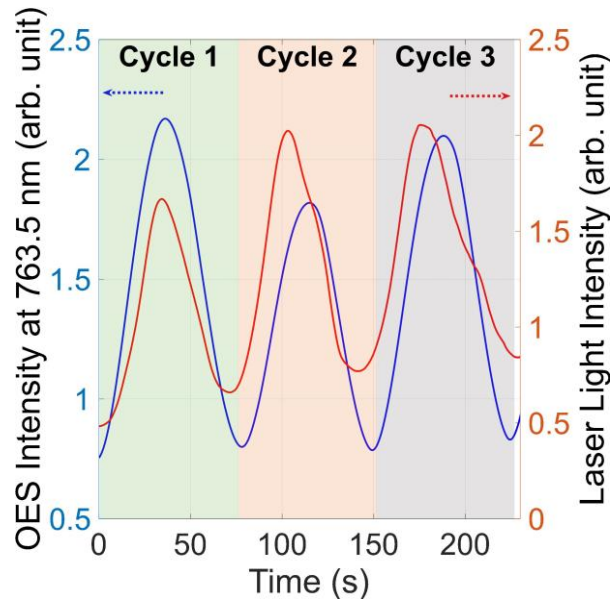

Figure 4.5: Optical emission spectroscopy (OES) at 763.5 nm and laser light intensity showing dusty plasma's cyclic behavior. Laser light intensity and OES intensity were measured from the red and blue region of Fig. 4.3b, respectively, over three cycles. (Adapted from [72].)

In experiment A, a green laser sheet (532 nm) was directed into the plasma to visualize the dust cloud in two dimension. A complementary metal oxide semiconductor (CMOS) camera recorded images of the cloud perpendicular to the laser sheet, up to a maximum of 50 frames per second. In Figure 4.3b, an image is shown. There was a recurrent appearance of a void, a dust-free region, in the central part of the plasma. This void exhibited cyclic expansion and contraction, mirroring the particle growth cycle. The laser light emission intensity of a region, highlighted in red was measured. This region covered some dust cloud and void in order to track the particle growth cycle. Concurrently, OES was employed to monitor the plasma, in the region highlighted in blue. The cyclic evolution of light emission and OES intensity were used to quantify the growth cycle. The resulting data, presented in Figure 4.5, revealed that the growth cycle was approximately 77 ± 9 and 77 ± 4 s, according to laser light and OES intensity respectively. The phase difference between the two sets of data potentially arises from experimental differences between the two measurements. For example, OES has a bigger field of view than the region analyzed for laser light intensity. Furthermore, the cyclic variation in the two measurements is attributed to distinct physical mechanisms. The changes in the laser light intensity occur due to fluctuations in particle radii. Meanwhile, changes in the OES intensity occur due to changes in the population of excited Ar.

### 4.1.3 Experiment B: Particle capture and size distribution

In experiment B, the plasma was energized for durations of either 10, 30, 50, 70, 90, 110, or 135 s to deposit nanoparticles on the substrate. SEM imaging was then employed to determine particle radii to correlate with the growth cycle, previously established as $\sim 77$ s in experiment A. Experiment B facilitated the analysis of particle radii from two cycles. When the plasma was energized and the growth cycle ended, particles were transported away from the plasma instead of depositing on the substrate. The sheath electric field deflected particles away from the electrodes and towards the plasma edges, as depicted in Fig. 4.4a. Conversely, when the plasma was de-energized, the absence of the sheath electric field allowed particles to deposit on the bottom electrode and substrate, as depicted in Fig. 4.4b. This enabled the collection and measurement of particle radii over time.

In Figure 4.6a, a linear progression in particles' radii from experiment B is observed as a function of plasma runtime. Subsequently, after a duration of $\sim 77$ s, the next cycle commenced. Particles from cycle 1 were still seen at 90 and 110 s, as shown in Fig. 4.6b. Thus, two size distributions were collected. By 120 s, the first cycle's particles were entirely cleared from the plasma, as confirmed by CMOS camera analysis. Therefore, the radii at 135 s were monodisperse. The first 10 s was dominated by creation of radicals, ions and species in the nucleation stage from the gas molecules. Because of the non-linear dynamics in this stage, we only considered the subsequent growth after 10 s, whereby the linear growth during was evident.

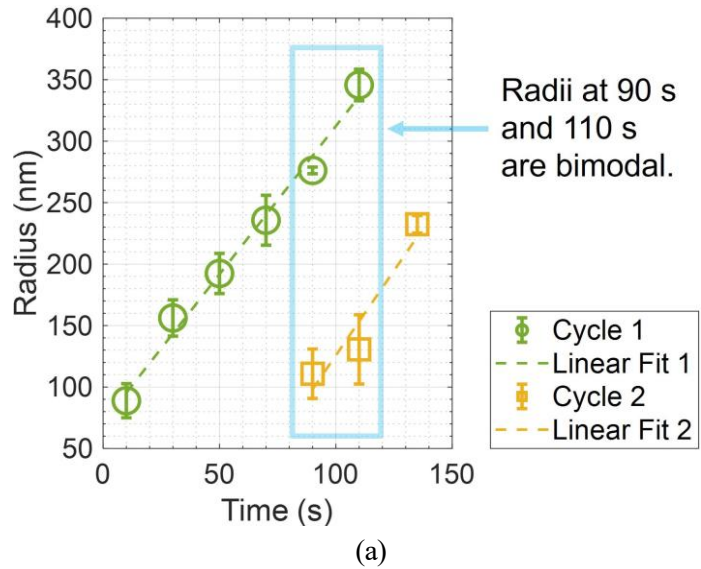

(a)

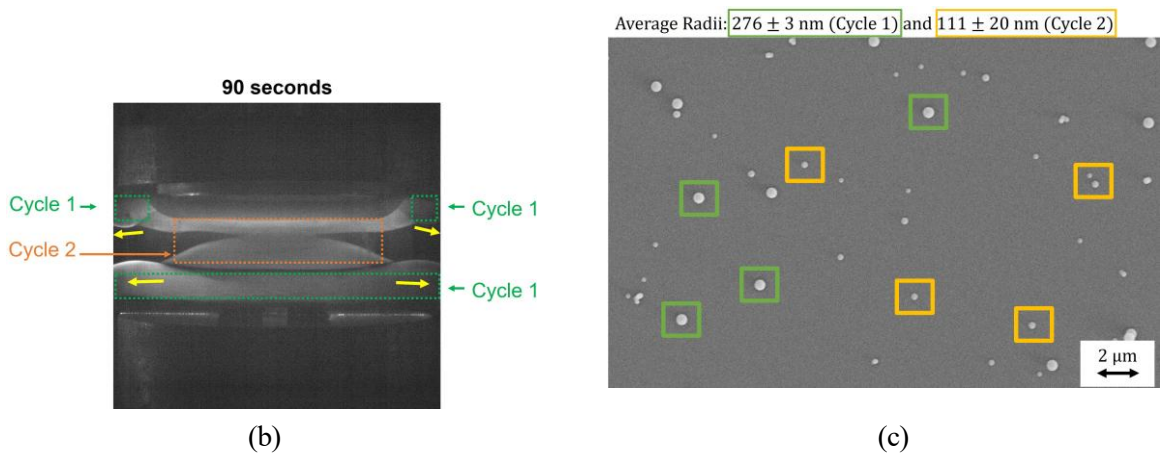

| (b) | (c) |

Figure 4.6: (a) Particles' radii distribution over time, with linear fit applied to both cycles. At 90 and 110 seconds, bimodal radii distributions were observed due to dust being present from two cycles. (b) Dust cloud at 90 seconds shows two cycles. Cycle 1 and 2 boxed in green and orange, respectively. Yellow arrows indicate dust from cycle 1 transporting away, similar to Fig. 4.4a. (c) SEM at 90 seconds showing bimodal radii distribution. Larger particles from cycle 1 and smaller particles from cycle 2 are boxed in green and yellow, respectively. (Adapted from [72].)

### 4.1.4   Experiment C: Nanoparticle Material Characterization

In experiment C, individual samples collected and annealed for two hours (h) at either 400, 600, 800, or 1000 ˚C. Samples were grown twenty times up to its maximum cycle time of 70 s and collected, in order to increase material density for characterization. XRD and Raman spectra, as illustrated in Fig. 4.7a and 4.7b, showed no peaks in the as-grown samples; hence they were amorphous.

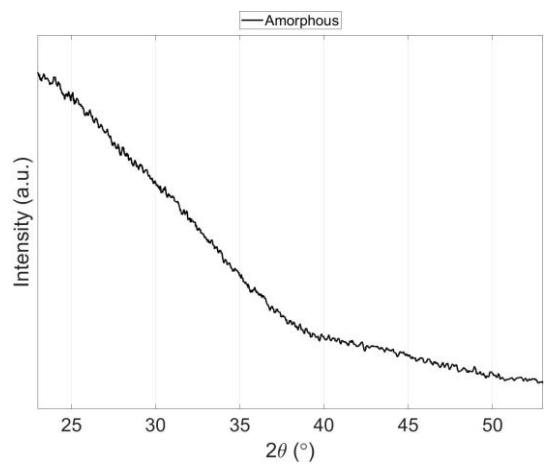

(a)

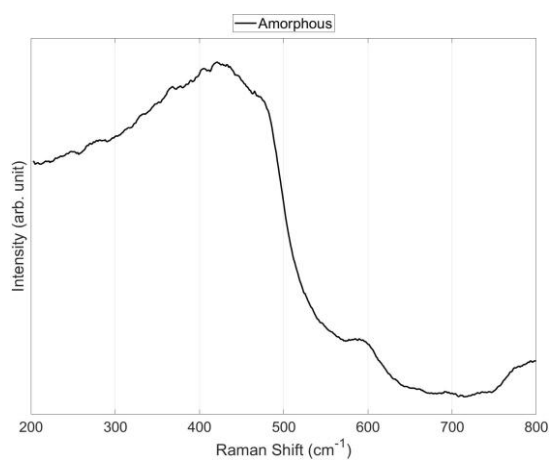

(b)

Figure 4.7: As-grown samples characterized by X-Ray Diffraction and Raman spectroscopy showed no peaks, suggesting amorphous nature.

At 400 °C, $TiO_2$ crystallized into anatase, and at 800 °C, it crystallized into rutile. At 600 °C, a mixed phase of anatase and rutile was observed. XRD, as displayed in Fig. 4.8a, of the annealed samples confirmed that the particles matched the known powder diffraction pattern for each phase. The Raman spectra, as displayed in Fig. 4.8b, closely resembled the known patterns for anatase and rutile [99], and as demonstrated in Fig.2.26b and 2.27b, respectively. The additional peaks, predicted from group theory for rutile and anatase at $\sim 827$ and $796 \ \mathrm{cm^{-1}}$ are shown in Fig. 4.8c and Fig. 4.8d respectively.

It has been observed that maintaining base pressure is important for best material synthesis. When the experiment was first conceived, the base pressure quickly rose to $\sim 10$ mTorr, instead of 3 mTorr. The material characterization of samples grown at those condition showed weaker signals and stronger noise, suggesting the presence of material disorder, as shown in SEM images, illustrated in Figure 4.10, were also used to calculate particles' radii before and after annealing. Monodisperse nanospheres were collected. These images clearly indicated a reduction in particles' radii following the annealing process. Particles that had been grown for 70 s prior to annealing exhibited an average radius of 235 ± 20 nm, while particles subjected to a 2 h annealing treatment at 400 and 800 ℃ displayed a reduced average radius of 171 ± 12 nm and 134 ± 19 nm, respectively. SEM of a 90 s sample, with bimodal size distributions is also shown.

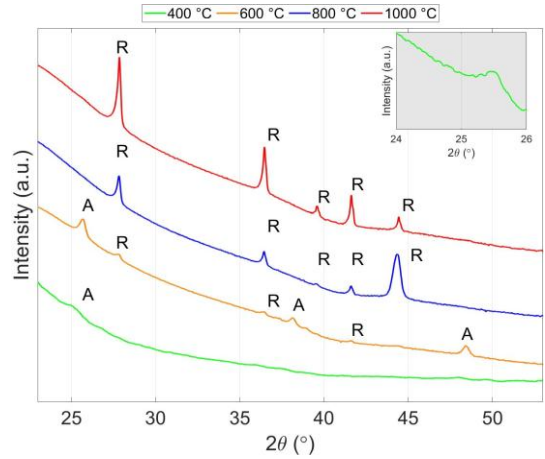

(a)

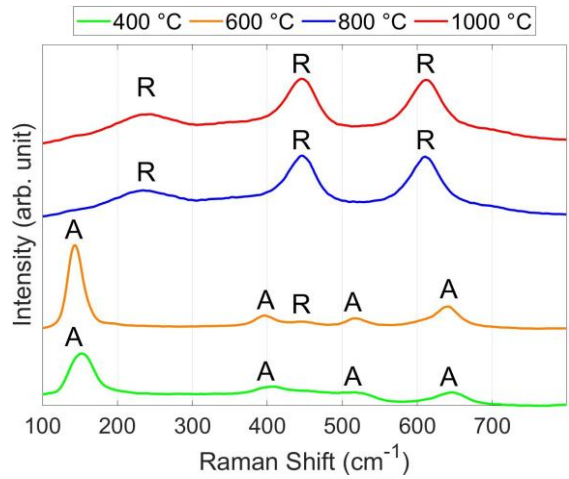

(b)

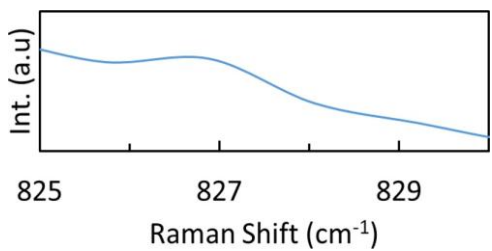

(c)

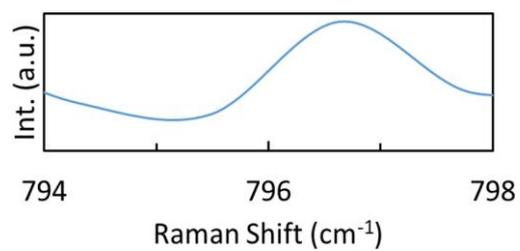

(d)

Figure 4.8: Characterization of particles grown for 70 seconds with 2 h anneal: (a) X-ray diffraction (XRD) showing anatase (A) at 400 °C and rutile (R) at 800 °C and 1000 °C, with mixed phases at 600 °C. Inset shows anatase peak at 400 °C. (b) Raman spectroscopy (Adapted from [72]). Additional Raman peaks for (c) rutile and (d) anatase not seen in (b).

Table 4.1: Impact of annealing on weight (%) of elements after 70 seconds growth, determined by Energy Dispersive X-ray Spectroscopy.

| Element | Weight (%) | | | | |
|---|---|---|---|---|---|
| | As-grown | 400 ℃ | 600 ℃ | 800 ℃ | 1000 ℃ |
| C K | 36.06 | 3.20 | 2.12 | 2.71 | 2.49 |
| O K | 26.43 | 42.32 | 48.60 | 44.37 | 50.34 |
| Si K | 31.32 | 39.07 | 28.18 | 34.79 | 15.23 |
| Ti K | 6.19 | 14.40 | 21.11 | 18.14 | 31.94 |
| Ti:O | 0.23 | 0.34 | 0.43 | 0.41 | 0.63 |

Fig 4.9. It is postulated that the higher base pressure, with the associated presence of the constituents of air such as oxygen, nitrogen, carbon dioxide, and water, could be possible sources of contamination that get incorporated into the forming nanoparticles. Once this was recognized, subsequent experiments were performed only for system base pressures below, 3 mTorr. Additionally, once the base pressure rose above that threshold, the vacuum chamber was disassembled and cleaned using (appendix C procedures) to remove built up contaminants.

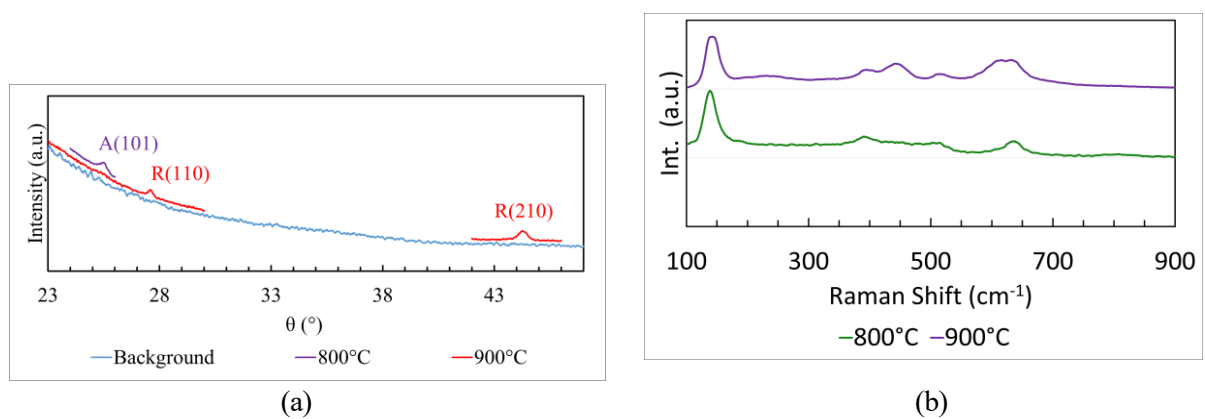

(a)  (b)

Figure 4.9: Particles grown at high base pressure characterized by (a) X-Ray Diffraction and (b) Raman spectroscopy.

Average Radii: 235 ± 20 nm

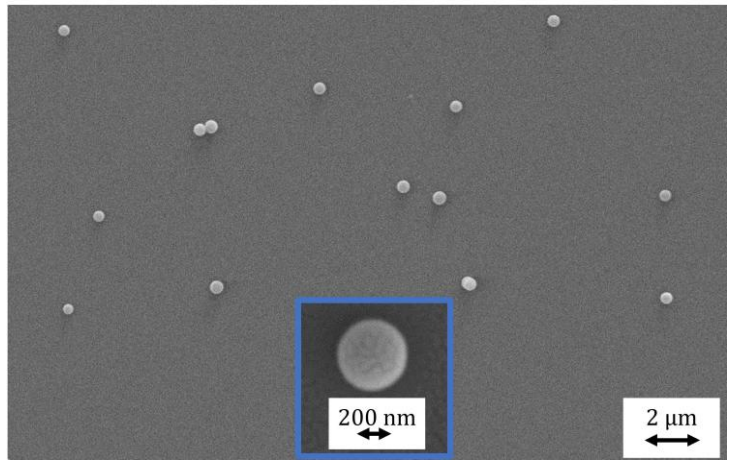

(a)

Average Radii: 171 ± 12 nm

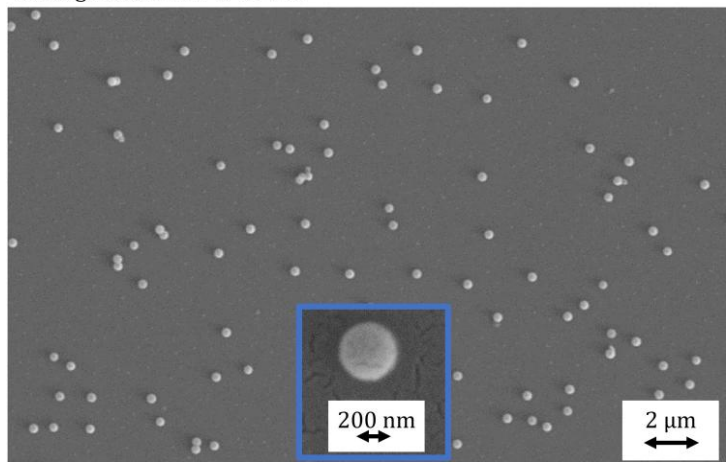

(b)

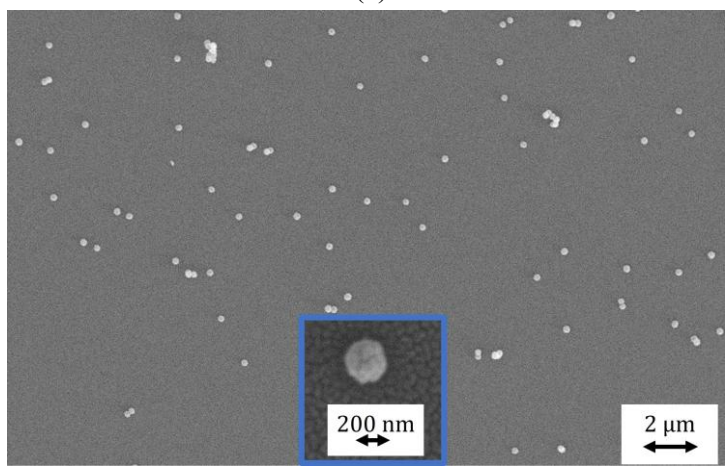

(c)

Figure 4.10: Scanning Electron Microscope images of nanoparticles collected on fused silica. (a) As-grown, (b) 400 °C anneal, and (c) 800 °C anneal. (Adapted from [72].)

Quantitative EDS analysis was performed on the samples. The findings, as summarized in Table 4.1, showed a decline in the weight percentage of carbon (C) and an increase in titanium (Ti) and oxygen (O) on the samples after the annealing. Furthermore, the ratio of Ti to O, as shown in the last row increases with annealing temperature, suggesting that even some O is lost after annealing. The combined SEM and EDS results suggest that both C and O were oxidized and desorbed from the nanoparticle during annealing. TTIP decomposed into molecules within the plasma, providing the source of Ti, O, and C in the nanoparticles. Fused silica substrates are the source of Si and additional O. In literature, PECVD experiments operating at similar experimental conditions as ours such as temperature, pressure, and LTPs, revealed that TTIP dissociates into $TiO_2$, propene ($C_3H_6$), and isopropanol ($C_3H_8O$) [135, 136]. The chemical pathway is elucidated in Eq. 4.1. Even though we were unable to monitor the breakdown of TTIP into its byproducts in our LTP in-situ, we suggest that a similar chemical pathway is possibly providing the source of the three aforementioned elements on the samples. Hydrogen has a low X-ray emission energy due to its low atomic number, making it difficult to detect via EDS.

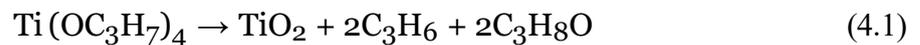

$$Ti\,(OC_3H_7)_4 \rightarrow TiO_2 + 2C_3H_6 + 2C_3H_8O \qquad (4.1)$$

The presence of C and O on the dusts' surface most likely arise from propene and iso-propanol, two by products of TTIP, in LTPs. To explain their potential presence and oxidation from the surface of the nanodust, it is essential to delve into the three key stages of plasma particle growth: nucleation, coagulation, and agglomeration [45, 53, 54, 56, 57]. During nucleation, plasma species such as radicals and ions are generated through interactions between the background plasma and reactive gases. These species subsequently engage in chemical reactions, forming clusters with sizes $\sim 1$ nm. During coagulation, these clusters collide and bond, thus creating larger clusters with dimensions $\sim 10$ nm. Coagulation is dominated by chemisorption of ions and clusters. Nucleation and coagulation processes occur within the first few hundred milliseconds and seconds, respectively [59]. The agglomeration stage, which can persist for several tens of seconds, involves the continued interaction of radicals and ions from the plasma, facilitating surface growth on the clusters and resulting in dust particles growing to several hundred nm. This is dominated by physisorption of ions and radicals on the clusters. MD simulation in literature has shown that in dusty plasma particle growth from chemical precursors, the ratio of chemisorption to physisorption decreased with increasing particle size [63]. Thus, surface growth on the dust most likely occurs through coulombic interactions [137]. This could potentially elucidate the presence of higher concentrations of C and O on the dust surface arising from either propene or isopropanol, which subsequently undergo oxidation during the annealing process.

## 4.2   Comparing the growth of titania and carbonaceous at  500 Gauss

The results presented in this chapter were previously published in *B. Ramkorun, et al, 2024 Plasma Sources Sci. Technol. 33 115004* [53]. This chapter will present that work with some additional text to explain its connection to the other studies conducted as part of this dissertation.

This section compares the growth cycles and spatial distribution of dust cloud for titania and carbonaceous dusty nanoparticles in capacitively coupled radiofrequency plasmas, with and without the presence of a weak magnetic field of approximately 500 Gauss. Findings on cycle time, growth rate, and spatial distribution of dust cloud are discussed. The growth of nanoparticles in these plasmas is cyclic, with particles reaching their maximum size and subsequently moving out of the plasma, followed by the generation of a new particle growth cycle [138,139]. Fig. 4.2 is a flowchart which summarizes the experimental procedures to grow dust, analyze the dusty plasma in-situ, collect the dusty nanoparticles for materials' characterization. Similarly to the previous experiment of growing and characterizing titania nanoparticles. The RF forward (reflected) power was 60 W (1 W) in the first 10 s and 30 W (1 W) during the remainder of the experimental time. In order to keep the experiment procedure consistent, the same power setting was used for the $Ar/C_2H_2$ plasma. In each individual plasma, the partial pressure of the precursor gas was constant, at $\sim 35$ mTorr. We assume that within each experiment, the plasma parameters are not changing due to variation in the initial density of precursors. Any changes that will eventually be detected should be due to the magnetic field.

### 4.2.1 Determining Cycle Time

The growth cycles were assessed through three distinct methods: firstly, by measuring the line intensity of Ar I at 760.5 nm via OES; secondly, by capturing images of laser light scattered from the dust cloud using a CMOS camera; and thirdly, by analyzing the dust size distribution for particles grown at different length of time. Each of these processes is described below as they are applied to this study.

### 4.2.2 Optical Emission Spectroscopy

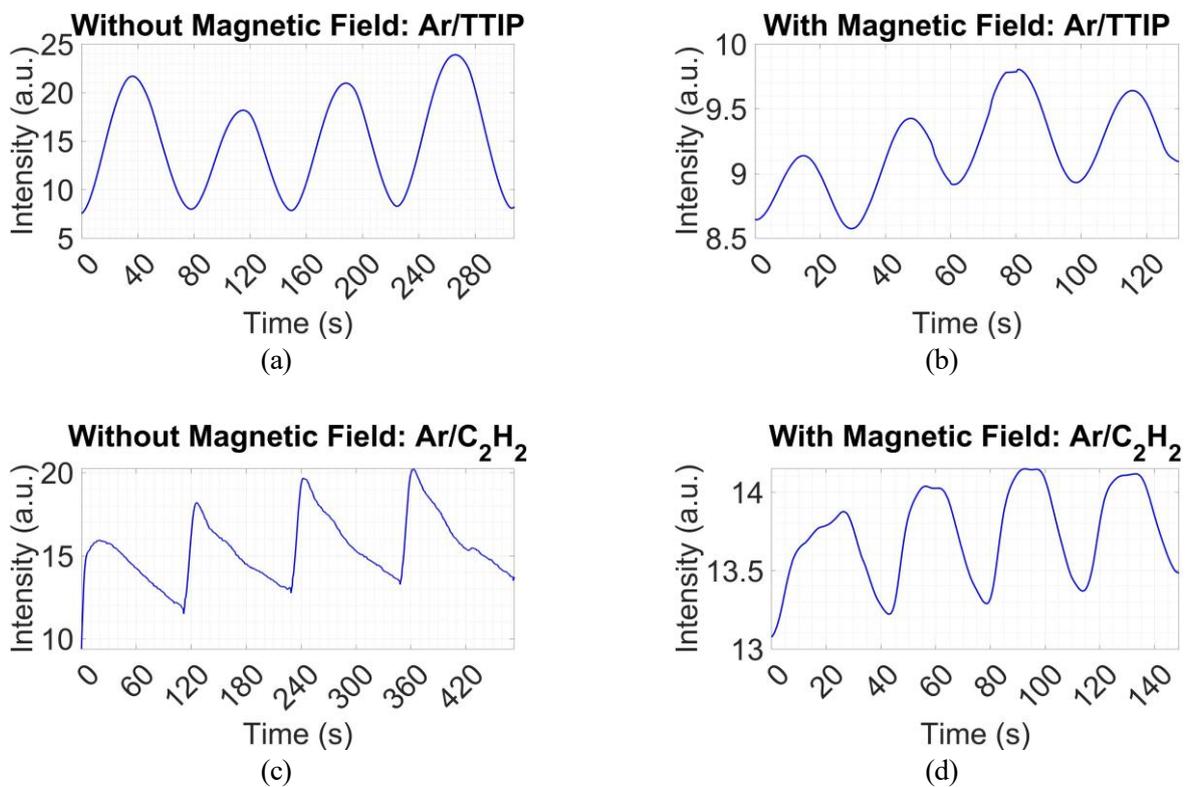

Figure 4.11: Cyclic variation of the intensity of Ar I (763.5 nm) line: Argon/Titanium iso-propoxide (Ar/TTIP) dusty plasma (b) without magnetic field ($77 \pm 4$ s) and (c) with magnetic field ($32\pm3$ s), Argon/Acetylene (Ar/$C_2H_2$) dusty plasma (d) without magnetic field ($115\pm1$ s) and (e) with magnetic field ($39 \pm 1$ s). (Adapted from [53].)

The timeframe of the OES was set from its integration time and number of data averaged. Data was collected roughly every 600 ms. The time when the plasma was turned on was set to zero. For the Ar/TTIP dusty plasma, without (with) the presence of the magnetic field, the measured cycle time was $77 \pm 4$ s ($32 \pm 3$ s), as shown in Fig. 4.11a and 4.11b respectively. For the Ar/$C_2H_2$ dusty plasma, without (with) the presence of the magnetic field, the cycle time was 115 $\pm$ 5 s (39 $\pm$ 1 s), as shown in Fig. 4.11c and 4.11d respectively. Studies have shown that the growth rate of dusty nanoparticles depends on the gas temperature of the reactive gas(es) [140–142]. Here, the gas temperature of $C_2H_2$ and TTIP are $\sim 20\,°C$ (room temperature), and $\sim 75\,°C$ respectively. Thus, the growth rate, and consequently the cycle time, of the two dusty plasma can not be directly compared to each other. Nevertheless, the individual gas temperatures are unchanged during the experiments both with and without the presence of the magnetic field. Therefore, we believe that the change in the cycle time is predominantly due to the presence of the magnetic field.

**Laser Light Scattering**

The laser light scattering from the dust cloud was recorded using a CMOS camera. Dust density waves, which can be used to calculate dust density and temperature, were detected at the edge of the dusty plasma, however, they were not further studied herein [107]. The time frame was set from the frame rate, whereby a frame was recorded every 20 ms, i.e 50 FPS. The first frame with plasma on was set to t = 0. In order to compare the different dust cloud to each other at the same time during their cycles, 4 images are chosen and shown. The region of the dust cloud between the electrodes are shown in Fig. 4.12, with each column being at $T_c/4$, $T_c/2$, $3T_c/4$, and $T_c$, where $T_c$ was the cycle time according to the OES.

Unlike OES which scans light emission across a volume, the laser light scattering occurs in a 2D plane. In order to consistently compare the cycle time, 30 squares measuring $\sim 100 \times 100$ pixels each were drawn on each dust cloud, 10 in 3 different rows as shown in the first column of Fig. 4.13. They were drawn to cover the entire region of the cloud between the electrodes where light scattering from dust particles is observed. Within each analysis box, the average light intensity was calculated from all of the pixels. A time series, spanning over three growth cycles, was then created by recording the light intensity as a function of time for each box. A numerical Fourier transform (NFT) of the time series was performed for each box to identify the dominant frequencies. The most occurring dominant frequency, i.e. mode frequency, measured from the 30 NFT was chosen to represent and calculate the growth cycles. Examples of the cyclic variation of the light emission intensity from the first row and third column (blue box) is shown in the second column of Fig. 4.13. The average intensity over 3 cycles is subtracted from the data before calculating NFT, hence the negative magnitudes. The X-axis of indicates the time calculated from the camera frames. The mode frequency from all 30 boxes is shown in the third column of Fig. 4.13. Moreover, the magnitude of the mode frequency in the 30 boxes superimposed on the dust cloud at T$_c$/2 is shown in the fourth column of Fig. 4.13. For the Ar/TTIP dusty plasma, without (with) the presence of the magnetic field, the mode frequency and cycle time were 0.0130 Hz (0.0315 Hz) and 77 s (32 s) respectively. For the Ar/C$_2$H$_2$ dusty plasma, without (with) the presence of the magnetic field, the mode frequency and cycle time were 0.0087 Hz (0.0264 Hz) and 115 s (38 s) respectively. The mode frequency and cycle time were rounded to two significant figures and they are in agreement with the values obtained from OES.

|  | **T$_c$/4** | **T$_c$/2** | **3T$_c$/4** | **T$_c$** |

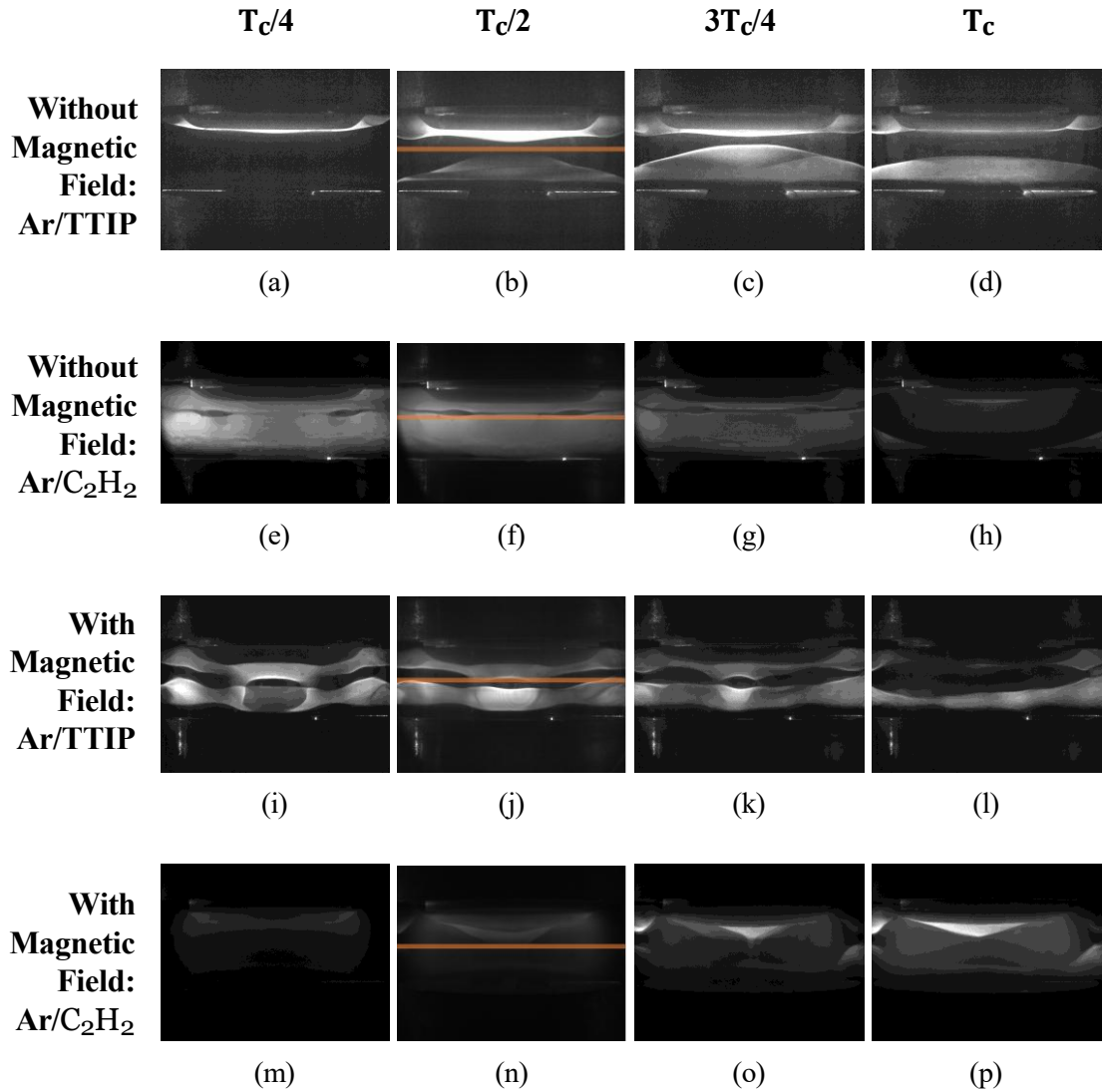

Figure 4.12: Camera images of the four different dust clouds (rows) at the same time (columns) within each experiment, T$_c$/4, T$_c$/2, 3T$_c$/4, and T$_c$, as labelled. The orange line is drawn at z = 12.5 mm of T$_c$/2, to compare the laser light scattering intensity in the mid-plane over 2 cycles in Fig. 4.15. (Adapted from [53].)

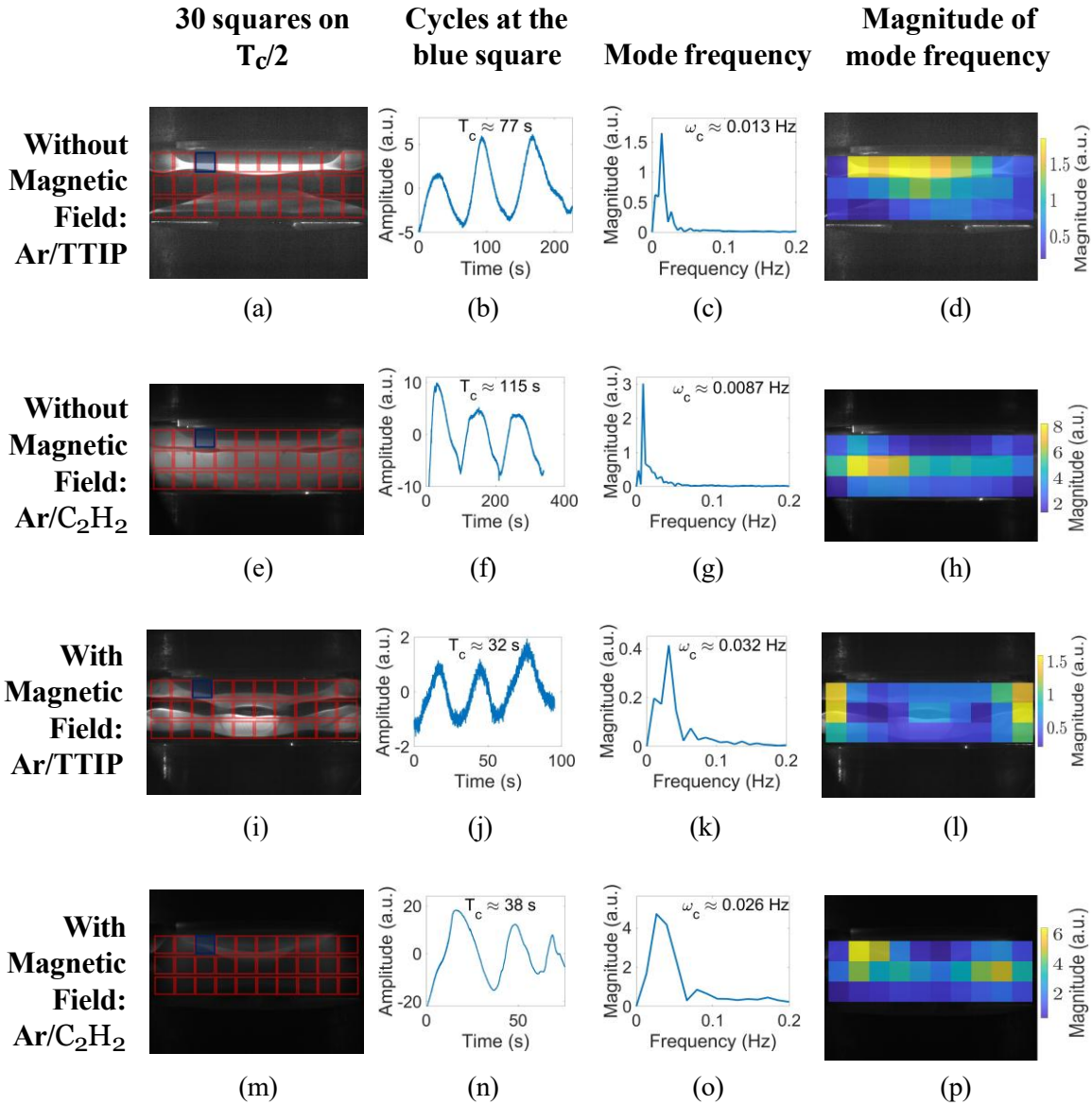

| 30 squares on $T_c/2$ | Cycles at the blue square | Mode frequency | Magnitude of mode frequency |

Figure 4.13: Each row represents an individual dusty plasma as labelled. The first column shows 30 squares of approximately $100 \times 100$ pixels, arranged in 3 rows and 10 columns, over the dust cloud at $T_c/2$. The second column shows the cyclic variation in light intensity at row 1 and column 3 (blue square). The third column shows the mode frequency obtained from all 30 squares. The fourth column shows the magnitude of the mode frequency in all 30 squares overlaid on the dust cloud at $T_c/2$. (Adapted from [53].)

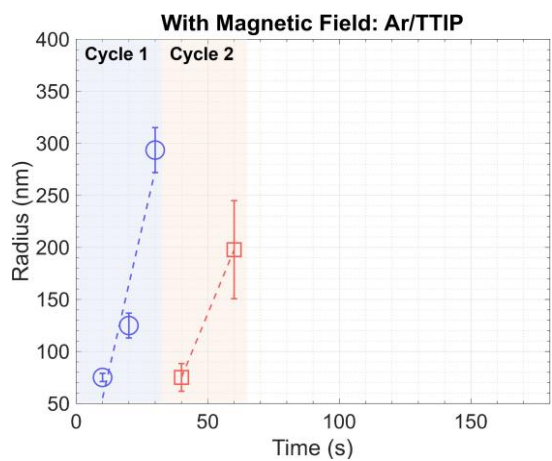

(a)

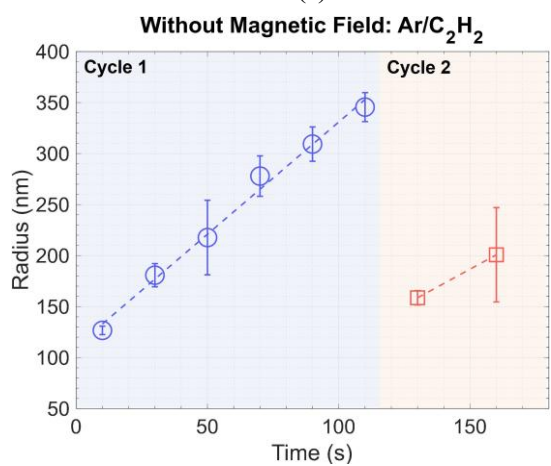

(b)

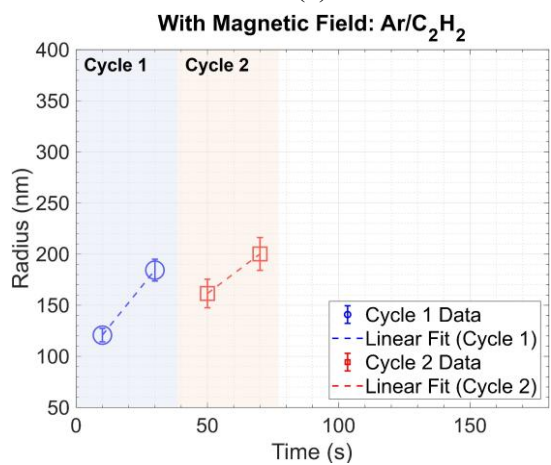

(c)

Figure 4.14: Linear growth of nanoparticles' radii within two growth cycles. Same for Ar/TTIP without magnetic field is shown in Fig. 4.6a. Legend of (d) applies to all the subfigures here. (a) Ar/TTIP with magnetic field. (b) Ar/$C_2H_2$ without magnetic field. (c) Ar/$C_2H_2$ with magnetic field. (Adapted from [53].)

## Size Distribution

In order to obtain the size of the nanoparticles, the individual experiments were ran for a certain instant of time and the nanoparticles subsequently were collected on a glass slide at the bottom electrode. The samples were then sputter coated with 15 nm of gold before being analyzed by SEM. The particles reached their maximum size by the end of each cycle. When a new cycle started, new particles initiated their growth. This growth exhibited a consistent linear progression in size throughout two cycle, as depicted in Figure 4.14. The time frame for the size distribution analysis was determined by referring to the frames from CMOS camera images. Each particle growth experiment was meticulously recorded and timed according to the number of frames, spanning from the activation of plasma to its deactivation.

At the end of the first cycle, the void expands, followed by a newer cycle which eventually replaces the precious one. Therefore, the size distribution of nanoparticles is gradually reset in the second cycle. During the growth of titania without magnetic field, dust from the previous cycle moved out of the plasma slowly, resulting in a bimodal size distribution at 90 and 110 s. Although a bimodal size distribution was not observed in the other three cases, it is possible that nanoparticles from the previous cycle were still present but were not collected on the substrate, located at the center of the electrodes. The maximum radius of the titania particles after 77 s were $235 \pm 20$ nm, however, the particles' radii grew to $355 \pm 12$ nm at 110 s, as shown in Fig. 4.6a. The carbonaceous particles reached their maximum radius of $346 \pm 14$ after 110 s, shortly after which the new cycle started, as shown in Fig. 4.14b. The maximum radii of titania and carbonaceous dust particles grown during the presence of magnetic field were $294 \pm 22$ nm and $200 \pm 16$ nm, respectively, as shown in Figs. 4.14a and 4.14c.

Table 4.2: Summary of the results from the two dusty plasma nanoparticle growth. With the magnetic field, the maximum radii are smaller and the cycle ends faster as measured by optical emission spectroscopy (OES) and numerical Fourier transform (NFT) of the dust cloud (adapted from [53]).

| Dust | Without magnetic field | | | With magnetic field | | |
|---|---|---|---|---|---|---|
| | Max Radius (nm) | Cycle Time (s) | | Max Radius (nm) | Cycle Time (s) | |
| | | OES | NFT | | OES | NFT |
| Titania | $355 \pm 12$ | $77 \pm 4$ | 77 | $294 \pm 22$ | $32 \pm 3$ | 32 |
| Carbonaceous | $346 \pm 14$ | $115 \pm 5$ | 115 | $200 \pm 16$ | $39 \pm 1$ | 38 |

Our are summarized in Table 4.2. The results show that the presence of a magnetic field leads to a reduction in both the maximum particle radius and the cycle time. For titania, the maximum radius decreased from $355 \pm 12$ nm to $294 \pm 22$ nm, and the cycle time shortened from $77 \pm 4$ s to $32 \pm 3$ s. Similarly, for carbonaceous particles, the maximum radius dropped from $346 \pm 14$ nm to $200 \pm 16$ nm, with the cycle time decreasing from $115 \pm 5$ s to $39 \pm 1$ s. At the end of their respective cycles without B-Field, the carbonaceous dust particles achieved a larger size compared to titania dust particles. This could be attributed to the equilibrium between two dominant forces that sustain dust particles in a levitated state during the process of agglomeration: the electric force and gravitational force. Carbonaceous particles, due to their lower mass density in contrast to titania, needed a greater mass to overcome the electric force before succumbing to gravitational pull and falling from the dust cloud.

### 4.2.3 Magnetic fields effects on the dust cloud

In addition to the faster growth cycles when the magnetic field is present, there is also a noticeable impact within the spatial distribution of the dust cloud in dusty plasma during the presence of the magnetic field. Specifically, the presence of the magnetic field resulted in a larger void, for the growth of carbonaceous particles as compared to the growth of the titania particles. Variations in chamber geometry are expected to influence the dust cloud. For example, there is a groove in the lower electrode to fit in the substrate for particle collection and a protrusion of the upper rf-electrode. However, this geometry is kept constant throughout all the experiments. Therefore, any changes in the dust cloud is not attributed to changes in geometry, but instead, to the presence of the magnetic field. Moreover, when the magnets are present, they are placed either above the top electrode or below the bottom electrode, so that they do not change the geometry of the surfaces which are directly in contact with the plasma and do not physically interfere with the dust cloud.

To visualize the overall radial distribution of the dust particles, the light intensity along the orange line of $T_C/2$ of Fig. 4.12 was summed over all frames for horizontal positions, in the range of $-37.5 \leq x$ (mm) $\leq 37.5$, positioned at the midplane between the two electrodes at z = 12.5 mm, as shown in Fig. 2.10a. The variation in intensity along the line, 50 pixels thick, represents the variation in spatial dust cloud distribution. The light intensity obtained was stacked vertically to show the temporal evolution of the dust cloud. Fig. 4.15a and 4.15b illustrate this for Ar/TTIP and Ar/$C_2H_2$ dusty plasmas without a magnetic field, while Fig. 4.15c and 4.15c show the same for Ar/TTIP and Ar/$C_2H_2$ dusty plasmas with a magnetic field, respectively. The intensity for each dusty was normalized to its highest number to enable a more direct comparison to be made between the particle growth of the two dust particle types. The temporal intensity profile appears to be similar for both kinds of dusty plasma without magnetic field and dissimilar during the presence of the magnetic field. Line intensity plots are shown in Figs. 4.15e and 4.15f which shows the overall spatial distribution of the grown dust particles over two growth cycles without and with the magnetic field. These plots serve as a quantitative comparison of the two dust clouds to reveal any similarities and differences. In the absence of a magnetic field, the radial intensity variation of the two dusty plasmas exhibits similarities. However, with a magnetic field present, the Ar/TTIP dusty plasma shows peaks at the centre and edges of the electrode, whereas the $Ar/C_2H_2$ dusty plasma exhibits a peak only at the edges and a reduced intensity, due to a void, in the middle of the plasma. This suggests that there was more titania dust in the middle of the plasma than there was carbonaceous dust; thus, there was a difference between the spatial distribution of the two dust clouds at 500 Gauss.

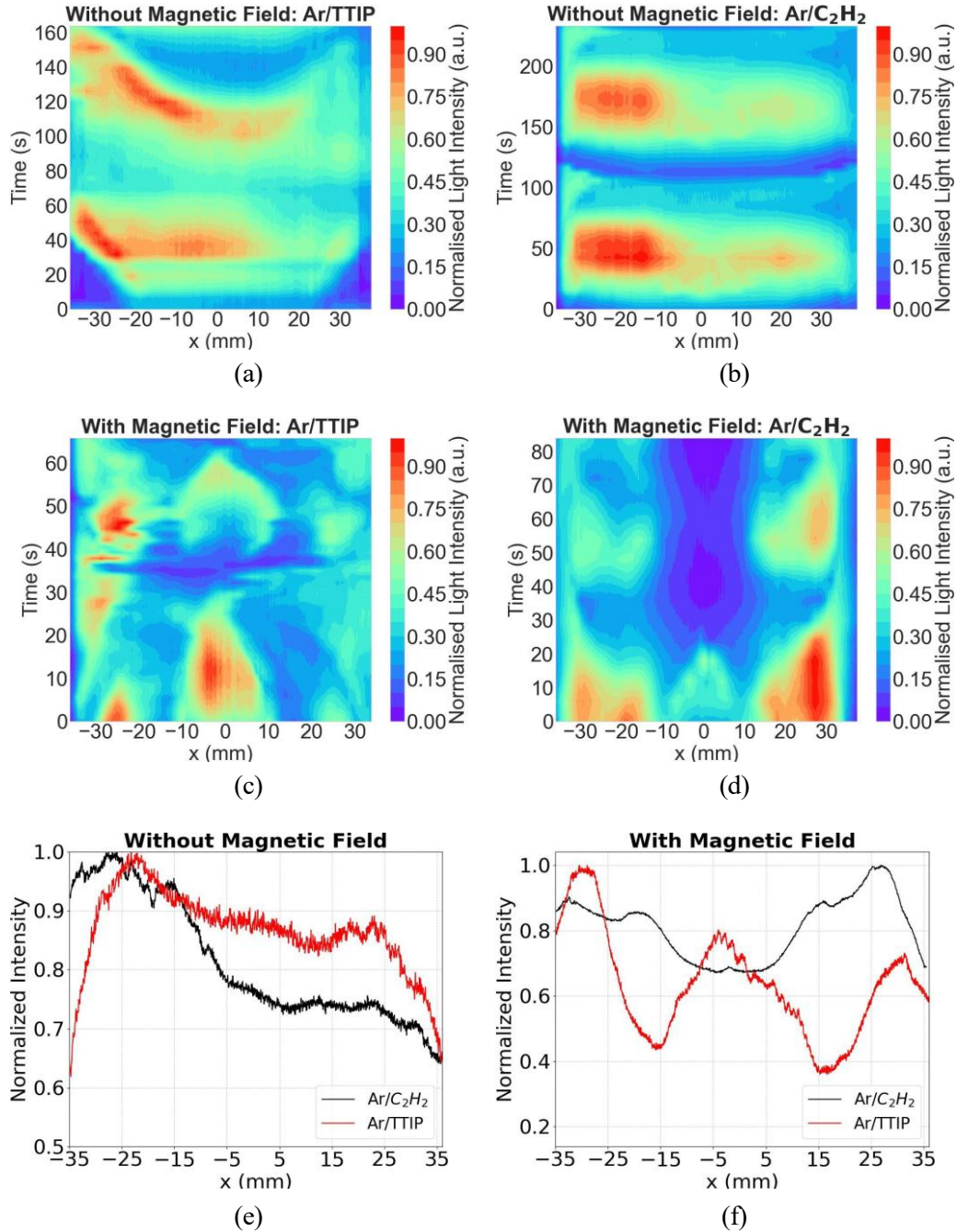

Figure 4.15: Temporal evolution of the light emission intensity from a line at z = 12.5 mm across the diameter of the dust cloud, as drawn in Fig. 4.12. Without magnetic field: (a) Ar/T-TIP and (b) Ar/$C_2H_2$. With magnetic field: (c) Ar/TTIP and (d) Ar/$C_2H_2$. Sum of intensity of both dusty plasmas from light emitted due to laser scattering across two growth cycles: (e) without magnetic field, (f) with magnetic field. (Adapted from [53].)

### 4.2.4 Langmuir probe analysis of background plasma

To understand why the dust cloud reacts differently to the presence of a magnetic field, it is important to explore changes in plasma conditions resulting from the presence of the magnets. Therefore, our investigation focuses on how the magnetic field has influenced plasma conditions, which can ultimately affect the spatial distribution of the dust cloud. Langmuir probes measurement of the plasmas' electron temperature and floating potential are shown in Fig. 4.16. There is a radial variation in floating potential as shown in Fig. 4.16a. However, there does not seem to be a radial variation of electron temperature. Even though these measurements were made in the absence of dust, we can suggest how the trend of the measurements will change. When the dust cloud has formed, the dust density is typically $\sim 10^{13}$ $m^{-3}$ [112]. This leads to an electron depletion in the background plasma due to the Havnes parameter being > 1 [143]. We expect the overall floating potential of the plasma to slightly increase during the presence of the dusty nanoparticles [106]. At 500 Gauss, the changes in the floating potential, and consequently the radial electric field in the plasma, is possibly contributing to the faster cycle time.

It is possible that the magnitude of several forces confining the particles in the plasma has decreased, thus leading to the faster cycle time. Additionally, it has already been shown that the electron temperature of particle growth as dusty plasma varies cyclically, in accordance with the dust growth cycles [79]. For the two cases considered, with and without a magnetic field, Fig. 4.16a clearly shows that the slope of the floating potential (in the absence of the dust), and by reasonable extension, the sign of the radial electric field change sign depending upon the presence of the magnet. If this change of sign persists when the dust is present, this could be interpreted as a change in the radial, electrostatic confining force on the negatively charged, growing dust particles.

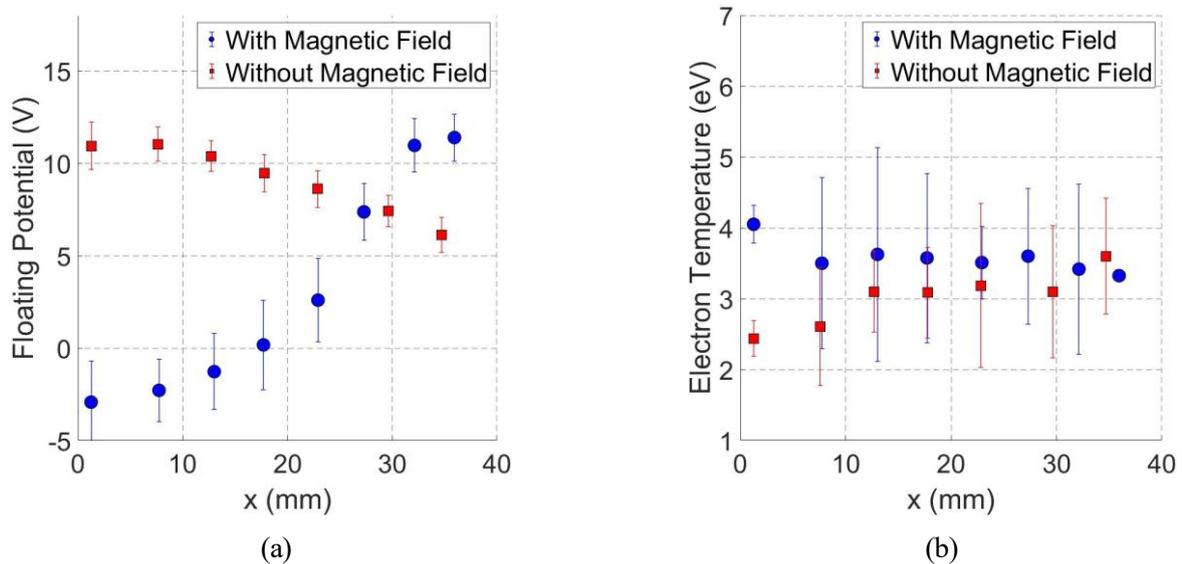

Figure 4.16: Background Argon plasma parameters measured with a Langmuir probe. Radial variation of (a) floating potential and (b) electron temperature at $z = 12.5$ mm, i.e., the mid-plane between the two electrodes. (Adapted from [53].)

Let us consider how this measurement of the background plasma parameter may be used to gain insights into the dynamics of the dust particle for the four cases discussed in this paper. Without a magnetic field, as indicated in Fig. 4.15 (a) - (c), the apparent growth of both the carbonaceous and titania particles appear to fill the plasma volume in a very similar manner. Although there is a left-right asymmetry (i.e., $x < 0$ mm vs. $x > 0$ mm) in the normalized light intensity, it is similar for both grown dust particle types without a magnetic field. By contrast, when evaluating the case with a magnetic field, as shown in Fig. 4.15 (d) - (f), there is a difference in the spatial distribution of the grown dust of the two particle types in the presence of a magnetic field. Under the assumption that both the carbonaceous and titania particles have the same negative charge, the change in the direction of the confining, radial electric field may contribute to the difference in the spatial distribution of the dust particles between the magnetized and unmagnetized cases. However, the change in the direction of the electric field is insufficient to describe how a difference could arise the two particle types.

We now consider the effects of the magnetic field on the dust. First, according to the Hall parameter, given by Eq. 3.22, a commonly employed metric for calculating the degree magnetisation of charged species in a plasma, only the electrons were magnetised at 500 Gauss. Ions and dust particles were not magnetised, due to their heavier masses. Therefore, there is no magnetisation due to charge accumulation on the dust that can explain the differences in spatial dust distribution of titania and carbonaceous nanoparticles. Second, there is a gradient in the radial direction of the magnetic field. According to Fig. 3.9b, the magnetic field, $B_z$ is constant in the region $-15 \leq x(mm) \leq 15$. However, there is a change in $B_z$ with respect to x outside that range until the edge of the electrodes. It is possible that the gradient is contributing to the differences seen between the spatial distribution of the two dusty plasma. A different study of the magnetic properties of the materials is necessary in the future in order to determine if they are contributing to the differences seen in the experiments due to the gradient.

## 4.3   Investigating the growth cycles of carbonaceous dust from $\sim 18$ - 1020 Gauss

The objective of this section is to grow dust in an $Ar/C_2H_2$ capacitively coupled plasma discharge and to conduct a comprehensive experimental investigation of particle growth behavior both in the absence and presence of a "weak" magnetic field, ranging from approximately 18 to 1020 Gauss. We measure the cycle time through the intensity variation in the OES of the background Ar plasma. Our findings indicate that the cycle time decreases rapidly with increasing magnetic field strength until the electrons become fully magnetized, after which the cycle time stabilizes at a minimum value.

Once we have determined cycle time from OES intensity variation, the particles were then grown and collected at four different time intervals within the first cycle: $T_c/4$, $T_c/2$, $3T_c/4$, and $T_c$, where $T_c$ represents the cycle time. We observe that the growth rate of nanoparticles during the initial cycle varies with the magnetic field strength [144]. Furthermore, OES of the background plasma reveals that the time to reach steady state upon plasma ignition increases with higher magnetic field strength. This delay may influence the chemistry and particle nucleation in the plasma during the initial seconds, ultimately affecting the particles' growth rate.

### 4.3.1 Cycle time

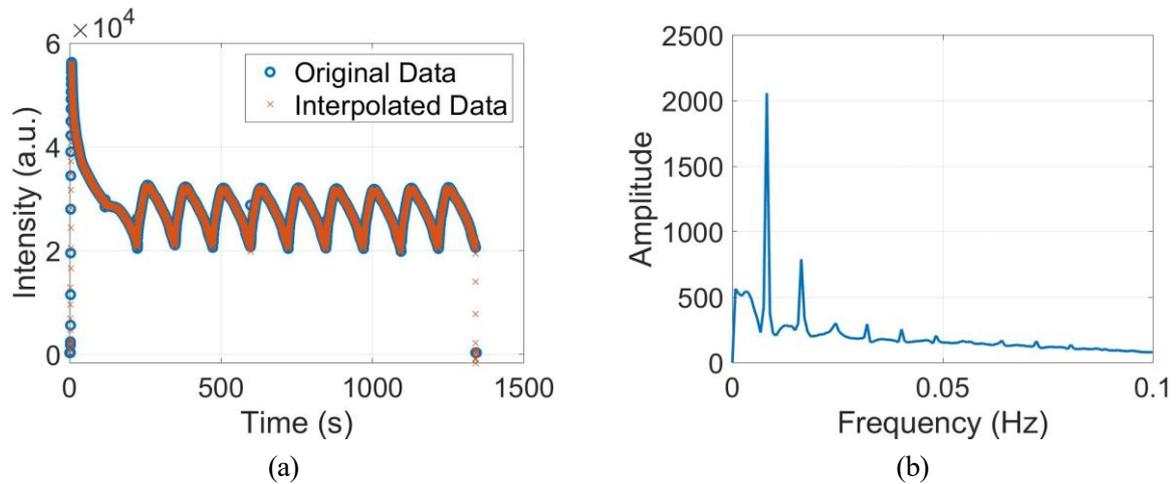

(a)  (b)

Figure 4.17: Example of temporal variaion in optical emission spectroscopy intensity and extracting cycle time from fast Fourier transform: (a) Temporal variation of 763.5 nm line intensity over time for carbonaceous dusty plasma at 0.5 Gauss. (b) Corresponding FFT showing cycle frequency. (Adapted from [53, 83].)

The cyclic variation of Ar I at 763.5 nm was measured for at least 8 cycles to determine the cycle time of the nano dusty plasma. An example of the cyclic variation of the OES intensity at 18 Gauss is shown in Figure 4.17a. The data was collected at slightly irregular time intervals of approximately 200 ms, and thus spline interpolation was used to interpolate the data at every 100 ms. From the interpolated data, a fast Fourier transform (FFT) was performed to determine the peak frequency of the cycles, as shown in Figure 4.17b.

At least three sets of data were collected and averaged for each magnetic field configuration to determine the cycle time as a function of magnetic field strength. The cycle time was calculated as the inverse of the peak frequency with maximum intensity. The results are shown in Figure 4.18. Initially, without any external magnetic field (0.5 Gauss from Earth's magnetic field), $T_c$ was 121 ± 7 s. This gradually decreased to 40 ± 3 s at 233 Gauss and continued to fluctuate between 38 and 47 s as the magnetic field strength increased from 335 to 1020 Gauss.

Overall, we observe that $T_c$ gradually decreased between 0.5 and 233 Gauss and remained relatively constant up to 1020 Gauss.

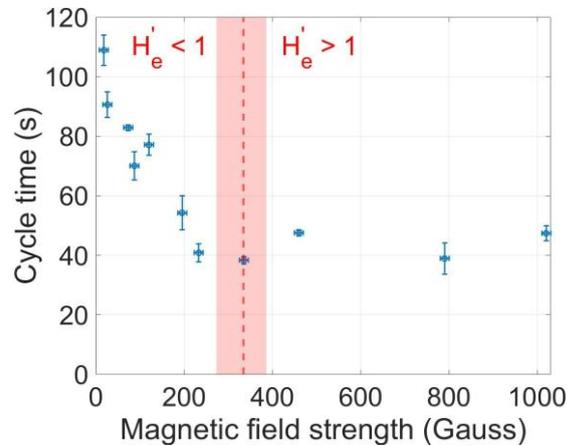

Figure 4.18: Cycle time (extracted from temporal intensity variation in optical emission spectroscopy) versus magnetic field, showing a decrease until 330 Gauss, after which it plateaus (Adapted from [83].)

### 4.3.2 Growth Rate

Once $T_c$ was established, dust was grown again and collected at four intervals within the first cycle: $T_c/4$, $T_c/2$, $3T_c/4$, and $T_c$. The goal was to collect the dust on the fused silica at the bottom electrode, image it using a scanning electron microscope (SEM), and determine the particle size distribution to assess the growth rate within the first cycle. Only clear and isolated particles were selected for size extraction. For example, Figure 4.19 shows nanoparticles collected at 1020 Gauss after growing for $T_c \sim 47$ s.

The results for the particle growth rate are shown in Fig. 4.20. We analyze the growth rate per cycle, meaning the growth rate of particles within a single cycle for each magnetic field configuration. We determined the growth rate within only the first cycle, because particles can exhibit a bimodal size distribution in subsequent cycles as discussed in sectio 4.1 in Fig. 4.6c [53, 72]. These results are shown in Fig. 4.20a, where we observe that the maximum particle radius ($\sim 250$ nm) and growth rate ($\sim 178$ nm/cycle), as determined by the slope of the graph, were achieved for particles grown without an external magnetic field.

In Fig. 4.20b, the particle radius is shown as a function of magnetic field strength for each section of the cycle time. The size decreases with increasing field strength up to about

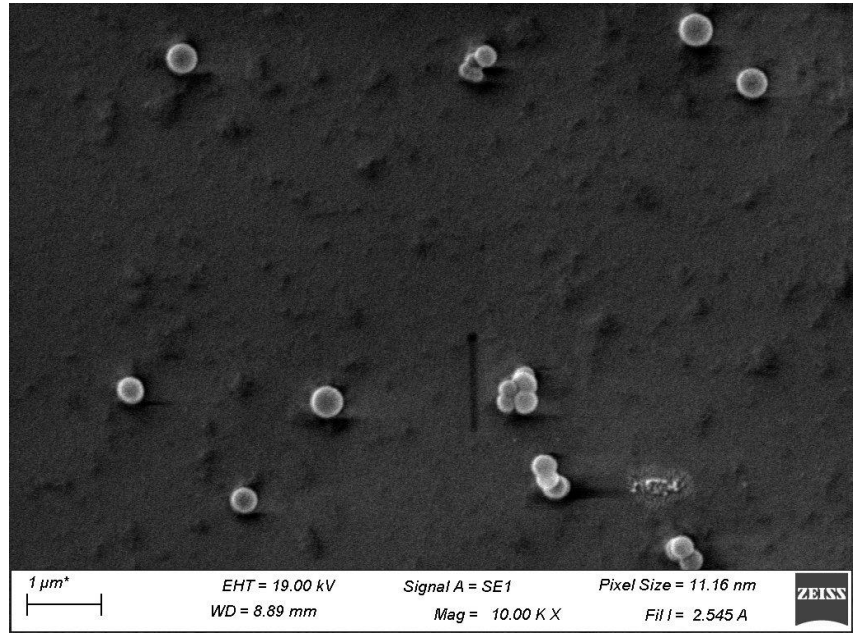

(a)

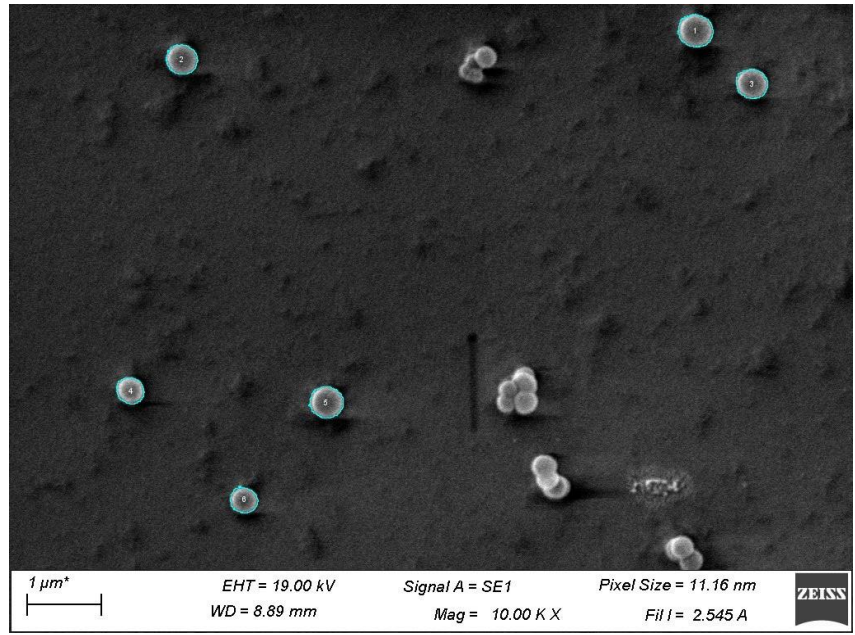

(b)

Figure 4.19: Scanning electron microscope (SEM) images of nanoparticles grown for $T_c = 47$ s at 1020 Gauss: (a) Original SEM image. (b) The same image with isolated particles highlighted for size distribution analysis. (Adapted from [83].)

233 Gauss, after which it slightly increases with further increases in magnetic field strength. In Fig. 4.20c, the growth rate, extracted from Fig. 4.20a, is plotted as a function of magnetic field strength. We observe that the growth rate initially decreases to approximately 30 nm/cycle at 233 Gauss but gradually increases again to approximately 82 nm/cycle at 1020 Gauss. These results indicate that the particle size and growth rate in nonthermal plasmas initially decrease in the presence of a magnetic field and can be further controlled by varying the field strength. The data in Fig. 4.20a and Fig. 4.20b are from one experiment, which was repeated twice in Fig. 4.20c, displaying similar growth patterns.

### 4.3.3 Electron hall parameter effects on the plasma

The red line in Fig. 4.18 represents B = 334 Gauss, where $H_e^{'} \approx 1$, assuming an electron temperature of 3 eV, based on our previous measurements for both magnetized and unmagnetized plasma [53]. However, variations in electron temperature between 2 and 4 eV would shift the magnetic field range for $H_e^{'} = 1$ from 273 to 386 Gauss, hence the red-colored band range of the magnetic field. To the right of the band ($H_e^{'} > 1$), electron dynamics are expected to be dominated by the presence of a magnetic field. To the left of the band ($H_e^{'} < 1$), electron dynamics are expected to be dominated by collisions. With increasing magnetic field strength beyond $H_e^{'} > 1$, the spatial distribution of electrons in the plasma will be governed by the magnetic field, while ions will remain unmagnetized. We observe that $T_c$ and particle radii gradually decrease until $H_e^{'} \approx 1$.

To interpret the cycle time measurements, it is necessary to summarize how the particle growth is altered. We start with an examination of the maximum size of the grown particles in a single cycle. The experimental results show: (a) that the maximum size obtained by the particles decreases with increasing magnetic field, and (b) there is a rapid reduction in the growth rate (in nm/cycle) with increasing magnetic field, up to B $\sim$ 250 Gauss. This combination of factors provides some initial evidence that the presence of the magnetic field has altered both the force balance on the particle that confines them in the plasma and the reaction rates that lead to particle formation.

In terms of the force balance, it is known that the zero order force balance that levitates the particles in the plasma is between determined by the electric and gravitational forces on the dust particles, as shown in Fig. 3.4. The important question is how might the presence of the magnetic field alter the distribution of the electric potential and, subsequently, the electric fields in the plasma. It is possible that the spatial distribution of electrons changed with the increasing magnetic field, thereby altering the plasma ambipolar electric field. Let us consider why the behavior of electrons might change in the presence of magnetic fields. The modified Hall parameter, $\left( H_e^{'} \right)$ given by Eq. 3.22, which is the ratio of the electron gyrofrequency ($\omega_e$) to the electron-neutral collision frequency ($\nu_{n,e}$), quantifies electron magnetization in a plasma exposed to magnetic fields. When $H_e^{'} > 1$, the electrons are magnetized, i.e. they are more likely to complete a full gyro-orbit around a magnetic field line before colliding with neutral species. It is possible that this changes the spatial distribution of electrons in the plasma, and thus altering the electric profile of the background plasma.

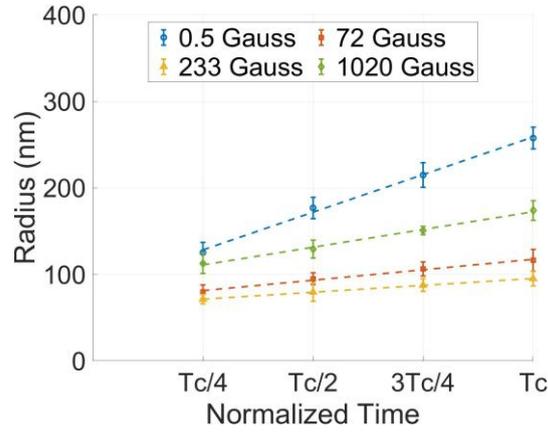

(a)

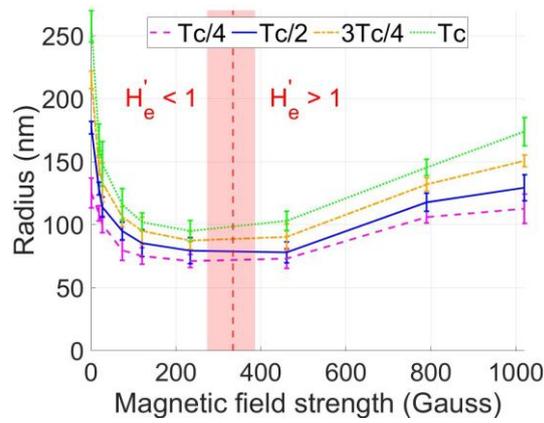

(b)

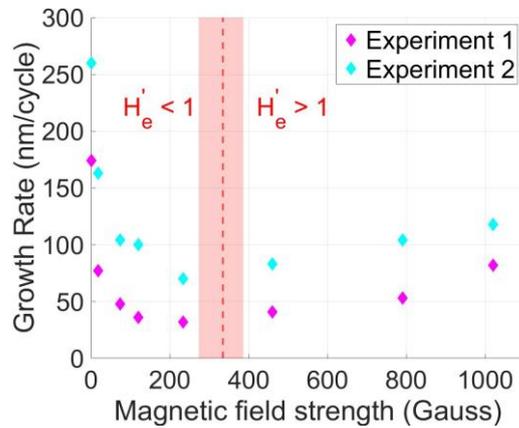

(c)

Figure 4.20: Particle size distribution from scanning electron microscope at four time intervals within one growth cycle at varying magnetic fields (Gauss). (a) Radius from a single cycle. (b) Radius as a function of magnetic field. (c) Growth rate (slope of (a)) as a function of magnetic field. (Adapted from [83].) The red band indicates the experimental regime for the modified electron hall parameter ($H_e' \approx 1$), with the electron dynamics dominated by either collisions in the plasma when ($H_e' < 1$ or by gyromotion when ($H_e' > 1$.

Tadsden et. al (2018) showed that a shorter growth period appeared to be due to weaker confinement caused by reduced dust charge at higher magnetic fields [145]. However, there could be other changes in the dusty plasma that could explain the observed phenomena. First, as electrons transition with increasing magnetic field strength, the electric potential of the plasma is modified. This alters the electric field and the force required for particle confinement, potentially explaining the reduced particle radii observed in the presence of magnetic fields. Second, the new spatial distribution of electrons might affect the charging of the dust particles. The particle charge is crucial for the force balance that levitates them in the plasma, and any changes in charging could be causing the reduced particle size and decreased cycle time ($T_c$). Third, increasing magnetic field strengths might affect the electron density in the dusty plasma. Dust acts as a sink for electrons, reducing both electron density and metastable argon density [146]. This reduction may impact particle coagulation and agglomeration during growth, thus explaining changes in the particle growth rate. Decoupling these three factors—changes in the electric field, dust charging, and electron density—is beyond the scope of this work. However, understanding their behavior is crucial and encouraged for further studies on particle growth in magnetized plasmas.

### 4.3.4   Magnetic Field Effects on OES of the dusty plasma

Electrons are crucial for material synthesis because they bombard precursor gas molecules, causing dissociation and ionization, which creates radicals and excited species that are highly reactive for material synthesis. Our experiments can monitor electron behavior via optical emission spectroscopy (OES). In the temporal variation of the OES intensity of Ar I at 763.5 nm, we observe that the first peak in the OES intensity of the dusty plasma is shifted to the right, as shown in Figure 4.21a. This suggests a prolonged time for electrons to populate metastable energy levels with increasing magnetic field. In contrast, the OES intensity of the clean plasma shows no consistent trend in the time required to reach the first OES peak, as shown in Figure 4.21b.

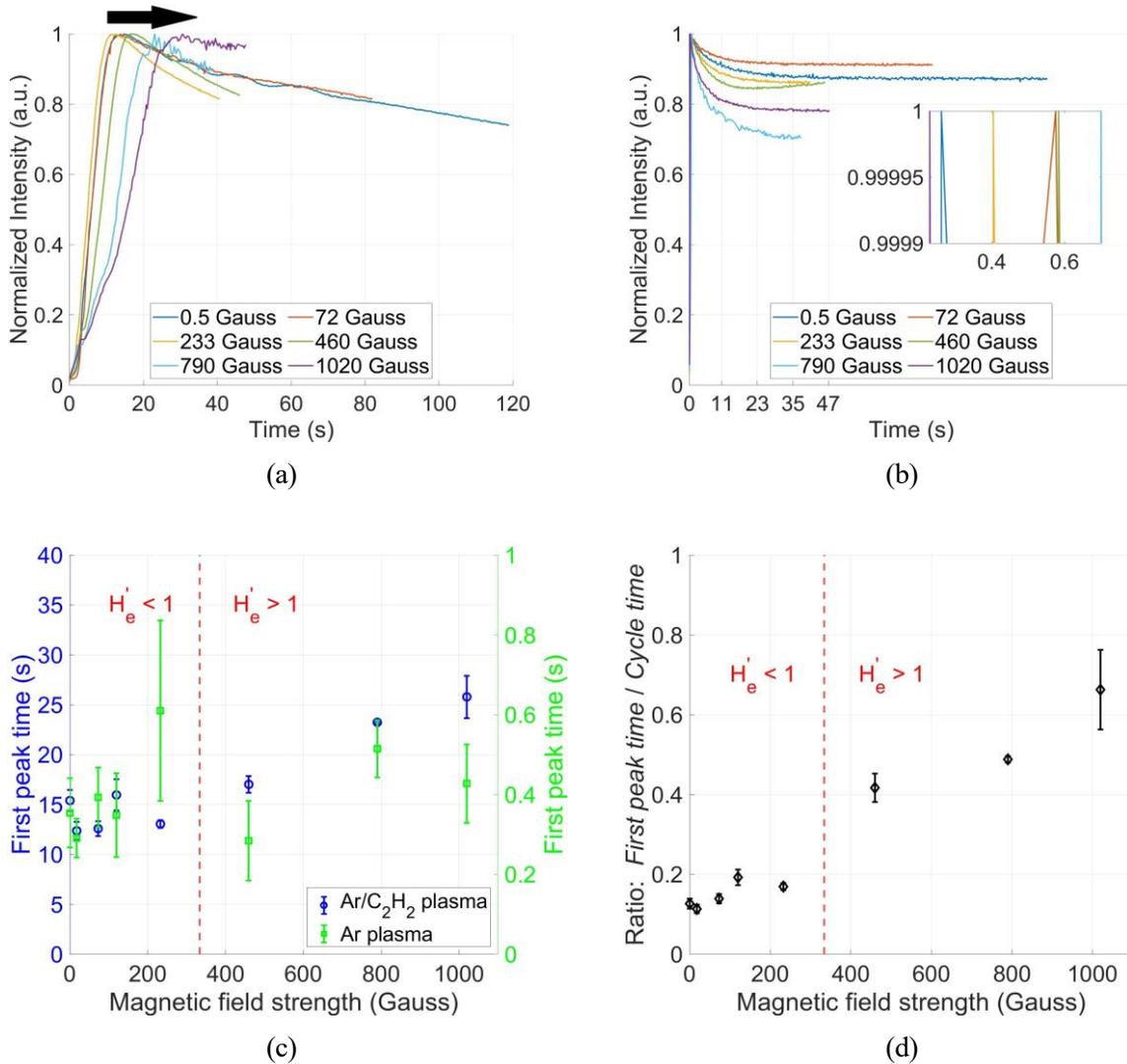

(a)

(b)

(c)

(d)

Figure 4.21: Temporal variation of first peak in optical emission spectroscopy (OES) intensity of Ar I at 763.5 nm for different magnetic field configurations. (a) In the dusty Ar/$C_2H_2$ plasma, the first OES peak shifts to the right with increasing magnetic field (see black arrow). (b) In clean Ar plasma, there is no noticeable trend in peak time shift. (c) The time from plasma ignition to the first peak for clean (blue) and dusty (green) plasma. The Ar plasma reaches the first peak in 0.4 s, whereas the Ar/$C_2H_2$ plasma shows an increasing delay with magnetic field strength for $H_e' > 1$. (d) The ratio of time needed to reach the first OES peak to the cycle time for the dusty plasma increases after $H_e' > 1$.

In Figure 4.21c, the time for the first peak is plotted for both types of plasma at various magnetic fields. For the clean Ar plasma, the time remained approximately 0.3 - 0.6 s across all field strengths. For the dusty plasma, the time was approximately 12 s when $H_e < 1$, but increased from approximately 17 s at 460 Gauss to about 25 s at 1020 Gauss. This increase in time also coincides with a slight increase in the growth rate of the particles from 460 to 1020 Gauss. Finally, in Figure 4.21d, the ratio of the time to the first peak to the cycle time for each dusty plasma is presented. It can be seen that with an increasing magnetic field, an increasing portion of the cycle time is spent before the first peak in OES. Since electrons are needed for both nucleation and to populate the metastable states before emitting photons for OES, it is possible that there is a competing effect between these two phenomena, resulting in a change in growth rate. OES data can be further improved by using a spectrometer with better time and wavelength resolution than ours.

## Chapter 5

## Summary and Future Directions

### 5.1  Summary

This dissertation reports on an investigation of the growth of carbonaceous and titania nanoparticles in unmagnetized plasmas and under conditions where the electrons are magnetized and the ions and charged dust particles are unmagnetized. In particular, the experimental results show that the presence of a magnetic field modifies the cyclical growth rate of the nanoparticles. The experiments show that the threshold of electron magnetization appears to be an important indicator for the evolution of the nanoparticles growth. The three main experimental results are summarized below:

### 5.1.1  Introducing titania dusty plasma

We have introduced a growth technique for $TiO_2$ via a TTIP metal-organic precursor that produces a dusty plasma. There was a growth cycle that lasted $\sim 77$ s, as measured by OES and CMOS camera images. Our experiments showed that $TiO_2$ particles can be grown to monodisperse sizes of $235 \pm 20$ nm within 70 s. SEM images revealed a linear growth rate in particles' radii over two growth cycles. Moreover, the samples after 70 s of the first cycle were collected twenty times and annealed to produce anatase and rutile phases. The results show room for future studies in particle growth using metal-organic precursors, characterization of particle chemistry and microstructure, and dusty plasma studies of interactions of particles with magnetic fields during growth. These findings open avenues for a rapid and controlled growth of titanium dioxide via dusty plasma.

### 5.1.2 Comparing titania to carbonaceus dusty plasma at 500 Gauss

We compared the growth of titania and carbonaceous nanoparticles as dusty plasmas, in the presence and absence of a weak magnetic field of 500 Gauss. Experiments performed in both acetylene (leading to carbonaceous particles) and TTIP (leading to titania particles) displayed a cyclic behaviour, with a faster cycle time at 500 Gauss. The cycle times were measured from OES of the plasma and a numerical Fourier transform analysis of light emission intensity of the dust cloud. The particles reached their maximum size at the end of the cycle. The spatial distribution of the dust cloud containing nanoparticles without magnetic fields appeared similar for both the acetylene and TTIP systems. However, during the presence of the magnetic field, the two dust clouds appeared differently. Measurements of background plasma parameters revealed changes in floating potential and subsequently the electric fields during the presence of the magnetic field. However, these changes in the electric fields were not enough to explain the differences between the two dust cloud during the presence of the magnetic field. Nevertheless, there was a gradient in the radial distribution of magnetic field, and it is possible that the two different materials responded differently to the changes in magnetic fields. Future studies need to investigate the material properties and whether the gradient is responsible for the differences in the spatial distribution of the dust cloud between titania and carbonaceous dusty plasmas.

### 5.1.3 Electron magnetization affects the growth cycle of carbonaceous dusty plasma

We successfully grew carbonaceous dusty nanoparticles in a capacitively coupled nonthermal plasma under the influence of weak magnetic fields ranging from 18 to 1020 Gauss. Particles grown without additional external magnetic fields (0.5 Gauss) exhibited the largest radii, longest cycle times, and fastest growth rates. These parameters gradually decreased until the electrons became magnetized $\left( H'e = 1 \right)$. Beyond this point, up to 1020 Gauss, the cycle time plateaued and the growth rate and particle size slightly increased, though not to the same extent as at 0.5 Gauss. The time for the OES to reach the first peak intensity increased for the dusty plasma, indicating changes in electron behavior in the presence of magnetic fields. These findings suggest that weak magnetic fields can be used to control material synthesis in nonthermal plasms, offering potential applications in various industrial processes such as semiconductor processing from reactive precursor gases.

## 5.2 Future Directions

Literature has shown that a weak magnetic field as low as 50 Gauss can affect the plasma potential, and hence the electric field in capacitively coupled Ar plasmas [147]. The smaller radii of nanoparticles grown during the presence of magnetic field suggest that there might be a reduction in the electric field. This is consistent with preliminary measurement of our electric field, however a full description the effect of magnetic field on the electric field will require a separate study. Towards the end of the growth cycle, the dominant forces acting on the nanoparticles are gravitational, electric and ion-drag. Therefore, a possible decrease in the electric force during the presence of the magnetic field is probably reducing the force balance, thus causing the faster growth cycle and reduced nanoparticles' size.

Initial measurements made by emissive probes and simulation made by COMSOL Multiphysics simulation suggest that the magnitude of electric field decreases with increasing pressure and increasing magnetic fields in the plasma chamber. However, this study requires further attention in the future to fully describe the plasma, and understand why particles grow at a decreasing cycle time with increasing magnetic fields. One of the major components that govern the forces' magnitude is the ambipolar electric field of the background plasma. We have some initial measurement of the electric field from the gradient of the plasma potential $\left( V_p \right)$ in order to determine whether its magnitude changes with the addition of an external magnetic field. This is done both experimentally using emissive probe measurements. We have an initial simulation from COMSOL multiphysics simulation for our geometry, however the addition of an external magnetic field requires further developments on the simulation. **Some initial results from emissive probe measurements and COMSOL multiphysics simulation are shown in this chapter, but further studies are required to obtain a complete understanding of the underlying plasma physics.**

### 5.2.1 Emissive Probe Measurement

We collected the plasma potential ($V_P$) from the emissive probe measurement of the background plasma and calculate the electric field from its gradient [148, 149]. The method of floating potential ($V_F$) saturation is used to collect spatial data in the x and z directions. There are various methods to collect $V_P$, but this method works well enough if there is no gradient in plasma electron temperature $T_e$ [91]. In fact, $T_e$ of this plasma has been calculated and reported to be 2 - 4 eV, with and without a magnetic field of 500 Gauss [53]. Hence we assume that there is no fluctuation and use the saturation in floating potential method to measure $V_P$. $V_F$ is measured in small increments of the heating current ($I_H$) until the former saturates. At that $I_H$, $V_F \approx V_P$. In reality, there should be a small and constant offset between the two potentials: $V_P = V_F + cT_e$, where c is a constant, but since we assume $T_e$ to be independent of the magnetic fields, they are both eliminated when calculating the gradient of the potential.

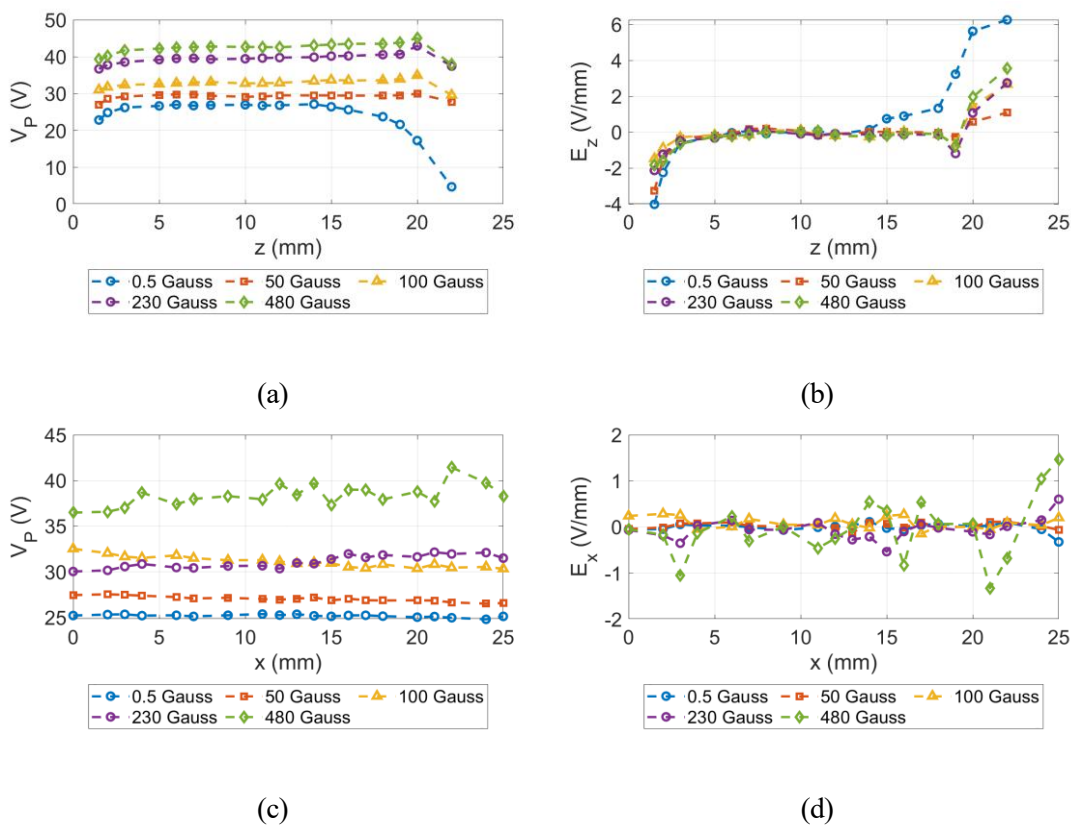

(a)  (b)

(c)  (d)

Figure 5.1: Emissive probe measurements: (a) Measured $V_P$ as a function of height, and (b) corresponding $E_z$, (c) Measured $V_P$ as a function of electrode length, and (d) corresponding $E_x$.

While these emissive probe measurements are preliminary, they suggest that the increased magnetic fields decrease the longitudinal electric field of the background plasma ($E_z$). However, the smaller particle size can be due to a decreased electric force which also depends on the dust charge $F_E = Q_d E$. There needs to be future work to determine whether $Q_d$ is decreasing for our nanoparticles growing in the magnetized plasma. Furthermore, the probe measurements are also perfomed at much lower pressure (20 milliTorr) than the nanoparticles' growth experiment (300 - 500 milliTorr). Unfornately, emissive probes can not reliably operate at higher pressures, because the increased collision will break the probe. This will require the probe to be replaced every few hours before even finishing a full scan of measurements (for example across z, or x). In order to make a better estimate of the electric field of the background plasma, we start some preliminary work in COMSOL Multiphysics simulation, discussed next.

### 5.2.2  1-D COMSOL Multiphysics Simulation

COMSOL Multiphysics simulation may offer a better estimate of the backgroud electric field which would then answer part of the question as to why the nanoparticles move away from the plasma faster during the presence of the background magnetic field. In this section, we introduce a 1-D plasma simulation.

A 1-D plasma simulation was conducted with argon gas. The setup included an electrode spacing of 25 mm, as seen in Fig. 5.2a. A non-uniform mesh was applied, with finer resolution near the walls to accurately capture the sheath region. This operated at room temperature (300 K). The system was powered by a 60 W radio frequency (RF) source at 13.56 MHz. The pressure in the chamber was varied between 0.1 and 0.7 Torr. The simulation determined key plasma properties such as electron density (Fig. 5.2b), electron temperature (Fig. 5.2c), and plasma potential (Fig. 5.2d). The electric field was calculated as $E_z = -dVp/dz$ (Fig. 5.2e).

This work still needs to be improved in the future to fully capture the geometry of our system, and to benchmark our experiment. **All results presented were obtained without magnetic fields, the introduction of which requires a 2-D simulation to accurately capture electrons' motion.** Furthermore, it is noted that we are obtaining plasma electron density of $\sim 10^{16} \text{m}^{-3}$ which is slightly higher than what is expected for LTPs. These simulation results currently show a trend in the changes of the plasma, but do not necessarily provide an accurate magnitude of the parameters. The simulation can be further improved when 2-D geometry are considered and might then provide better estimates of the magnitudes of parameters in the plasma.

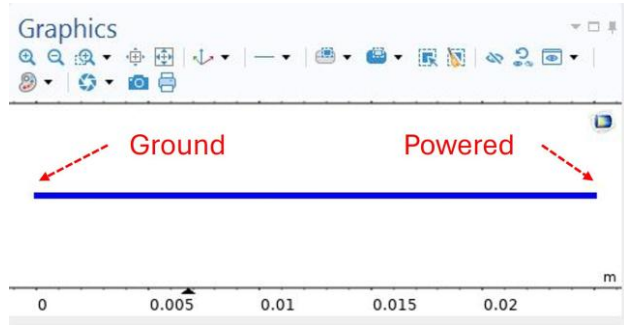

(a) 1-D set up

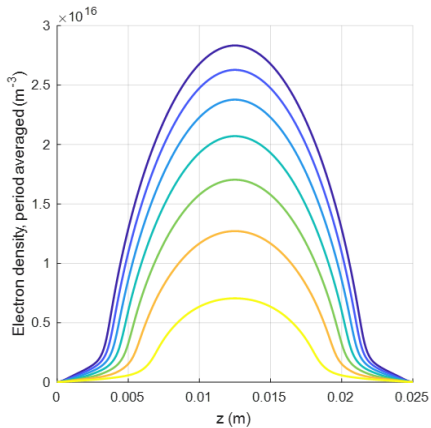

(b) 1-D Electron Density

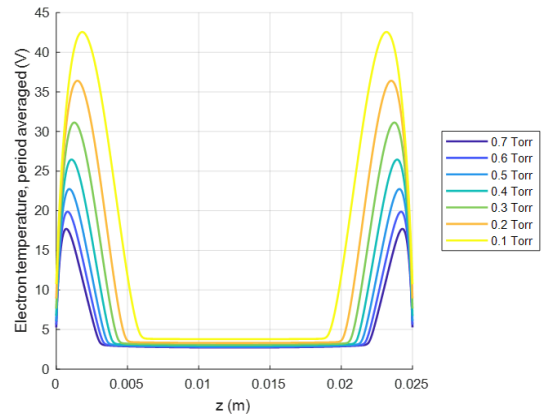

(c) 1-D Electron Temperature

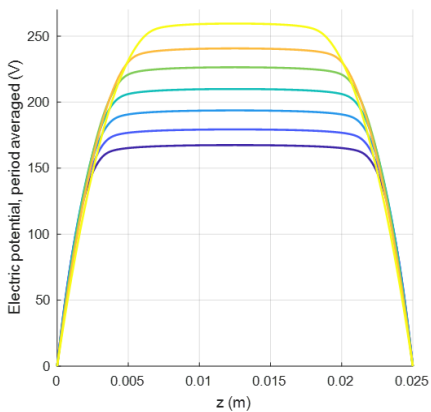

(d) 1-D Plasma Potential

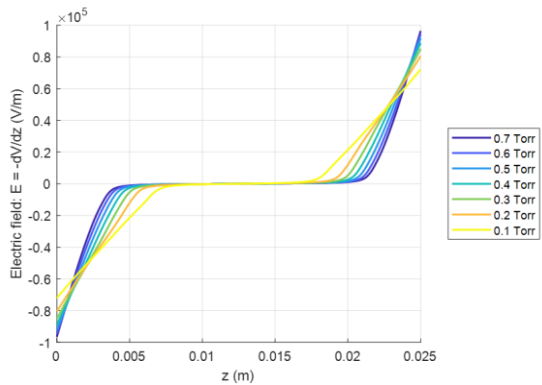

(e) 1-D Electric Field

Figure 5.2: Summary of COMSOL multiphysics simulation in 1-D for the CCP used in this dissertation.

# References

[1] B. B. Reynolds, Q. R. Weinberg, A. A. Witt, A. T. Newton, H. R. Feiler, B. Ramkorun, D. B. Clayton, P. Couture, J. E. Martus, M. Adams, et al. Quantification of dti in the pediatric spinal cord: application to clinical evaluation in a healthy patient population. *American Journal of Neuroradiology*, 40(7):1236–1241, 2019.

[2] B. Ramkorun, B. Reynolds, S. Smith, P. Couture, and A. Bhatia. Diffusion tensor imaging of the spine in pediatric patients. *Medical Imaging 2018: Biomedical Applications in Molecular, Structural, and Functional Imaging*, 10578:528–534, 2018.

[3] B. Reynolds, S. By, A. Witt, Q. Weinberg, B. Ramkorun, B. Harrington, J.M. Ndolo, D. Pastakia, A. Esbenshade, J.C. Wellons, et al. Oncologic treatment effects in the pediatric spinal cord assessed by diffusion tensor imaging. In *PEDIATRIC BLOOD & CANCER*, volume 65, pages S653–S653. WILEY 111 RIVER ST, HOBOKEN 07030-5774, NJ USA, 2018.

[4] B. Ramkorun, S. Smith, B. Reynolds, P. Couture, and A. Bhatia. Medical imaging physics: Quantitative imaging and analysis of the pediatric spinal cord to detect pathologies. *APS March Meeting Abstracts*, V46:007, 2018.

[5] L. Couëdel, D. Artis, M. P. Khanal, C. Pardanaud, Stéphane Coussan, S. LeBlanc, T. Hall, E. Thomas Jr., U. Konopka, M. Park, et al. Influence of magnetic field strength on nanoparticle growth in a capacitively-coupled radio-frequency ar/c2h2 discharge. *Plasma Res. Express*, 1(015012):1–18, 2019.

[6] M. Menati, S. Williams, B. Rasoolian, E. Thomas Jr., and U. Konopka. Formation of turing patterns in strongly magnetized electric discharges. *Commun. Phys.*, 6(221):1–12, 2023.

[7] S. Jaiswal, M. Menati, L. Couëdel, V. H. Holloman, V. Rangari, and E. Thomas Jr. Effect of growing nanoparticle on the magnetic field induced filaments in a radio-frequency ar/c2h2 discharge plasma. *Jpn. J. Appl. Phys.*, 59(SHHC07):1–6, 2020.

[8] K. Yi, D. Liu, X. Chen, J. Yang, D. Wei, Y. Liu, and D. Wei. Plasma-enhanced chemical vapor deposition of two-dimensional materials for applications. *Acc. Chem. Res.*, 54(4):1011–1022, 2021.

[9] B. Ramkorun, K. Chakrabarty, S. Harris, and S. Catledge. Effects of substrate bias on current for varying argon intensity in a low-pressure hydrogen/argon plasma in chemical vapor deposition. *Bulletin of the American Physical Society*, 65, 2020.

[10] B. Ramkorun, K. Chakrabarty, and S. A. Catledge. Effects of direct current bias on nucleation density of superhard boron-rich boron carbide films made by microwave plasma chemical vapor deposition. *Mater. Res. Express*, 8(046401):1–12, 2021.

[11] N. A. Shepelin, Z. P. Tehrani, N. Ohannessian, C. W. Schneider, D. Pergolesi, and T. Lippert. A practical guide to pulsed laser deposition. *Chem. Soc. Rev.*, 52(7):2294–2321, 2023.

[12] S. B. S. Heil et al. Deposition of tin and hfo$_2$ in a commercial 200mm plasma atomic layer deposition reactor. *J. Vac. Sci. Technol. A*, 25(5):1357–1366, 2007.

[13] H. B. Profijt, S. E. Potts, M. C. M. Van de Sanden, and W. M. M. Kessels. Plasma-assisted atomic layer deposition: basics, opportunities, and challenges. *J. Vac. Sci. Technol. A*, 29(5):1–26, 2011.

[14] P. K. Shukla. A survey of dusty plasma physics. *Phys. Plasma*, 8(5):1791–1803, 2001.

[15] P. J. Bruggeman, C. Canal, et al. The 2022 plasma roadmap: low temperature plasma science and technology. *J. Phys. D: Appl. Phys.*, 55(373001):1–55, 2022.

[16] J. Beckers, J. Berndt, D. Block, M. Bonitz, P. J. Bruggeman, Lénaïc Couëdel, G. L. Delzanno, Y. Feng, R. Gopalakrishnan, F. Greiner, et al. Physics and applications of dusty plasmas: The perspectives 2023. *Phys. Plasma*, 30(120601):1–51, 2023.

[17] S. V. Vladimirov and K. Ostrikov. Dynamic self-organization phenomena in complex ionized gas systems: new paradigms and technological aspects. *Phys. Rep.*, 393(3-6):175–380, 2004.

[18] V. N. Tsytovich et al. Self-organized dusty structures in a complex plasma under microgravity conditions: Prospects for experimental and theoretical studies. *Phys. Usp.*, 58(2):150–166, 2015.

[19] Y. L. Song, F. Huang, Z. Y. Chen, Y. H. Liu, and M. Y. Yu. Self-organization and oscillation of negatively charged dust particles in a 2-dimensional dusty plasma. *Phys. Lett. A*, 380(7-8):886–895, 2016.

[20] R. E. Boltnev, M. M. Vasiliev, E. A. Kononov, and O. F. Petrov. Self-organization phenomena in a cryogenic gas discharge plasma: Formation of a nanoparticle cloud and dust–acoustic waves. *J. Exp. Theor. Phys.*, 126:561–565, 2018.

[21] W.-T. Juan, Z.-H. Huang, J.-W. Hsu, Y.-J. Lai, and I. Lin. Observation of dust coulomb clusters in a plasma trap. *Phys. Rev. E*, 58(6):R6947–R6950, 1998.

[22] S. Ratynskaia, C. Knapek, K. Rypdal, S. Khrapak, and G. Morfill. Statistics of particle transport in a two-dimensional dusty plasma cluster. *Phys. Plasmas*, 12(022302):1–11, 2005.

[23] A. Schella, T. Miksch, A. Melzer, J. Schablinski, D. Block, A. Piel, H. Thomsen, P. Ludwig, and M. Bonitz. Melting scenarios for three-dimensional dusty plasma clusters. *Phys. Rev. E*, 84(056402):1–11, 2011.

[24] S. Nunomura, D. Samsonov, and J. Goree. Transverse waves in a two-dimensional screened-coulomb crystal (dusty plasma). *Phys. Rev. Lett.*, 84(22):5141–5144, 2000.

[25] P. K. Shukla and A. A. Mamun. Dust-acoustic shocks in a strongly coupled dusty plasma. *IEEE Trans. Plasma Sci.*, 29(2):221–225, 2001.

[26] Y. Nakamura and A. Sarma. Observation of ion-acoustic solitary waves in a dusty plasma. *Phys. Plasmas*, 8(9):3921–3926, 2001.

[27] E. Thomas Jr., R. Fisher, and R. L. Merlino. Observations of dust acoustic waves driven at high frequencies: Finite dust temperature effects and wave interference. *Phys. Plasmas*, 123701(1–6), 2007.

[28] S. I. Popel, S. I. Kopnin, A. P. Golub', G. G. Dol'nikov, A. V. Zakharov, L. M. Zelenyi, and Yu. N. Izvekova. Dusty plasma at the surface of the moon. *Sol. Syst. Res.*, 47:419–429, 2013.

[29] M. Horányi, T. W. Hartquist, O. Havnes, D. A. Mendis, and G. E. Morfill. Dusty plasma effects in saturn's magnetosphere. *Rev. Geophys.*, 42(RG4002):1–20, 2004.

[30] J.-E. Wahlund, M. André, A. I. E. Eriksson, M. Lundberg, M. W. Morooka, M. Shafiq, T. F. Averkamp, D. A. Gurnett, G. B. Hospodarsky, W. S. Kurth, et al. Detection of dusty plasma near the e-ring of saturn. *Planet. Space Sci.*, 57(14-15):1795–1806, 2009.

[31] M. W. Morooka, J.-E. Wahlund, A. I. E. Eriksson, W. M. Farrell, D. A. Gurnett, W. S. Kurth, A. M. Persoon, M. Shafiq, M. André, and M. K. G. Holmberg. Dusty plasma in the vicinity of enceladus. *J. Geophys. Res. Space Phys.*, 116(A12221):1–15, 2011.

[32] D. A. Mendis and M. Rosenberg. Cosmic dusty plasma. *Annu. Rev. Astron. Astrophys.*, 32(1):419–463, 1994.

[33] L. Boufendi and A. Bouchoule. Industrial developments of scientific insights in dusty plasmas. *Plasma Sources Sci. Technol.*, 11(3A):A211–A218, 2002.

[34] S. Lee, X. Zhang, T. McKnight, B. Ramkorun, H. Wang, V. Gopalan, J. M. Redwing, and T. N. Jackson. Low-temperature processed beta-phase in$_2$se$_3$ ferroelectric semiconductor thin film transistors. *2D Mater.*, 9(025023):1–9, 2022.

[35] S. I. Krasheninnikov, R. D. Smirnov, and D. L. Rudakov. Dust in magnetic fusion devices. *Plasma Phys. Control. Fusion*, 53(083001):1–54, 2011.

[36] A. Bouchoule. Dusty plasmas. *Phys. World*, 6(8):47–51, 1993.

[37] S. Jaiswal, T. Hall, S. LeBlanc, R. Mukherjee, and E. Thomas. Effect of magnetic field on the phase transition in a dusty plasma. *Phys. Plasmas*, 24(113703):1–7, 2017.

[38] S. Jaiswal and E. Thomas Jr. Melting transition of two-dimensional complex plasma crystal in the dc glow discharge. *Plasma Res. Express*, 1(025014):1–11, 2019.

[39] N. Chaubey, J. Goree, S. J. Lanham, and M. J. Kushner. Positive charging of grains in an afterglow plasma is enhanced by ions drifting in an electric field. *Phys. Plasma*, 28(103702):1–14, 2021.

[40] L. Boufendi and A. Bouchoule. Particle nucleation and growth in a low-pressure argon-silane discharge. *Plasma Sources Sci. Technol.*, 3:252–267, 1994.

[41] E. Kovacevic, I. Stefanovic, J. Berndt, and J. Winter. Infrared fingerprints and periodic formation of nanoparticles in ar/c2h2 plasmas. *J. Appl. Phys.*, 93(5):2924–2930, 2003.

[42] E. Kovacevic, J. Berndt, I. Stefanovic, H.-W. Becker, C. Godde, T. Strunskus, J. Winter, and L. Boufendi. Formation and material analysis of plasma polymerized carbon nitride nanoparticles. *J. Appl. Phys.*, 105(104910):1–8, 2009.

[43] C. Pattyn, E. Kovacevic, S. Hussain, A. Dias, T. Lecas, and J. Berndt. Nanoparticle formation in a low pressure argon/aniline rf plasma. *Appl. Phys. Lett.*, 112(013102):1–5, 2018.

[44] V. Garofano, R. Berard, S. Boivin, C. Joblin, K. Makasheva, and L. Stafford. Multi-scale investigation in the frequency domain of ar/hmdso dusty plasma with pulsed injection of hmdso. *Plasma Sources Sci. Technol.*, 28(055019):1–16, 2019.

[45] E. Kovacevic, J. Berndt, T. Strunskus, and L. Boufendi. Size dependent characteristics of plasma synthesized carbonaceous nanoparticles. *J. Appl. Phys.*, 112(013303):1–5, 2012.

[46] A. Bouchoule and L. Boufendi. Particulate formation and dusty plasma behaviour in argon-silane rf discharge. *Plasma Sources Sci. Technol.*, 2(3):204–213, 1993.

[47] Y. Watanabe and M. Shiratani. Growth kinetics and behavior of dust particles in silane plasmas. *Jpn. J. Appl. Phys.*, 32(6S):3074–3080, 1993.

[48] B. Ganguly, A. Garscadden, J. Williams, and P. Haaland. Growth and morphology of carbon grains. *J. Vac. Sci. Technol., A*, 11(4):1119–1125, 1993.

[49] A. Garscadden, B.N. Ganguly, P.D. Haaland, and J. Williams. Overview of growth and behaviour of clusters and particles in plasmas. *Plasma Sources Sci. Technol.*, 3(3):239–245, 1994.

[50] J. Cao and T. Matsoukas. Deposition kinetics on particles in a dusty plasma reactor. *J. Appl. Phys.*, 92(5):2916–2922, 2002.

[51] F. Galli and U. R. Kortshagen. Charging, coagulation, and heating model of nanoparticles in a low-pressure plasma accounting for ion–neutral collisions. *IEEE Trans. Plasma Sci.*, 38(4):803–809, 2009.

[52] B. Chutia, T. Deka, Y. Bailung, S. K. Sharma, and H. Bailung. A nanodusty plasma experiment to create extended dust clouds using reactive argon acetylene plasmas. *Phys. Plasmas*, 28(063703):1–7, 2021.

[53] B. Ramkorun, G. Chandrasekhar, V. Rangari, S. C. Thakur, R. Comes, and E. Thomas Jr. Comparing growth of titania and carbonaceous dusty nanoparticles in weakly magnetised capacitively coupled plasmas. *Plasma Sources Sci. Technol.*, 33(115004):1–13, 2024.

[54] A. A. Fridman, L. Boufendi, T. Hbid, B. V. Potapkin, and A. Bouchoule. Dusty plasma formation: Physics and critical phenomena. theoretical approach. *J. Appl. Phys.*, 79:1303–1314, 1996.

[55] D. Samsonov and J. Goree. Instabilities in a dusty plasma with ion drag and ionization. *Phys. Rev. E*, 59(1):1047–1058, 1999.

[56] I. Stefanovic, E. Kovacevic, J. Berndt, and J. Winter. Hα emission in the presence of dust in an ar-c2h2 radio-frequency discharge. *New J. Phys.*, 5(39):1–12, 2003.

[57] T. J. M. Donders, T. J. A. Staps, and J. Beckers. Characterization of cyclic dust growth in a low-pressure, radio-frequency driven argon-hexamethyldisiloxane plasma. *J. Phys. D: Appl. Phys.*, 55(39):1–15, 2022.

[58] S. Hong, J. Berndt, and J. Winter. Growth precursors and dynamics of dust particle formation in the ar/ch$_4$ and ar/c$_2$h$_2$ plasmas. *Plasma Sources Sci. Technol.*, 12(46):46–52, 2002.

[59] L. Ravi and S. L. Girshick. Coagulation of nanoparticles in a plasma. *Phys. Rev. E*, 79(026408):1–9, 2009.

[60] C. Cui and J. Goree. Fluctuations of the charge on a dust grain in a plasma. *IEEE Trans. Plasma Sci.*, 22(2):151–158, 1994.

[61] J. Goree. Charging of particles in a plasma. *Plasma Sources Sci. Technol.*, 3(400):400–406, 1994.

[62] S. L. Girshick. Particle nucleation and growth in dusty plasmas: On the importance of charged-neutral interactions. *J. Vac. Sci. Technol., A*, 38(011001):1–6, 2020.

[63] X. Shi, P. Elvati, and A. Violi. On the growth of si nanoparticles in non-thermal plasma: physisorption to chemisorption conversion. *J. Phys. D: Appl. Phys.*, 54(365203):1–10, 2021.

[64] Y. A. Mankelevich, M. A. Olevanov, and T. V. Rakhimova. Dust particle coagulation mechanism in low-pressure plasma: rapid growth and saturation stage modeling. *Plasma Sources Sci. Technol.*, 17(015013):1–9, 2008.

[65] R. Bingham and V. N. Tsytovich. New mechanism of dust growth and gravitation-like instabilities in astrophysical plasmas. *Astron. Geophys.*, 376(3):L43–L47, 2001.

[66] M. S. Sodha, S. Misra, S. K. Mishra, and S. Srivastava. Growth of embryonic dust particles in a complex plasma. *J. Appl. Phys.*, 107(103307):1–7, 2010.

[67] I. B. Denysenko, E. von Wahl, M. Mikikian, J. Berndt, S. Ivko, H. Kersten, E. Kovacevic, and N. A. Azarenkov. Plasma properties as function of time in ar/c$_2$h$_2$ dust-forming plasma. *J. Phys. D: Appl. Phys.*, 53(135203):1–12, 2020.

[68] P. K. Shukla and B. Eliasson. Colloquium: Fundamentals of dust-plasma interactions. *Rev. Mod. Phys.*, 81(1):25–44, 2009.

[69] R. L. Merlino. *Dusty plasmas and applications in space and industry*, volume 81. Transworld Research Network, Kerala, India, 2006.

[70] J. Beckers. *Dust particle(s)(as) diagnostics in plasmas*. Phd thesis, Technische Universiteit Eindhoven, 2011. Applied Physics and Science Education.

[71] F. M. J. H. Van de Wetering, R. J. C. Brooimans, S. Nijdam, J. Beckers, and G. M. W. Kroesen. Fast and interrupted expansion in cyclic void growth in dusty plasma. *J. Phys. D: Appl. Phys.*, 48(035204):1–13, 2015.

[72] B. Ramkorun, S. Jain, A. Taba, M. Mahjouri-Samani, M. E. Miller, S. C. Thakur, E. Thomas Jr., and R. B. Comes. Introducing dusty plasma particle growth of nanospherical titanium dioxide. *Appl. Phys. Lett.*, 124(144102):1–7, 2024.

[73] H. Feng, Y. Mao-Fu, W. Long, and J. Nan. Voids in an experimental dusty plasma system. *Chin. Phys. Lett.*, 21(1):121–124, 2004.

[74] M. Mikikian, L. Boufendi, A. Bouchoule, H. M. Thomas, G. E. Morfill, A. P. Nefedov, V. E. Fortov, et al. Formation and behaviour of dust particle clouds in a radio-frequency discharge: results in the laboratory and under microgravity conditions. *New J. Phys.*, 5(19):1–12, 2003.

[75] M. Hundt, P. Sadler, I. Levchenko, M. Wolter, H. Kersten, and K. Ostrikov. Real-time monitoring of nucleation-growth cycle of carbon nanoparticles in acetylene plasmas. *J. Appl. Phys.*, 109(123305):1–7, 2011.

[76] J. Winter, J. Berndt, S. K. Hong, E. Kovacevic, I. Stefanovic, and O. Stepanovic. Dust formation in ar/ch4 and ar/c2h2 plasmas. *Plasma Sources Sci. Technol.*, 18(034010):1–8, 2009.

[77] B. Despax, F. Gaboriau, H. Caquineau, and K. Makasheva. Influence of the temporal variations of plasma composition on the cyclic formation of dust in hexamethyldisiloxane-argon radiofrequency discharges: Analysis by time-resolved mass spectrometry. *AIP Adv.*, 6(105111):1–20, 2016.

[78] B. Despax, K. Makasheva, and H. Caquineau. Cyclic powder formation during pulsed injection of hexamethyldisiloxane in an axially asymmetric radiofrequency argon discharge. *J. Appl. Phys.*, 112(9):1–10, 2012.

[79] V. Garofano, L. Stafford, B. Despax, R. Clergereaux, and K. Makasheva. Cyclic evolution of the electron temperature and density in dusty low-pressure radio frequency plasmas with pulsed injection of hexamethyldisiloxane. *Appl. Phys. Lett.*, 107(183104):1–5, 2015.

[80] F. Tuinstra and J.L. Koenig. Raman spectrum of graphite. *J. Chem. Phys.*, 53(3):1126–1130, 1970.

[81] B. Kwiecińska, I. Suárez-Ruiz, C. Paluszkiewicz, and S. Rodriques. Raman spectroscopy of selected carbonaceous samples. *Int. J. Coal Geol.*, 84(3-4):206–212, 2010.

[82] G. S. Oehrlein and S. Hamaguchi. Foundations of low-temperature plasma enhanced materials synthesis and etching. *Plasma Sources Sci. Technol.*, 27(023001):1–21, 2018.

[83] B. Ramkorun, S. C. Thakur, R. B. Comes, and E. Thomas Jr. Electron magnetization effects on carbonaceous dusty nanoparticles grown in $ar/c_2h_2$ capacitively coupled nonthermal plasma. *arXiv preprint arXiv:2504.21217*, 2025.

[84] G. Norlen. Wavelengths and energy levels of ar i and ar ii based on new interferometric measurements in the region 3400–9800 Å. *Phys. Scr.*, 8(6):249–268, 1973.

[85] W. L. Wiese, J. W. Brault, K. Danzmann, V. Helbig, and M. Kock. Unified set of atomic transition probabilities for neutral argon. *Phys. Rev. A*, 39(5):2461–2471, 1989.

[86] A. Kramida, Yu. Ralchenko, J. Reader, and NIST ASD Team. NIST Atomic Spectra Database (ver. 5.10), [Online]. Available: `https://physics.nist.gov/asd` [2023, September 12]. National Institute of Standards and Technology, Gaithersburg, MD., 2022.

[87] J. Nash, J. Powell, B. Ramkorun, S.C. Thakur, and E. Thomas. Investigating the effect of biasing of a hollow cylindrical electrode in the alexis plasma device. *APS Division of Plasma Physics Meeting Abstracts*, CP11(043), 2023.

[88] B. Ramkorun, K. Chakrabarty, and S. Catledge. Effect of argon flow and dc bias on optical emission spectroscopy of argon/hydrogen plasmas. *Bulletin of the American Physical Society*, 64, 2019.

[89] I. H. Hutchinson. Principles of plasma diagnostics. *Plasma Phys. Controlled Fusion*, 44(12):2603–2603, 2002.

[90] C. Cho, S. Kim, Y. Lee, I. Seong, W. Jeong, Y. You, M. Choi, and S. You. Determination of plasma potential using an emissive probe with floating potential method. *Materials*, 16(2762):1–11, 2023.

[91] J. P. Sheehan, Y. Raitses, N. Hershkowitz, I. Kaganovich, and N. J. Fisch. A comparison of emissive probe techniques for electric potential measurements in a complex plasma. *Phys. Plasma*, 18(073501):1–9, 2011.

[92] P. Li, N. Hershkowitz, and G. Severn. Building langmuir probes and emissive probes for plasma potential measurements in low pressure, low temperature plasmas. *J. Visualized Exp.*, 171(e61804):1–23, 2021.

[93] C. Suryanarayana and M. G. Norton. Practical aspects of x-ray diffraction. *X-ray Diffraction: A Practical Approach*, pages 63–94, 1998.

[94] C. Kittel. *Solid state physics*, volume 3. Shell Development Company Emeryville, 1955.

[95] C. G. Pope. X-ray diffraction and the bragg equation. *J. Chem. Educ.*, 74(1):129–131, 1997.

[96] J. C. Mauro. *Materials kinetics: transport and rate phenomena*. Elsevier, 2020.

[97] N. Chaubey and J. Goree. Controlling the charge of dust particles in a plasma afterglow by timed switching of an electrode voltage. *J. Phys. D: Appl. Phys.*, 56(375202):1–9, 2023.

[98] M. S. Dresselhaus, G. Dresselhaus, and A. Jorio. *Group theory: application to the physics of condensed matter*. Springer Science & Business Media, 2007.

[99] U. Balachandran and N. Eror. Raman spectra of titanium dioxide. *J. Solid State Chem.*, 42(3):276–282, 1982.

[100] D. A. Gurnett and A. Bhattacharjee. *Introduction to plasma physics: with space and laboratory applications*. Cambridge University Press, 2005.

[101] O. Havnes, G. E. Morfill, and C. K. Goertz. Plasma potential and grain charges in a dust cloud embedded in a plasma. *J. Geophys. Res.: Space Phys.*, 89(A12):10999–11003, 1984.

[102] E. C. Whipple, T. G. Northrop, and D. A. Mendis. The electrostatics of a dusty plasma. *J. Geophys. Res.: Space Phys.*, 90(A8):7405–7413, 1985.

[103] O. Havnes, T. K. Aanesen, and F. Melandsø. On dust charges and plasma potentials in a dusty plasma with dust size distribution. *J. Geophys. Res.: Space Phys.*, 95(A5):6581–6585, 1990.

[104] E. Thomas Jr. and M. Watson. Charging of silica particles in an argon dusty plasma. *Phys. Plasmas*, 7(8):3194–3197, 2000.

[105] E. Thomas Jr. Driven dust acoustic waves with thermal effects: Comparison of experiment to fluid theory. *Phys. Plasmas*, 17(043701):1–8, 2010.

[106] A. Petersen, O. H. Asnaz, B. Tadsen, and F. Greiner. Decoupling of dust cloud and embedding plasma for high electron depletion in nanodusty plasmas. *Commun. Phys.*, 5(308):1–8, 2022.

[107] B. Tadsen, F. Greiner, S. Groth, and A. Piel. Self-excited dust-acoustic waves in an electron-depleted nanodusty plasma. *Phys. Plasmas*, 22(113701):1–9, 2015.

[108] V. E. Fortov, A. G. Khrapak, S. A. Khrapak, V. I. Molotkov, and O. F. Petrov. Dusty plasmas. *Phys. Usp.*, 47(5):447–492, 2004.

[109] M. S. Barnes, J. H. Keller, J. C. Forster, J. A. O'Neill, and D. K. Coultas. Transport of dust particles in glow-discharge plasmas. *Phys. Rev. Lett.*, 68(3):313–316, 1992.

[110] I. H. Hutchinson. Collisionless ion drag force on a spherical grain. *Plasma Phys. Controlled Fusion*, 48:185–202, 2006.

[111] S. A. Khrapak, A. V. Ivlev, G. E. Morfill, and H. M. Thomas. Ion drag force in complex plasmas. *Phys. Rev. E*, 66(046414):1–4, 2002.

[112] B. Tadsen, F. Greiner, and A. Piel. On the amplitude of dust-density waves in inhomogeneous dusty plasmas. *Phys. Plasmas*, 24(033704):1–5, 2017.

[113] H. Zhang, B. Chen, J. F. Banfield, and G. A. Waychunas. Atomic structure of nanometer-sized amorphous $tio_2$. *Phys. Rev. B: Condens. Matter*, 78(214106):1–12, 2008.

[114] M. Ortner and L. G. C. Bandeira. Magpylib: A free python package for magnetic field computation. *SoftwareX*, 11(100466):1–7, 2020.

[115] A. Caciagli, R. J. Baars, A. P. Philipse, and B. W. M. Kuipers. Exact expression for the magnetic field of a finite cylinder with arbitrary uniform magnetization. *J. Magn. Magn. Mater.*, 456:423–432, 2018.

[116] N. Derby and S. Olbert. Cylindrical magnets and ideal solenoids. *Am. J. Phys.*, 78(3):229–235, 2010.

[117] E. Thomas Jr., U. Konopka, D. Artis, B. Lynch, S. Leblanc, S. Adams, R. L. Merlino, and M. Rosenberg. The magnetized dusty plasma experiment (mdpx). *J. Plasma Phys.*, 81(345810206):1–21, 2015.

[118] S. Williams, S. Chakraborty Thakur, M. Menati, and E. Thomas Jr. Experimental observations of multiple modes of filamentary structures in the magnetized dusty plasma experiment (mdpx) device. *Phys. Plasmas*, 29(012110):1–10, 2022.

[119] B. Ramkorun, S. Jain, A. Taba, M. Mahjouri-Samani, S. C. Thakur, R. B. Comes, and E. Thomas. Growth and analysis of titanium dioxide dusty microparticle in capacitively coupled rf-plasmas. *2023 IEEE International Conference on Plasma Science (ICOPS)*, O-1.6.6(251):1–1, 2023.

[120] S. Williams. *Filament morphology in highly magnetized capacitively coupled low temperature plasmas*. PhD thesis, Auburn University, 2023.

[121] L. C. Pitchford, L. L. Alves, K. Bartschat, S. F. Biagi, M. C. Bordage, A. V. Phelps, C. M. Ferreira, G. J. M. Hagelaar, W. L. Morgan, S. Pancheshnyi, et al. Comparisons of sets of electron–neutral scattering cross sections and swarm parameters in noble gases: I. argon. *J. Phys. D: Appl. Phys.*, 46(334001):1–19, 2013.

[122] S. A. Khrapak. Practical expression for an effective ion-neutral collision frequency in flowing plasmas of some noble gases. *J. Plasma Phys.*, 79(6):1123–1124, 2013.

[123] E. Thomas Jr. Modeling of particle transport in the magnetized dusty plasma experiment. *Japan. J. Plasma Fusion Res.*, 93:586–595, 2017.

[124] F. F. Chen et al. *Introduction to plasma physics and controlled fusion*, volume 2nd ed. Springer, 1984.

[125] A. Di Paola, M. Bellardita, and L. Palmisano. Brookite, the least known $tio_2$ photocatalyst. *Catalysts*, 3(1):36–73, 2013.

[126] D. Reyes-Coronado, G. Rodr´ıguez-Gattorno, M. E. Espinosa-Pesqueira, C. Cab, R. D. De Coss, and G. Oskam. Phase-pure tio$_2$ nanoparticles: anatase, brookite and rutile. *Nanotechnology*, 19(145605):1–10, 2008.

[127] J. Zhang, P. Zhou, J. Liu, and J. Yu. New understanding of the difference of photo-catalytic activity among anatase, rutile and brookite tio$_2$. *Phys. Chem. Chem. Phys.*, 16(38):20382–20386, 2014.

[128] S. S. Huang and J.-S. Chen. Comparison of the characteristics of TiO$_2$ films prepared by low-pressure and plasma-enhanced chemical vapor deposition. *J. Mater. Sci.: Mater. Electron.*, 13:77–81, 2002.

[129] H. H. Nguyen, D. J. Kim, D. W. Park, and K. S. Kim. Effect of initial precursor concentration on tio2 thin film nanostructures prepared by pcvd system. *J. Energy Chem.*, 22:375–381, 2013.

[130] B. R. Stoner, G.-H. M. Ma, S. D. Wolter, and J. T. Glass. Characterization of bias-enhanced nucleation of diamond on silicon by in vacuo surface analysis and transmission electron microscopy. *Phys. Rev. B*, 45:11067–11084, 1992.

[131] M. Kuhr, S. Reinke, and W. Kulisch. Nucleation of cubic boron nitride (c-bn) with ion-induced plasma-enhanced cvd. *Diamond Relat. Mater.*, 4:375–380, 1995.

[132] B. Ramkorun. An investigation of ion bombardment from microwave plasma chemical vapor deposition during bias enhanced nucleation of boron-rich materials. Master's thesis, The University of Alabama at Birmingham, 2020.

[133] S. Groth, F. Greiner, B. Tadsen, and A. Piel. Kinetic mie ellipsometry to determine the time-resolved particle growth in nanodusty plasmas. *J. Phys. D: Appl. Phys.*, 48(465203):1–9, 2015.

[134] J. Berndt, S. Hong, E. Kovacevic, I. Stefanovic, and J. Winter. Dust particle formation in low pressure ar/ch4 and ar/c2h2 discharges used for thin film deposition. *Vacuum*, 71(3):377–390, 2003.

[135] C. P. Fictorie, J. F. Evans, and W. L. Gladfelter. Kinetic and mechanistic study of the chemical vapor deposition of titanium dioxide thin films using tetrakis-(isopropoxo)-titanium (iv). *J. Vac. Sci. Technol. A*, 12(4):1108–1113, 1994.

[136] K.-H. Ahn, Y.-B. Park, and D.-W. Park. Kinetic and mechanistic study on the chemical vapor deposition of titanium dioxide thin films by in situ FT-IR using TTIP. *Surf. Coat. Technol.*, 171(1-3):198–204, 2003.

[137] J. Winter. Dust in fusion devices—a multi-faceted problem connecting high-and low-temperature plasma physics. *Plasma Phys. Controlled Fusion*, 46(12B):B583–B592, 2004.

[138] B. Ramkorun, S. Thakur, S. Jain, R. Comes, and E. Thomas Jr. Growth and analysis of carbonaceous and metallic microparticles using capacitively coupled rf plasmas. In *APS Division of Plasma Physics Meeting Abstracts*, volume 2022, pages GP11–068, 2022.

[139] B. Ramkorun, S. C. Thakur, C. Royer, R. Comes, and E. Thomas. Comparing the growth of carbonaceous and titanate dust particles in capacitively coupled rf plasmas at 500 gauss. *Bulletin of the American Physical Society*, 2023.

[140] W. W. Stoffels, M. Sorokin, and J. Remy. Charge and charging of nanoparticles in a $sih_4$ rf-plasma. *Faraday Discuss.*, 137:115–126, 2008.

[141] J. Beckers, W. W. Stoffels, and G. M. W. Kroesen. Temperature dependence of nucleation and growth of nanoparticles in low pressure $ar/ch_4$ rf discharges. *J. Phys. D: Appl. Phys.*, 42(155206):1–10, 2009.

[142] J. Beckers and G. M. W. Kroesen. Surprising temperature dependence of the dust particle growth rate in low pressure $ar/c_2h_2$ plasmas. *Appl. Phys. Lett.*, 99(181503):1–3, 2011.

[143] O. Havnes, C. K. Goertz, G. E. Morfill, E. Grün, and W. Ip. Dust charges, cloud potential, and instabilities in a dust cloud embedded in a plasma. *J. Geophys. Res.: Space Phys.*, 92(A3):2281–2287, 1987.

[144] B. Ramkorun. Growth of tio2 dusty microparticles in a magnetized plasma. *MagNatUS Meeting 2023*, 2023(37), 2023.

[145] B. Tadsen, F. Greiner, and A. Piel. Probing a dusty magnetized plasma with self-excited dust-density waves. *Phys. Rev. E*, 97(033203):1–7, 2018.

[146] V. Sushkov, A.-P. Herrendorf, and R. Hippler. Metastable argon atom density in complex argon/acetylene plasmas determined by means of optical absorption and emission spectroscopy. *J. Phys. D: Appl. Phys.*, 49(425201):1–11, 2016.

[147] M. J. Kushner. Modeling of magnetically enhanced capacitively coupled plasma sources: Ar discharges. *J. Appl. Phys.*, 94(3):1436–1447, 2003.

[148] B. Ramkorun, E. Williamson, D. Jose, R. Comes, S. C. Thakur, and E. Thomas. Investigating the cycle time of dusty plasma nanoparticles' growth during the presence of magnetic fields. *APS Division of Plasma Physics Meeting Abstracts*, GP12(093), 2024.

[149] D. Jose, R. Kumar, B. Ramkorun, E. Thomas, S. C. Thakur, and R. Gopalakrishnan. Mapping and analyzing plasma parameter distribution in low-temperature plasma using collecting and emitting probes. *APS Division of Plasma Physics Meeting Abstracts*, B008(003), 2024.

**Appendices**

# Appendix A

## Computer Aided Design (CAD) for the Two Electrodes

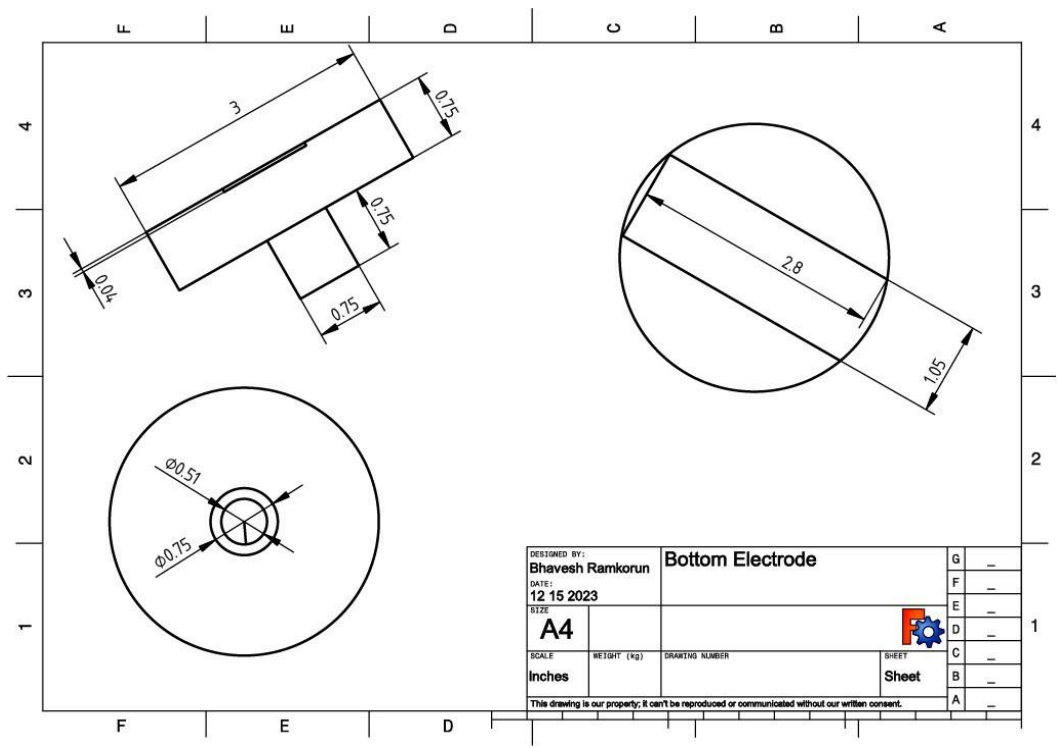

(a)

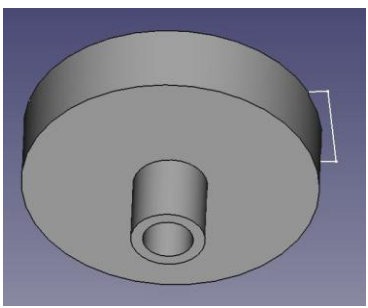

(b)

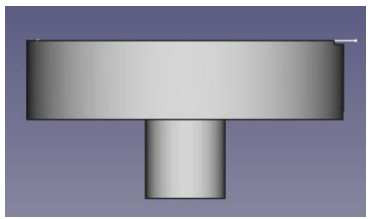

(c)

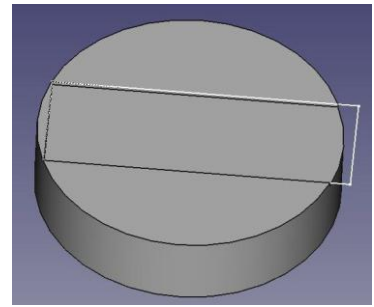

(d)

Figure A.1: Computer-Aided Design (CAD) design for the bottom electrode, where materials are collected.

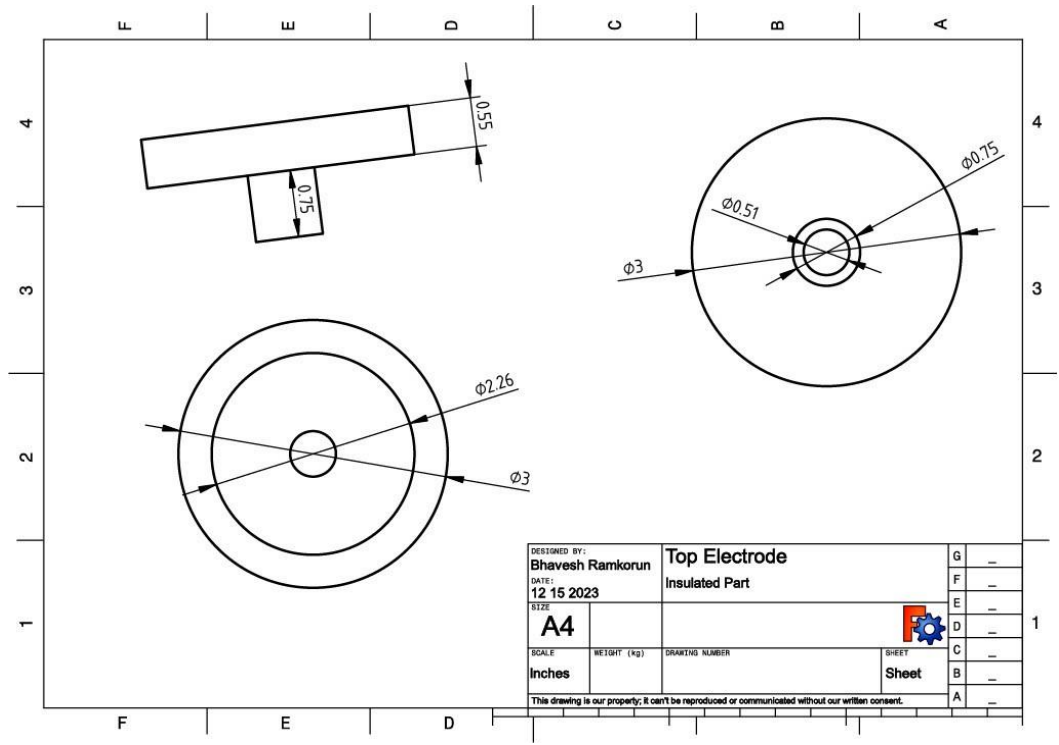

(a)

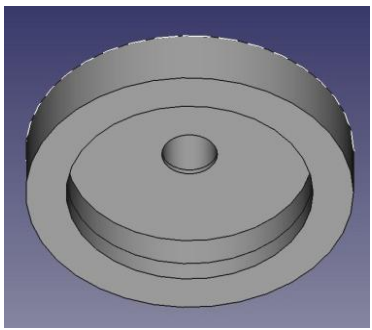

(b)

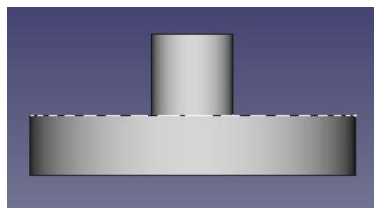

(c)

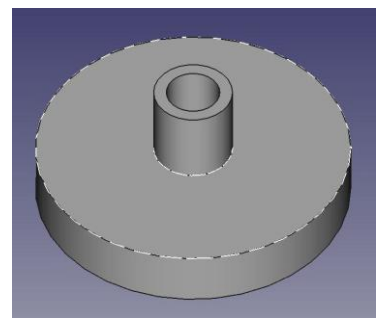

(d)

Figure A.2: Computer-Aided Design (CAD) design for the top electrode, where power is supplied to the plasma.

# Appendix B

## Measuring Pressure Using Data Acquisition (DAQ)

### B.1   Convert voltage to pressure in milliTorr

| True Total Pressure | | N$_2$ |
|---|---|---|
| 0 | mTorr | 0.3751 |
| **0.1** | **mTorr** | **0.3759** |
| 0.2 | mTorr | 0.3768 |
| 0.5 | mTorr | 0.3795 |
| **1** | **mTorr** | **0.3840** |
| 2 | mTorr | 0.3927 |
| 5 | mTorr | 0.4174 |
| **10** | **mTorr** | **0.4555** |
| 20 | mTorr | 0.5226 |
| 50 | mTorr | 0.6819 |
| **100** | **mTorr** | **0.8780** |
| 200 | mTorr | 1.1552 |
| 500 | mTorr | 1.6833 |
| **1** | **Torr** | **2.2168** |

Figure B.1: Pressure gauge is calibrated to Nitrogen. Using data from KJLC, the N$_2$ voltage (2nd column) is converted into real pressure up to 1 torr (1000 milli Torr)

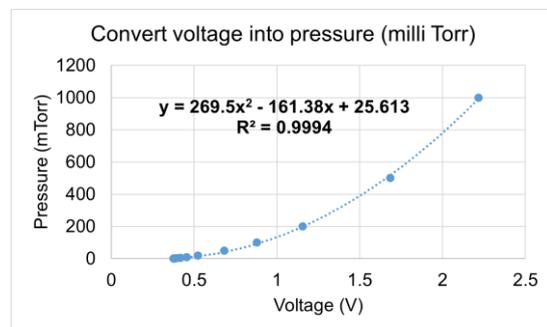

Figure B.2: Using a second order polynomial to extract curve of best fit for pressure reading

## B.2 Block Diagram

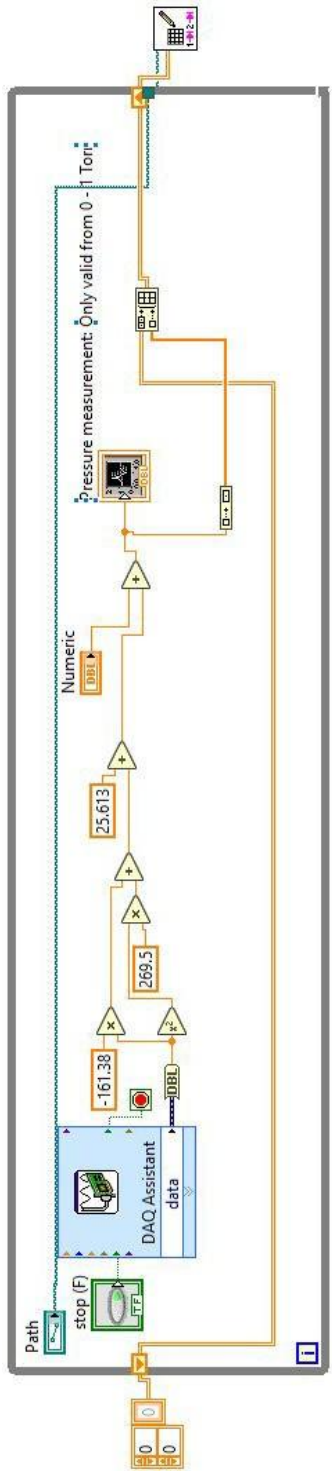

Figure B.3: Block diagram from LabVIEW interface for pressure input using a data acquisition system (DAQ). The second order equation from B.2 is used.

## B.3 Interface

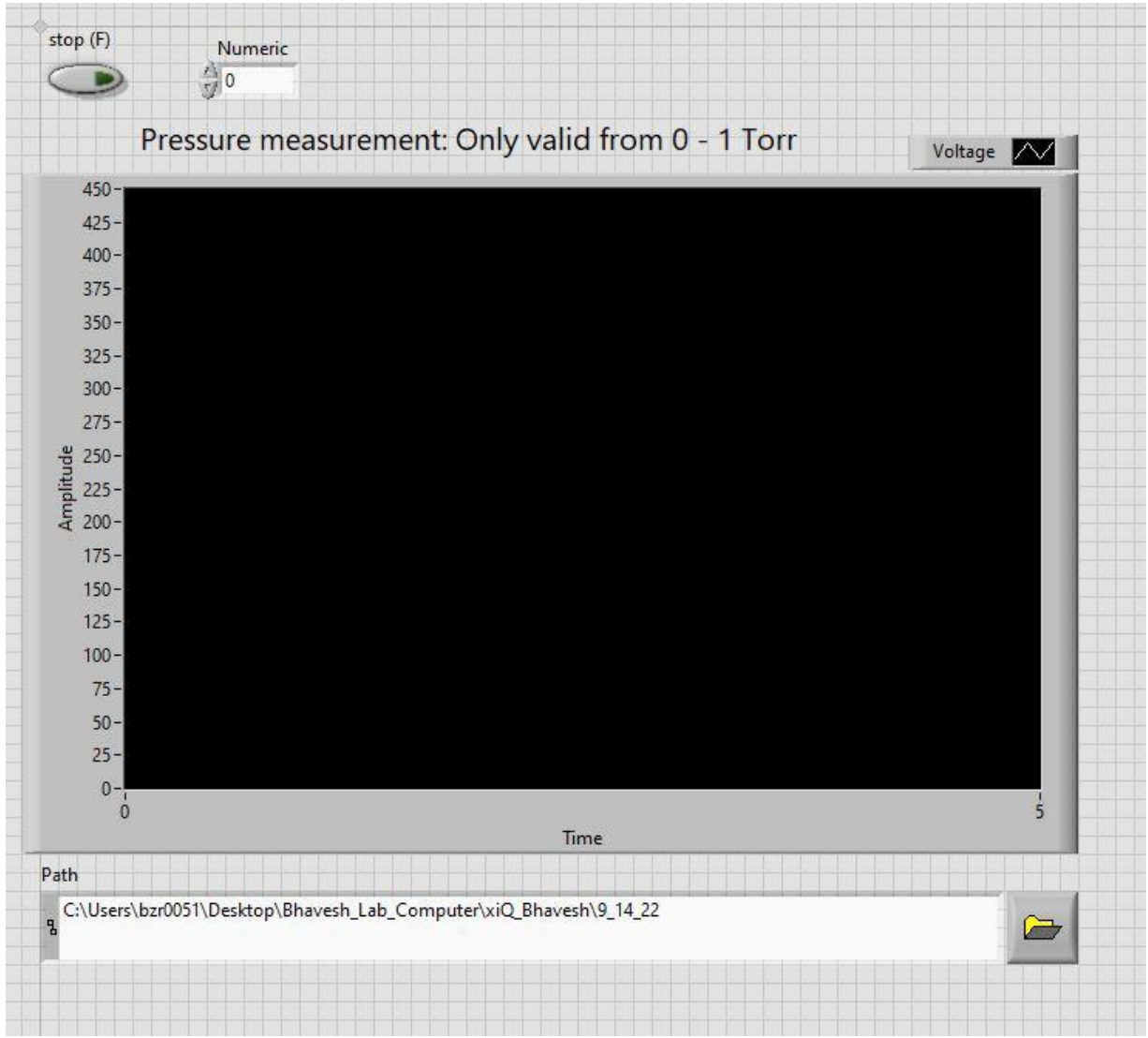

Figure B.4: User interface from LabVIEW interface for pressure input using a data acquisition system (DAQ). The data is only valid up to 1 Torr due to the limited value input in B.2.

# Appendix C

# Cleaning the Vacuum Chamber and Parts

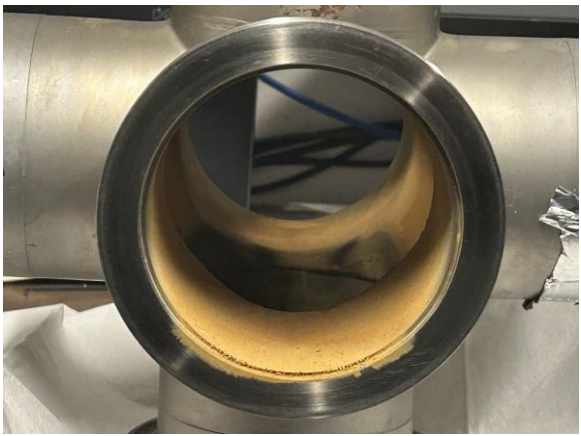

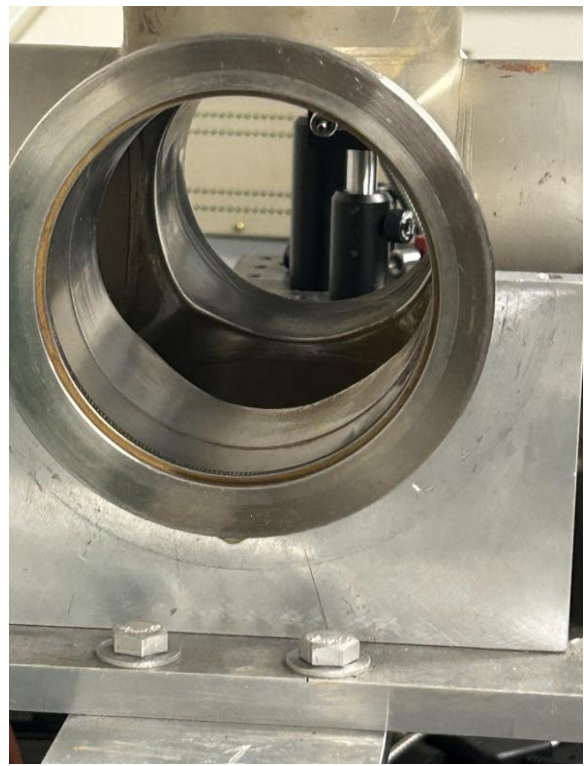

(a) **BEFORE**　　　　　　　　　(b) **AFTER**

Figure C.1: Comparison of vacuum chamber before and after cleaning with very fine sandpaper.

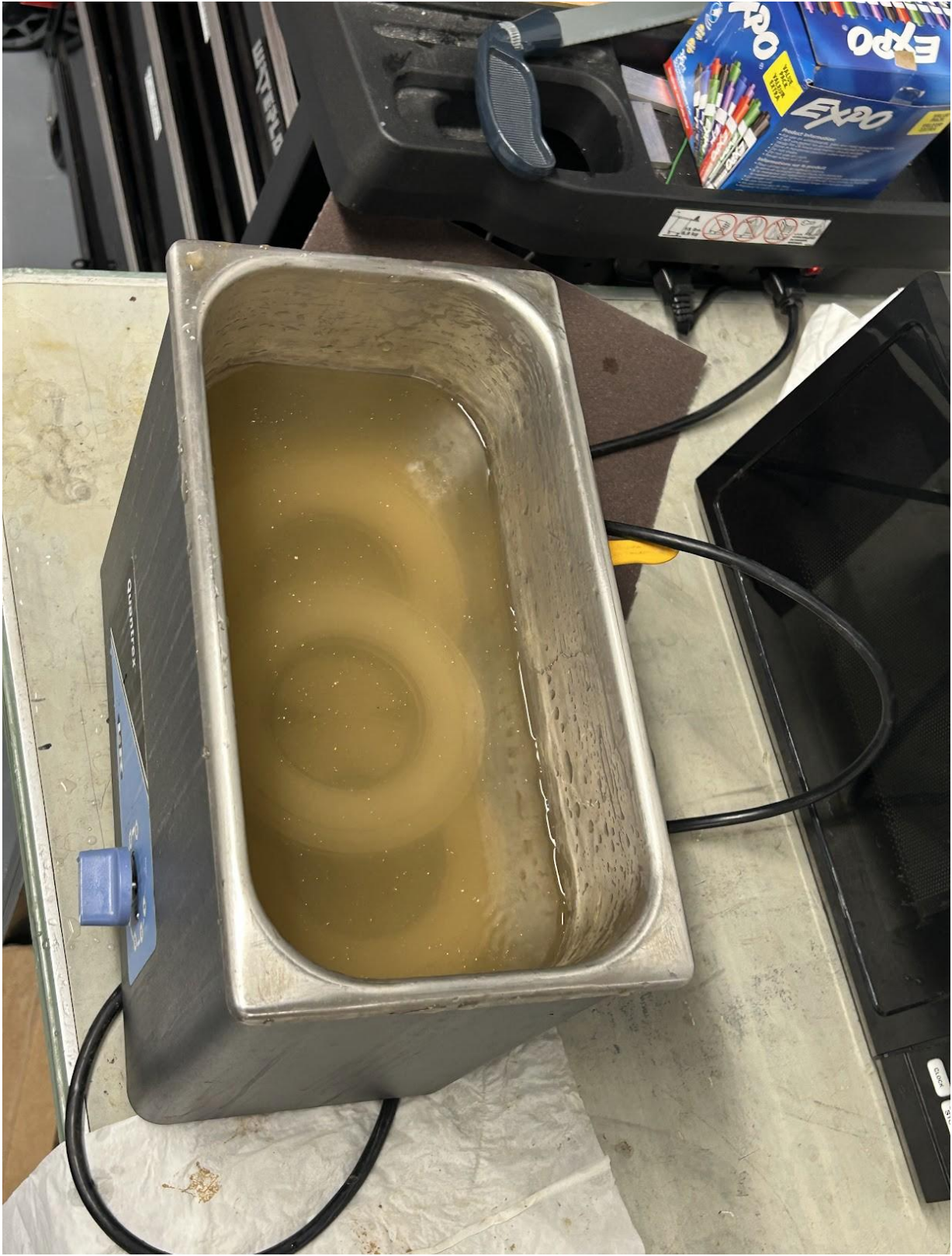

Figure C.2: Several vacuum parts can be dropped in the sonicator for cleaning. For example, here the windows are being cleaned.

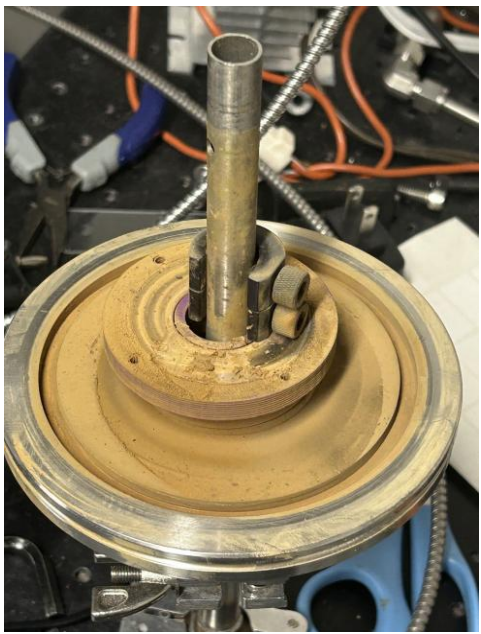 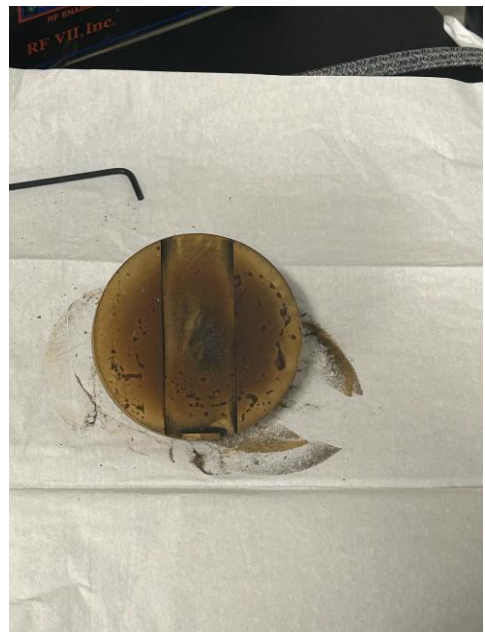

(a) **BEFORE**                    (b) **BEFORE**

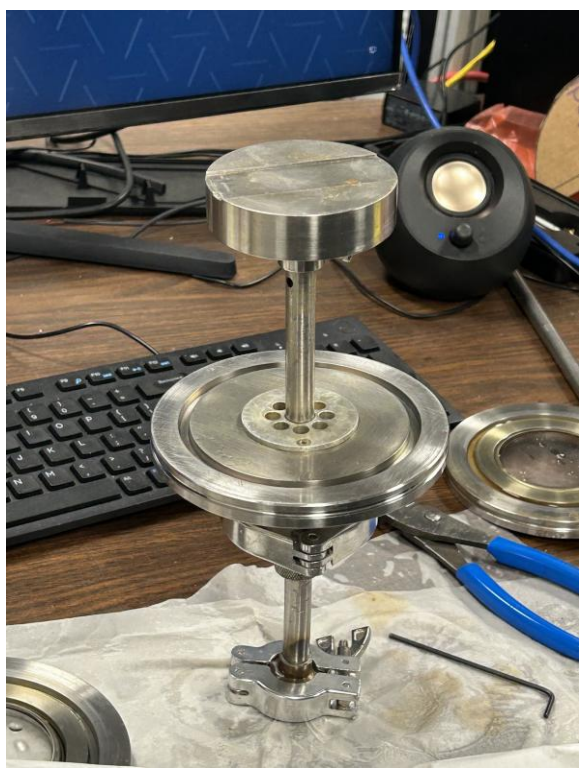

(c) **After**

Figure C.3: Comparison of (a) and (b) electrodes before and (c) after cleaning in the sonicator.

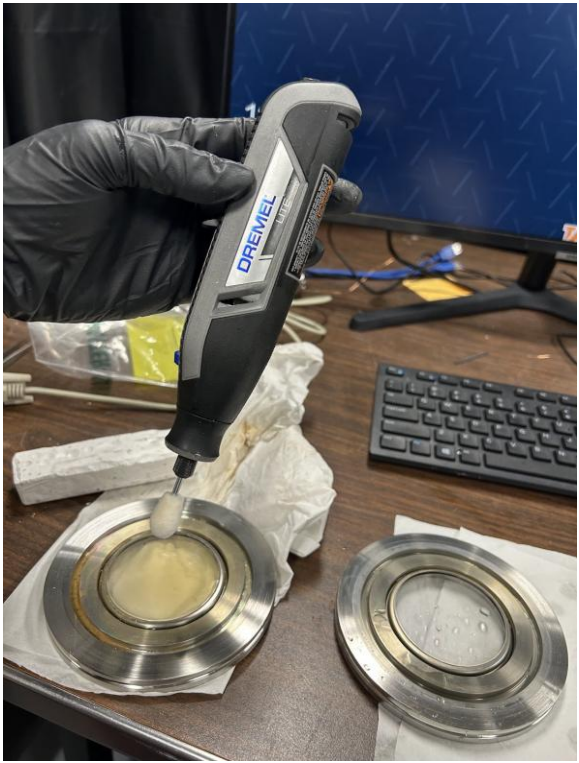 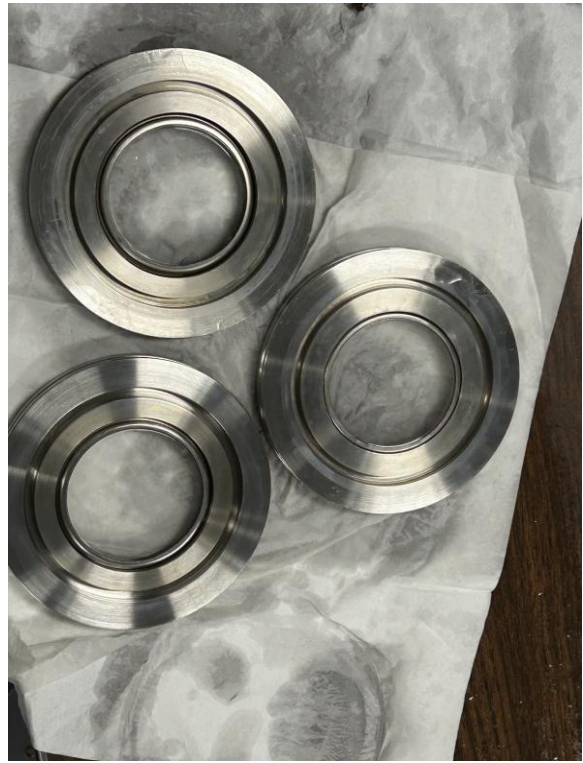

(a) **BEFORE**                    (b) **AFTER**

Figure C.4: Comparison of windows before and after cleaning with DREMEL Lite rotary tool.

# Appendix D

## Simulating Magnetic Fields in Python using Magpylib

```python
# -*- coding: utf-8 -*-
"""
Created on Tue Apr 11 08:45:54 2023

@author: bzr0051
"""

import numpy as np
import matplotlib.pyplot as plt
import magpylib as magpy
from magpylib.magnet import CylinderSegment

# Inputs in imperial
Br_Gauss = 13200  # gauss
OD_inch = 2
ID_inch = 1
thick_inch = 0.25

r_measurement = 0  # inch
z = 12.5
```

```python
# Converting to metric
r = r_measurement * 25.4  # mm
OD_mm = OD_inch * 25.4
ID_mm = ID_inch * 25.4
h = thick_inch * 25.4
Br_Tesla = Br_Gauss / 10000  # gauss
Br_mT = int(Br_Tesla * 1000)  # millitesla
r1 = ID_mm / 2
r2 = OD_mm / 2
bottom_mag_location = -22
bottom_2 = bottom_mag_location - 8
electrode_spacing = 25
top_mag_location = -bottom_mag_location + electrode_spacing
top_2 = top_mag_location + 8

# Sensor position and spacing
position = np.linspace((-40, 0, z), (40, 0, z), 150)  # this is
    the x-axis values
sensor = magpy.Sensor(position)

# Creating rings with CylinderSegment
ring1 = CylinderSegment(magnetization=(0, 0, Br_Gauss), dimension
    =(r1, r2, h, 0, 360), position=(0, 0, bottom_mag_location))
ring3 = CylinderSegment(magnetization=(0, 0, Br_Gauss), dimension
    =(r1, r2, h, 0, 360), position=(0, 0, bottom_2))
ring2 = CylinderSegment(magnetization=(0, 0, Br_Gauss), dimension
    =(r1, r2, h, 0, 360), position=(0, 0, top_mag_location))
ring4 = CylinderSegment(magnetization=(0, 0, Br_Gauss), dimension
    =(r1, r2, h, 0, 360), position=(0, 0, top_2))

# Superposition of two rings B-field
```

```python
Field = magpy.Collection(ring1, ring2, ring3, ring4)
net = sensor.getB(Field).T[2]  # Bx = 0, By = 1, Bz = 2

# Line plot (lineplot)
fig1 = plt.figure(figsize=(9, 7))  # Adjusted height for
    proportionality
lineplot = fig1.add_subplot(111)  # Single plot for the line plot

# Plot field strength
lineplot.plot(position[:, 0], net, ls='--')  # position[:, 0] 0:
    x, 1: y, 2: z
lineplot.grid(color='.9')
lineplot.tick_params(axis='x', labelsize=16)
lineplot.tick_params(axis='y', labelsize=16)

#  Set  axis  labels  and  font  size
lineplot.set_xlabel('x (mm)', fontsize=16)
lineplot.set_ylabel('Field Strength at z = 12.5 mm (Gauss)',
    fontsize=16)

# Save or show the line plot
plt.tight_layout()
plt.savefig("field_strength_plot.png", dpi=300)  # Save plot as a
    PNG file
plt.show()

# Create positions for 2D vectors
xs = np.linspace(-45, 45, 100)
zs = np.linspace(-125, 150, 100)
posis = [[x, r, z] for z in zs for x in xs]
```

```python
# Calculate field and amplitude
B = [Field.getB(pos) for pos in posis]
Bs = np.array(B).reshape([100, 100, 3])  # Reshape
Bamp = np.linalg.norm(Bs, axis=2)

# 2D jet plot (jetplot)
fig2 = plt.figure(figsize=(9, 7))  # Adjusted height to make the
    figure taller
jetplot = fig2.add_subplot(111)  # Single plot for the 2D jet
    plot

# Amplitude plot on jetplot
im = jetplot.pcolor(xs, zs, Bamp, cmap='jet', vmin=-400, vmax
    =1400)

# Plot field lines on jetplot
X, Z = np.meshgrid(xs, zs)
U, V = Bs[:, :, 0], Bs[:, :, 2]
jetplot.streamplot(X, Z, U, V, color='k', density=1)

# Add a colorbar
cbar = fig2.colorbar(im, ax=jetplot)
cbar.set_label('Field Strength', fontname='Arial', fontsize=16)

# Customize colorbar tick labels
cbar.ax.tick_params(labelsize=16)  # Set fontsize for tick labels
for label in cbar.ax.get_yticklabels():
    label.set_fontname('Arial')  # Set fontname for tick labels

# Set axis labels and font size
jetplot.set_xlabel('x (mm)', fontsize=16)
```

```python
jetplot.set_ylabel('z (mm)', fontsize=16)

# Increase font size of tick labels
jetplot.tick_params(axis='x', labelsize=16)
jetplot.tick_params(axis='y', labelsize=16)

# Add a black rectangle for "Top Electrode"
rectangle_top = plt.Rectangle((-37.5, 25), 75, 19, color='black',
    alpha=0.65)  # Rectangle with no transparency
jetplot.add_patch(rectangle_top)

# Add white label inside the rectangle (centered)
jetplot.text(0, 34.5, "Top Electrode", fontsize=16, fontname='
  Arial', color='white',
            ha='center', va='center')  # Center label with white
                color

# Add a  rectangle for "Bottom Electrode"
rectangle_bottom = plt.Rectangle((-37.5, -19), 75, 19, color='
  black', alpha=0.65)  # Rectangle with no transparency
jetplot.add_patch(rectangle_bottom)

# Add white label inside the bottom rectangle (centered)
jetplot.text(0, -9.5, "Bottom Electrode", fontsize=16, fontname='
  Arial', color='white',
            ha='center', va='center')  # Center label with white
                color

# Save or show the 2D jet plot
plt.tight_layout()
```

```
plt.savefig("2D_jet_plot.png", dpi=300)  # Save plot as a PNG
    file
plt.show()
```

## Appendix E

## Fast Fourier Transform of Intensity as a Function of Time

```matlab
clear all
clc
close all;

%% Step 1: Import data from excel
% Load data from Excel file
filename = 'Intensity_vs_time_excel_document.xlsx';
sheetname = 'Sheet1';

% Read the columns from the sheet
data = readtable(filename, 'Sheet', sheetname);

% Extract time and data columns
time = data{:, 1}; % time (seconds) is in the first column
data_values = data{:, 2}; % intensity data is in the second
    column

%% Step 2: Interpolate the data
% Define new time points with constant intervals (0.1 seconds)
new_time = min(time):0.1:max(time);

% Perform spline interpolation
```

```matlab
new_data = interp1(time, data_values, new_time, 'spline');

% Plot original and interpolated data
figure;
plot(time, data_values, 'o', 'LineWidth', 2, 'DisplayName', '
    Original Data'); % Original Data
hold on;
plot(new_time, new_data, 'x', 'DisplayName', 'Interpolated Data (
    Spline)');% Interpolated Data

% Set font properties for axis labels and legend
xlabel('Time (s)', 'FontSize', 20, 'FontName', 'Arial');
ylabel('Intensity (a.u.)', 'FontSize', 20, 'FontName', 'Arial');
legend('show', 'FontSize', 20, 'FontName', 'Arial');

% Set font properties for tick marks
set(gca, 'FontSize', 20, 'FontName', 'Arial');

grid on;
hold off;

%% Step 3: Fast Fourier Transform (FFT) on new data
N = length(new_data); % Length of the signal
Fs = 1 / (new_time(2) - new_time(1)); % Sampling frequency

% Remove the average of all the signal to eliminate DC level
new_data_detrended = new_data - mean(new_data);

% Compute the FFT of the detrended data
Y = fft(new_data_detrended);
```

```matlab
% Frequency vector
frequencies = linspace(0, Fs, N);

% Find peak frequency with maximum amplitude
[~, max_index] = max(abs(Y(1:N/2))); % Only consider half the
    spectrum (positive frequencies)
peak_frequency = frequencies(max_index);

% Calculate magnitude at peak frequency
magnitude_at_peak_frequency = abs(Y(max_index)) / N;

% Display peak frequency and its magnitude
disp(['Peak Frequency: ', num2str(peak_frequency), ' Hz']);
disp(['Magnitude at Peak Frequency: ', num2str(
    magnitude_at_peak_frequency)]);

% Plot the FFT result
figure;
plot(frequencies(1:N/2), abs(Y(1:N/2))/N, 'LineWidth', 2); % Plot
    only the first half (positive frequencies)
xlabel('Frequency (Hz)', 'FontSize', 20, 'FontName', 'Arial');
xlim([0 0.1]); % Limit x-axis to 0.1 Hz
ylabel('Amplitude', 'FontSize', 20, 'FontName', 'Arial');
title('FFT of Interpolated Data', 'FontSize', 20, 'FontName', '
    Arial');
set(gca, 'FontSize', 20, 'FontName', 'Arial');
grid on;
```

## Appendix F

## Plotting the Magnitude of Forces in the Dusty Plasma

```matlab
clear all

close all

clc

% Constants
epsilon_0 = 1e-12 * 8.854;   % Vacuum permittivity [F/m]

k_B = 1.38e-23;              % Boltzmann constant [J/K]

e = 1.602e-19;               % Elementary charge [C]

g = 9.8;                     % Gravity [m/s^2]

% Plasma parameters
T_e = 3 * 11500;             % Electron temperature [K]

T_i = 300;                   % Ion temperature [K]

m_e = 9.11e-31;              % Electron mass [kg]

m_i = 6.6e-26;               % Ion mass [kg]

n_e = 1e15;                  % Electron density [m^-3]

n_d = 1e13;                  % Dust density [m^-3]

E = 120;                     % Electric field [V/m]

% Dust densities
rho_c = 1300;                % Carbonaceous nanoparticles [kg/m^3]

rho_t = 3700;                % Amorphous titania nanoparticles [kg
    /m^3]
```

```matlab
% Radius range
r = logspace(-9, -6, 1000);

% Floating potential phi_s(r)
sqrt_term = sqrt((m_i * T_e) / (m_e * T_i));
phi_s_sol = zeros(size(r));

for i = 1:length(r)
    f = @(phi_s) (1 + (4 * pi * epsilon_0 * r(i) * n_d * phi_s) /
        (e * n_e)) * ...
                 sqrt_term * exp(e * phi_s / (k_B * T_e)) - ...
                 (1 - (e * phi_s / (k_B * T_i))); % Numerically
                    phi_s from nonlinear equation
    phi_s_sol(i) = fzero(f, -1); % Store phi_s(r) for later to
        solve electric force
end

% Electric force
lambda_De = sqrt(epsilon_0 * k_B * T_e / (e^2 * n_e));
                % Electron Debye Length
E_eff = E .* (1 + (r ./ lambda_De) ./ (3 * (1 + r ./ lambda_De)))
   ;  % Effective Electric Field
F_e = 4 * pi * epsilon_0 .* phi_s_sol .* E_eff .* r;
                    % Electric Force

% Gravitational forces
F_g_c = (4/3) * pi * rho_c * g .* r.^3;      % Carbonaceous dust
F_g_t = (4/3) * pi * rho_t * g .* r.^3;      % Titania dust

% Find intersections
```

```matlab
[~, idx_c] = min(abs(abs(F_e) - F_g_c));
[~, idx_t] = min(abs(abs(F_e) - F_g_t));

r_c = r(idx_c);
r_t = r(idx_t);
r_c_nm = r_c * 1e9;
r_t_nm = r_t * 1e9;

% Plot
figure;
loglog(r*1e9, F_g_c, 'b-', 'LineWidth', 2);
hold on;
loglog(r*1e9, F_g_t, 'g-', 'LineWidth', 2);
loglog(r*1e9, abs(F_e), 'r-', 'LineWidth', 2);

xlabel('Particle radius (nm)');
ylabel('Force (N)');
legend({'Gravitational (Carbonaceous)', 'Gravitational (Titania)'
    , 'Electric'}, ...
        'Location', 'best');
grid on;
set(gca, 'FontSize', 12, 'XScale', 'log', 'YScale', 'log');

% Convert data coords to normalized figure coords
ax = gca;
pt1 = data2norm(ax, r_c*1e9, F_g_c(idx_c));
pt2 = data2norm(ax, r_t*1e9, F_g_t(idx_t));

% Arrow offset
offset = 0.05; % arrow distance from point
```

```matlab
% First arrow: Carbonaceous, from southeast (bottom-right)
annotation('textarrow', ...
    [pt1(1)+offset, pt1(1)], [pt1(2)-offset, pt1(2)], ...
    'String', sprintf('˜ %.0f nm', r_c_nm), ...
    'FontSize', 12, 'Color', 'k', 'LineWidth', 1.5, ...
    'HorizontalAlignment', 'center', 'VerticalAlignment', 'top');

% Second arrow: Titania, from northwest (top-left)
annotation('textarrow', ...
    [pt2(1)-offset, pt2(1)], [pt2(2)+offset, pt2(2)], ...
    'String', sprintf('˜ %.0f nm', r_t_nm), ...
    'FontSize', 12, 'Color', 'k', 'LineWidth', 1.5, ...
    'HorizontalAlignment', 'center', 'VerticalAlignment', 'bottom
        ');

%%
% Helper function to convert log data to normalized figure
    coordinates
function norm_coords = data2norm(ax, x, y)
    ax_pos = ax.Position;
    x_log = log10(x);
    y_log = log10(y);
    x_lim = log10(ax.XLim);
    y_lim = log10(ax.YLim);

    x_norm = (x_log - x_lim(1)) / diff(x_lim);
    y_norm = (y_log - y_lim(1)) / diff(y_lim);

    norm_x = ax_pos(1) + x_norm * ax_pos(3);
    norm_y = ax_pos(2) + y_norm * ax_pos(4);
```

```
    norm_coords = [norm_x, norm_y];
end
```

## Appendix G

## Controlling RF Power Supply via RS-232

## G.1    Power Supply ON for pre-Determined Time

```python
# -*- coding: utf-8 -*-
"""
Created on Fri Oct  6 10:44:05 2023

@author: bzr0051
"""

import serial
import time

Tc = 46 #The cycle time
s= 1 #quarter of Tc (1: 1/4 *Tc, 2: 2/4 *Tc, 3: 3/4 *Tc, 4: 4/4 *
    Tc)

# Define the COM port and baud rate (adjust as needed)
com_port = 'COM5'
baud_rate = 9600  # Adjust to match your device's baud rate

try:
    # Open the serial port
```

```python
    ser = serial.Serial(com_port, baud_rate, timeout=1)

    # Send the "HI" command
    ser.write(b'HI\r\n')
    time.sleep(1)

    # Send the "P 60" command
    ser.write(b'P_60\r\n')
    time.sleep(1)

    # Send the "RFON" command
    ser.write(b'RFON\r\n')
    time.sleep(s*Tc/4) #Power is on for s/4 x Cycle time

    # Send the "RFOFF" command to turn off power for 5 seconds
    ser.write(b'RFOFF\r\n')
    time.sleep(1)  # Power off

    # Send the "BYE" command
    ser.write(b'BYE\r\n')
    time.sleep(1)

    # Read and print the response from the device if needed
    response = ser.read(100)  # Read up to 100 bytes (adjust as
        needed)
    print("Response:", response.decode('utf-8'))

except serial.SerialException as e:
    print("Serial_port_error:", str(e))
finally:
```

```python
    # Close the serial port when done
    ser.close()
```

## G.2    Pulse Power Supply

```python
# -*- coding: utf-8 -*-
"""
Created on Fri Oct  6 10:44:05 2023

@author: bzr0051
"""

import serial
import time

# Define the COM port and baud rate (adjust as needed)
com_port = 'COM5'
baud_rate = 9600  # Adjust to match your device's baud rate

try:
    # Open the serial port
    ser = serial.Serial(com_port, baud_rate, timeout=1)

    # Send the "HI" command to enter RS-232 mode
    ser.write(b'HI\r\n')
    time.sleep(0.2)

    #set the power to 30 W
    ser.write(b'P_30\r\n')
    time.sleep(0.2)
```

```python
    # Initialize a count variable
    count = 0

    # Loop to perform the operation 20 times
    for _ in range(10):
        # Increment the count
        count += 1

        # Send the "RFON" command to turn on power for 300
            milliseconds
        ser.write(b'RFON\r\n')
        print(f"Power_turned_ON,_Count:_{count}")
        time.sleep(70)  # Power on for 70 seconds

        # Send the "RFOFF" command to turn off power for 5
            seconds
        ser.write(b'RFOFF\r\n')
        time.sleep(10)  # Power off for 10 seconds

except serial.SerialException as e:
    print("Serial_port_error:", str(e))
finally:
    # Send the "BYE" command to exit RS-232 mode
    ser.write(b'BYE\r\n')
    time.sleep(1)

    # Close the serial port when done
    ser.close()

print("The_end")
```